\documentclass[11pt]{article}
\usepackage{amsfonts}

\usepackage{graphics}
\usepackage{indentfirst}
\usepackage{cite}
\usepackage{latexsym}
\usepackage{amsmath}
\usepackage{amssymb}
\usepackage[dvips]{epsfig}
\usepackage{amscd}

\newtheorem{theorem}{Theorem}[section]
\newtheorem{remark}{Remark}[section]
\newtheorem{definition}{Definition}[section]
\newtheorem{lemma}[theorem]{Lemma}
\newtheorem{pro}{Proposition}[section]
\newtheorem{cor}[theorem]{Corollary}

\renewcommand{\div}{ {\rm div }  }
\newcommand{\tu}{1}

\newcommand{\re}{\rho_{0,\eta}}

\newcommand{\tet}{\te_{0,\eta}}

\newcommand{\rd}{ \rho^{\de,\eta} }
\newcommand{\ud}{ u^{\de,\eta} }
\newcommand{\td}{ \theta^{\de,\eta}}

\newcommand{\htn}{ \hat\rho^{\de,\eta} }
\newcommand{\htu}{ \hat u^{\de,\eta} }
\newcommand{\hte}{\hat \theta^{\de,\eta}}

\newcommand{\hn}{  \rho^{\de} }
\newcommand{\hu}{   u^{\de} }
\newcommand{\he}{  \theta^{\de}}

\newcommand{\na}{\nabla }

\newcommand{\pa}{\partial}
\renewcommand{\r}{\mathbb{R}}

\newcommand{\bi}{\bibitem}

\newcommand{\dis}{\displaystyle}
\newcommand{\ia}{\int_0^T}

\newcommand{\bl}{\begin{lemma}}
\newcommand{\el}{\end{lemma}}
\newcommand{\et}{\end{theorem}}
\newcommand{\ga}{\gamma}

\newcommand{\te}{\theta}

\newcommand{\al}{\alpha}
\newcommand{\de}{\delta}
\newcommand{\ve}{\varepsilon}
\newcommand{\la}{\label}
\newcommand{\dl}{\Delta}
\newcommand{\p}{p(\rho)  }

\newcommand{\ka}{\kappa}

\newcommand{\bn}{\begin{eqnarray}}
\newcommand{\en}{\end{eqnarray}}
\newcommand{\bnn}{\begin{eqnarray*}}
\newcommand{\enn}{\end{eqnarray*}}

\newcommand{\bnnn}{\begin{eqnarray*}}
\newcommand{\ennn}{\end{eqnarray*}}

\newcommand{\ba}{\begin{aligned}}
\newcommand{\ea}{\end{aligned}}
\newcommand{\be}{\begin{equation}}
\newcommand{\ee}{\end{equation}}
\def\O{\r^3}
\def\p{\partial}
\def\norm[#1]#2{\|#2\|_{#1}}

\def\o{\omega}

\newcommand{\no}{\nonumber\\}
\newcommand{\ep}{\varepsilon}
\newcommand{\n}{\rho}
\newcommand{\si}{\sigma}

\def\la{\label}

\def\na{\nabla}
\def\on{\bar\rho}

\def\tn{1}

\def\tp{\tilde{P}}

\makeatletter      
\@addtoreset{equation}{section}
\makeatother       

 \title{  Global   Classical and Weak Solutions to the Three-Dimensional Full
Compressible Navier-Stokes System with Vacuum and Large
Oscillations \thanks{This research   is partially supported  by SRF for ROCS, SEM, and
National
Science Foundation of China under  grant 10971215.
 Email: xdhuang@ustc.edu.cn (X. Huang), ajingli@gmail.com (J. Li).
 }}

\author{Xiangdi H{\small UANG}$^{a}$,  Jing L{\small I}$^{b}$  \\[3mm] {\normalsize $^a$ Department of Mathematics,} \\
{\normalsize University of Science and Technology of China, Hefei
230026, P. R. China} \\[2mm]
{\normalsize $^b$ Institute of Applied Mathematics, AMSS,} \\ {\normalsize \&   Hua Loo-Keng Key Laboratory of Mathematics,}\\
{\normalsize  Chinese Academy of Sciences,    Beijing 100190,
P. R. China}
 }

\date{}
\begin{document}
 \maketitle

 \begin{abstract}
We establish the global existence and uniqueness of classical
solutions to the three-dimensional full  compressible Navier-Stokes
system with smooth initial data which are of small energy  but
possibly large oscillations  where  the initial density is allowed
to vanish.  Moreover, for the  initial   data, which may be
discontinuous and contain  vacuum states, we also obtain the global
existence of weak solutions. These results generalize previous ones
on classical and  weak solutions for initial density being strictly
away from vacuum, and are the first for global classical  and  weak
solutions which may have large oscillations and can contain vacuum
states.

 \end{abstract}

\section{Introduction}

The motion of compressible viscous, heat-conductive, Newtonian polytropic fluid   occupying a  spatial domain
$\Omega\subset \r^3$ is governed by the following full  compressible Navier-Stokes system:
 \be \la{a0}  \begin{cases}\n_t+{\rm div} (\n u)=0,\\
 (\n u)_t+{\rm div}(\n u\otimes u)-\mu\Delta u-(\mu+\lambda)\na({\rm
div} u)+\na P=0,\\ (\n E)_t+\div (\n Eu+Pu)=\Delta
\left(\ka \te+\frac{1}{2}\mu |u|^2\right) +\mu \div(u\cdot \na u
)+\lambda \div(u\div u ).
\end{cases}\ee
Here $t\ge 0$ is time, $x\in \Omega$ is the spatial coordinate, and
 $\n, u=\left(u^1,u^2,u^3\right)^{\rm tr},e, P(\n,e) ,$ and
$\te $ represent respectively the fluid density, velocity, specific
internal energy, pressure, and  absolute temperature, and
$E=e+\frac{1}{2}|u|^2$ is the specific total energy.
 The constant
viscosity coefficients $\mu$ and $\lambda$  satisfy the physical
restrictions:
 \be\la{h3} \mu>0,\quad 2 \mu + 3\lambda\ge 0.
\ee The equations (\ref{a0}) then express respectively the
conservation of mass, the balance of momentum, and the balance of
energy  under internal pressure, viscosity forces, and the
conduction of thermal energy. We study the ideal   polytropic fluids
so that $P$ and $e$ are given by the state equations:  \be
P(\n,e)=(\ga-1)\n e=R\n \te, \quad e=\frac{R\theta}{\ga-1} ,   \ee
where
  $\ga>1$  is the
adiabatic constant, and $ R, \ka$ are both positive
constants.

 Let $\Omega=\r^3 $ and    $\tilde\n,\tilde\te$   both be fixed positive
constants.
 We look for the solutions, $(\n(x,t),u(x,t),\te(x,t)),$ to the Cauchy problem
for (\ref{a0}) with  the far field behavior:\be\la{h1}
(\n,u,\te)(x,t)\rightarrow (\tilde\n,0,\tilde\te) , \quad\mbox{ as
}\, |x|\rightarrow\infty,\,\, t>0,\ee and initial data: \be \la{hh2}
(\rho,\n u,\n \te)(x,{t=0})=(\rho_0,\n_0 u_0, \n_0\te_0)(x),\quad
x\in \r^3,  \ee with $\n_0  \ge 0,\te_0 \ge 0.$  Note  here that for
classical solutions, \eqref{a0} can be rewritten as
 \be \la{a1}  \begin{cases}\n_t+{\rm div} (\n u)=0,\\
\n (u_t+  u\cdot \na u )=\mu\Delta u+(\mu+\lambda)\na({\rm div}
u)-\na P  ,\\ \frac{R}{\ga-1}\n  ( \te_t+u\cdot\na \te)=\ka\Delta
\te-P\div u+\lambda (\div u)^2+2\mu |\mathfrak{D}(u)|^2,
\end{cases}\ee where $ \mathfrak{D}(u)$ is the deformation tensor:
\bnn   \mathfrak{D}(u)  = \frac{1}{2}(\nabla u + (\nabla u)^{\rm
tr}). \enn Moreover, for classical solutions,  we  replace the
initial condition (\ref{hh2})  with \be \la{h2} (\rho,u,
\te)(x,{t=0})=(\rho_0,u_0,\te_0),\quad x\in \r^3.  \ee

There is a lot  of literature  on the large time existence and
behavior of solutions to (\ref{a0}). The one-dimensional problem
with strictly positive initial density and temperature has been
studied extensively by many people, see \cite{kazh01,Kaz,akm} and
the references therein. For the multi-dimensional case, the local
existence and uniqueness of classical solutions are known in
\cite{Na,se1}  in the absence of vacuum.  Recently, for the case
that the initial density need not be positive and may vanish in open
sets,  Cho-Kim \cite{choe1} obtained the   local existence and
uniqueness of strong solutions.  The global classical solutions were
first obtained by Matsumura-Nishida \cite{M1} for initial data close
to a non-vacuum equilibrium in some Sobolev space $H^s.$ In
particular, the theory requires that the solution has small
oscillations from a uniform non-vacuum state so that the density is
strictly away from   vacuum and the gradient of the density remains
bounded uniformly in time. Later, Hoff \cite{Hof1} studied the
global weak solutions with strictly positive initial density and
temperature for discontinuous initial data. On the other hand, in
the presence of vacuum, this issue becomes much more complicated.
Concerning viscous compressible fluids in a barotropic regime, where
the state of these fluids at each instant $t>0 $ is completely
determined by the density $\n=\n(x,t)$ and the velocity $u=u(x,t),$
the pressure $P$ being an explicit function of the density,    the
major breakthrough is due to Lions \cite{L1} (see also  Feireisl
\cite{F1,feireisl1}), where he obtained global existence of weak
solutions, defined as solutions with finite energy,  when the
pressure $P(\n)=a\n^\ga (a>0,\ga>1)$  with suitably large  $\ga.$
The main restriction on initial data is that the initial energy  is
finite, so that the density vanishes at far fields, or even has
compact support.
 Recently, Huang-Li-Xin \cite{hulx} established the global existence and uniqueness of classical
solutions to the Cauchy problem for the isentropic compressible Navier-Stokes equations in three-dimensional space with smooth initial data which are of small energy  but possibly large oscillations; in particular, the initial density is allowed to vanish, even has compact support. This result can be regarded as uniqueness and
 regularity theory of Lions-Feireisl's weak solutions  in \cite{L1,F1,feireisl1} with small initial energy.

However, the global
well-posedness of classical solutions, even the global existence  of weak solutions to (\ref{a0})  remains completely open in the presence of vacuum. For specific pressure laws   excluding the perfect gas
equation of state, the question of existence of so-called
``variational" solutions in dimension $d\ge 2$ has been recently
 addressed
in \cite{feireisl,feireisl1}, where the temperature equation is
satisfied only as an inequality  which justifies the notion of
variational solutions. Let us emphasize that this work is the very
first attempt towards the existence of weak solutions to the full
compressible Navier-Stokes system for large initial data with
vacuum. Recently, for   a very particular form of the viscosity
coefficients depending on the density, Bresch-Desjardins
\cite{bd} obtained global stability of weak solutions. It  is worth
noting  here that   {Xin} \cite{X1} first established remarkable
blow-up results, which
  show  that in the case that the initial density has compact support, any smooth solution to the Cauchy problem of the full compressible Navier-Stokes
system without heat conduction blows up in finite time.  See also the recent generalizations to the
cases for the full compressible Navier-Stokes system with heat conduction (\cite{cj})
and for non-compact but rapidly decreasing at far field initial densities (\cite{R}).

Motivated by our previous work  on the isentropic compressible Navier-Stokes equations \cite{hulx},
  we try to look for the global existence of classical and weak solutions   to the three-dimensional full  compressible Navier-Stokes system \eqref{a0}; in particular, the initial density is allowed to vanish.

Before stating the main results, we explain the notations and
conventions used throughout this paper.  We denote
$$\int fdx=\int_{\r^3}fdx.$$   For $1\le p\le \infty $ and
 integer $k\ge 0,$   we adopt the   simplified notations for
 the standard homogeneous and inhomogeneous Sobolev spaces as
follows:
   \bnnn \begin{cases} L^p=L^p(\r^3),\quad W^{k,p}=W^{k,p}(\r^3),\quad
   H^k=W^{k,2},\\
  D^1   = \left. \left\{u\in L^6 \,\right|
 \|{\nabla u}\|_{L^2}<\infty \right\},\quad  D^{1,p}  =
   \left.\left\{u\in
L^1_{loc}(\r^3)\,\right|\|{\nabla  u}\|_{L^p}<\infty\right\} .
 \end{cases}\ennn

Without loss of generality, we    assume that $\tilde\n=\tilde\te=1.$ We  define  the initial energy  $C_0$  as follows:
\be\la{e}\ba  C_0\triangleq &\frac{1}{2}\int\n_0
|u_0|^2dx+R \int  \left( \n_0\log {\n_0}-\n_0+1 \right)dx\\
&+\frac{R}{\ga-1}\int  \n_0\left(\te_0- \log {\te_0} -1
\right) dx+\frac{R}{2(\ga-1)}\int  \n_0 (\te_0-1)^2 dx.\ea\ee
Then the first  main result  in this paper  can be stated  as follows:
\begin{theorem}\la{th1}   For  given numbers $M>0$ (not necessarily
small), $q\in (3,6),$
  and $\on> 2,$
suppose that the initial data $(\n_0,u_0,\te_0)$ satisfies \be
\la{co3} \n_0-1\in H^2\cap W^{2,q}, \quad  u_0 \in H^2 ,\quad
\te_0-1\in H^2 , \ee \be
 \la{co4} 0\le\inf\rho_0\le\sup\rho_0<\bar{\rho},\quad \inf\te_0\ge 0, \quad \|\na u_0\|_{L^2} \le M,
   \ee  and the compatibility conditions: \be
\la{co2}-\mu \Delta u_0-(\mu+\lambda)\na\div u_0+R\na (\n_0\te_0)
=\sqrt{\n_0} g_1,\ee \be\la{co1} \ka\Delta \te_0+\frac{\mu}{2}|\na
u_0+(\na u_0)^{\rm tr}|^2+\lambda (\div u_0)^2=\sqrt{\n_0}g_2,\ee
with  $g_1,g_2\in L^2. $ Then there exists a positive constant $\ve$
depending only
 on $\mu,$ $\lambda,$ $ \ka,$ $ R,$ $ \ga,$  $\on,$ and $M$ such that if
 \be
 \la{co14} C_0\le\ve,
   \ee  the Cauchy problem  (\ref{a1}) (\ref{h1}) (\ref{h2})
  has a unique global classical solution $(\rho,u,\te)$ in
   $\r^3\times(0,\infty)$ satisfying
  \be\la{h8}
  0\le\rho(x,t)\le 2\bar{\rho},\quad \te(x,t)\ge 0,\quad x\in \r^3,\, t\ge 0,
  \ee
 \be
   \la{h9}\begin{cases}
   \rho-1 \in C([0,T];H^2\cap W^{2,q}),\quad ( u,\te-1) \in C([0,T];H^2), \\
  u \in   L^\infty(\tau,T;H^3\cap W^{3,q}),\quad  \te-1 \in   L^\infty(\tau,T;H^4), \\
  (u_t,\te_t)\in
L^{\infty}(\tau,T;H^2)\cap H^1(\tau,T;H^1), \end{cases} \ee
   and the following large-time behavior:
  \be\la{h11}
  \lim_{t\rightarrow \infty}\left( \|\n(\cdot,t)-1\|_{L^p} +\|\na u(\cdot,t)\|_{L^r }+\|\na \te(\cdot,t)\|_{L^r }\right)=0,  \ee
   with any    \be\la{eq1} 0<\tau<T<\infty,\quad p\in  (2 ,\infty),\quad r\in [2,6).
     \ee

\end{theorem}

The next result of this paper will treat the weak solutions due to
the fact that discontinuous solutions are fundamental both in the
physical theory of nonequilibrium thermodynamics as well as in the
mathematical theory of inviscid models for compressible fluids. To
begin with, we   give the definition of weak solutions.
\begin{definition}\la{def} We say that $(\n,u,E=\frac{1}{2}|u|^2+\frac{R}{\ga-1}\te)$ is a weak solution to Cauchy problem (\ref{a0}) (\ref{h1}) (\ref{hh2})
 provided that $$\n-1\in L^\infty_{\rm
loc}([0,\infty);L^2\cap L^\infty(\r^3)),\quad u ,\te-1 \in L^2 (
0,\infty; H^1(\r^3)),$$ and that for all test functions $\psi\in
\mathcal{D}(\r^3\times(-\infty,\infty)),$ \be \la{def1}
\int_{\r^3}\n_0\psi(\cdot,0)dx+\int_0^\infty\int_{\r^3}\left(
\n\psi_t+\n u\cdot\na\psi\right) dxdt=0,\ee   \be\la{def2}\ba&
\int_{\r^3}\n_0u^j_0\psi(\cdot,0)dx+\int_0^\infty\int_{\r^3}\left(
\n u^j\psi_t+\n u^ju\cdot\na\psi+P(\n,\te)\psi_{x_j}\right)
dxdt\\&-\int_0^\infty\int_{\r^3}\left(  \mu\na
u^j\cdot\na\psi+(\mu+\lambda)(\div u)\psi_{x_j}\right) dxdt=0,\quad
j=1,2,3, \ea\ee

\be \la{def3}\ba &
\int_{\r^3}\left(\frac{1}{2}\n_0|u_0|^2 +\frac{R}{\ga-1}\n_0\te_0 \right)\psi(\cdot,0)dx\\ &= \int_0^\infty\int_{\r^3}\left(\n E\psi_t+ (\n E  +P)u\cdot\na\psi\right) dxdt\\ &\quad -\int_0^\infty\int_{\r^3}\left(\ka\na
\te+\frac{1}{2}\mu \na(|u|^2) +\mu  u\cdot \na u
 +\lambda  u\div u   \right)\cdot\na \psi dxdt.\ea\ee

\end{definition}
We also define \be \la{hj1}\dot f\triangleq f_t+u\cdot\nabla f,\quad
G\triangleq(2\mu + \lambda)\div u - R(\n\te-1) ,\quad\o
\triangleq\na\times u, \ee which are  the material derivative of
$f,$ the effective viscous flux, and the vorticity respectively. We
now state our second main result as follows:
\begin{theorem}\la{th2}   For  given numbers $M>0$ (not necessarily
small),    and $\on>2,$ there exists a   positive constant  $\ve$
  depending only
 on $\mu,$ $\lambda,$ $ \ka,$ $ R,$ $ \ga,$  $\on,$ and $M$ such that
  if  the initial data $(\n_0,u_0,\te_0)$ satisfies (\ref{co4})
   and
 \be
 \la{cco14}  C_0\le\ve,
   \ee with $C_0$  as in (\ref{e}), there is a global weak solution $(\rho,u,E=\frac{1}{2}|u|^2+\frac{R}{\ga-1}\te)$ to the Cauchy problem  (\ref{a0}) (\ref{h1}) (\ref{hh2})
    satisfying
    \be\la{hq1}
  \n-1\in C([0,\infty);L^2\cap L^p), \quad (\n u,\,\n |u|^2,\,\n (\te-1 ))\in C([0,\infty);H^{-1}) ,
  \ee \be\la{hq2}u\in C((0,\infty);L^2 )  ,\quad \te-1\in C((0,\infty); W^{1,r}),
  \ee \be \la{hq3}u(\cdot,t),\,\,\o(\cdot,t),\,\,G(\cdot,t),  \,\,\na\te(\cdot,t)     \in H^1,\quad t>0,\ee
  \be\la{hq4}
   \rho\in [0,2\bar{\rho}] \quad \mbox{ \rm a.e.}, \quad \te\ge 0 \quad\mbox{ \rm a.e.},
  \ee
 and the following large-time behavior:
\be \la{lar1} \lim_{t\rightarrow \infty}\left( \|\n(\cdot,t)-1\|_{L^p}+\|u(\cdot,t)\|_{L^p\cap L^\infty}+\|\na \te(\cdot,t)\|_{L^r }\right)=0,\ee  with any $p,r$ as in \eqref{eq1}. In addition, there exists some  positive constant $C$ depending  only on $ \mu,$ $\lambda,$ $ \ka,$ $ R,$ $ \ga,$  $\on, $ and $M,$ such that, for  $\si(t)\triangleq\min\{1,t\},$  the following estimates hold
  \be\la{hq5}\ba  \sup_{t\in (0,\infty)} \| u \|_{H^1}  +\int_0^\infty\int\left| (\n u)_t+{\rm div}(\n u\otimes u)\right|^2dxdt\le C,\ea\ee
  \be\la{hq7}\ba &\sup_{t\in (0,\infty)} \int\left((\n-1)^2+\n |u|^2+\n(\te-1)^2\right)dx  \\&\quad   +\int_0^\infty\left( \|\na u\|^2_{L^2}+ \|\na \te\|_{L^2}^2 \right)dt\le CC_0^{1/4}, \ea\ee
   \be\la{hq8}\ba &\sup_{t\in (0,\infty)}\left(\si^2\|\na u \|^2_{L^6}+\si^4\|\te-1\|^2_{H^2}\right) \\&  +\int_0^\infty\left( \si^2\|u_t\|^2_{L^2}+\si^2\|\na \dot u\|_{L^2}^2+\si^4\|\te_t\|^2_{H^1}\right)dt\le CC_0^{1/8}. \ea\ee
Moreover, $(\n,u,\te)$ satisfies $(\ref{a1})_3$ in the weak form, that is, for any test function  $\psi\in
\mathcal{D}(\r^3\times(-\infty,\infty)),$
\be\la{vu019}\ba &\frac{R}{\ga-1} \int\n_0 \te_0\psi(\cdot,0) dx+\frac{R}{\ga-1}\int_0^\infty\int\n \te \left(\psi_t+u \cdot\na\psi
\right)dxdt\\ & =  \ka \int_0^\infty\int \na\te \cdot\na\psi dxdt+R\int_0^\infty\int  \n\te  \div u \psi dxdt\\&\quad -\int_0^\infty\int \left(\lambda(\div u)^2+2\mu   |\mathfrak{D}(u)|^2  \right) \psi dxdt.\ea\ee

\end{theorem}

The following  Corollary \ref{th3}, whose proof can be found in     \cite[Theorem 1.2]{hulx},  shows that  we can obtain from (\ref{h11}) the
following large time behavior of the gradient of the density when
vacuum states appear initially,  which is completely in contrast to the classical
theory (\cite{M1}).
\begin{cor} [\cite{hulx}]\la{th3}
In addition to the conditions  of Theorem \ref{th1}, assume further
that there exists some point $x_0\in \r^3$ such that $\rho
_0(x_0)=0.$ Then  the unique global    classical solution
$(\rho,u,\te)$ to the Cauchy problem (\ref{a1}) (\ref{h1}) (\ref{h2})
obtained in Theorem \ref{th1} has to blow up as $t\rightarrow
\infty,$ in the sense that for any $r>3,$
$$\lim\limits_{t\rightarrow \infty}\|\nabla \rho(\cdot,t)
\|_{L^r }=\infty.$$
\end{cor}

A few remarks are in order:

\begin{remark} It follows from (\ref{h9}) that, for any
 $0<\tau<T<\infty,$
\be \la{hk1} (\n-\tn,\,\, \na\n,\,\,  u,\,\, \te-\tu ) \in C(\overline{\r^3} \times[0,T] ),\ee
and \be  \la{hk2} \na u,\,\, \na^2u\in C( [\tau,T];L^2 )\cap L^\infty(\tau,T;  W^{1,q})\hookrightarrow  C(\overline{\r^3} \times[\tau,T] ),\ee
which together with ${\it (\ref{a1})_1}$ and (\ref{hk1}) gives
\be  \la{hk3} \n_t\in   C(\overline{\r^3}\times[\tau,T] ).\ee
Similarly, we deduce from (\ref{h9}) that
\bnn\na \te,\,\, \na^2\te\in C( [\tau,T];H^1 )\cap L^\infty(\tau,T;
  H^2)\hookrightarrow  C(\overline{\r^3}\times[\tau,T] ),\enn which
  combining with (\ref{hk1})--(\ref{hk3}) thus shows that  the
   solution $(\n,u,\te)$ obtained in Theorem \ref{th1} is  in fact a classical
    one to  the Cauchy problem  (\ref{a1}) (\ref{h1}) (\ref{h2})
     in $\r^3\times (0,\infty).$ Although  it has small energy, yet
its oscillations could be arbitrarily large. In particular, initial vacuum states are allowed.
 \end{remark}

\begin{remark} Theorem \ref{th1}   is the first result concerning the global existence of classical   solutions with vacuum to the  full compressible Navier-Stokes system. Moreover, the conclusions in Theorem \ref{th1} generalize the classical theory of Matsumura-Nishida (\cite{M1}) to the case of large oscillations since in this case, the requirement of small
energy, (\ref{co14}), is equivalent to smallness of the mean-square norm of $(\n_0-1,u_0,\te_0-1).$ In addition, the initial density is allowed to vanish and the initial temperature  may be zero.    However, although the large-time
asymptotic behavior (\ref{h11}) is similar to that in \cite{M1},
yet our solution may contain vacuum states, whose appearance leads
to the large time blowup behavior stated in Corollary \ref{th3},
this is in sharp contrast to that in \cite{M1} where the
gradients of the density are suitably small uniformly for all
time.
\end{remark}

\begin{remark}It is worth noting that
 the conclusions in
Theorem \ref{th1} rely heavily on the positivity of both $\tilde\n$ and $\tilde\te$ in (\ref{h1}) which prevents  the density from being compactly supported or  decaying  for large values of the spatial variable $x.$   Indeed, any smooth solution will blow up in finite time  provided that it has compactly supported   initial density (\cite{X1,cj})  or that it  and   its spatial derivatives decay fast enough for large values of the spatial variable $x$(\cite{R}). Therefore, it would be interesting to study the global existence and large time asymptotic
behavior of solutions for the case that   initial data decay  slowly enough for large values of the spatial variable $x.$ This is left for the future.
\end{remark}

\begin{remark} It should be noted here that Theorem \ref{th2} is  the first result concerning the
global existence of  weak solutions to (\ref{a0}) in the presence of vacuum and extends the
global weak solutions  of Hoff  (\cite{Hof1}) to the case of
large oscillations and non-negative initial density.   Moreover, the initial temperature is allowed to be zero.
\end{remark}

\begin{remark} It follows from (\ref{hq5}) and Sobolev's embedding theorem that $u$ and $\theta$ are  in fact
  H\"older continuous away from $t=0,$ that is, for any $0<\tau<\infty,$
 \be \la{hq6} \sup_{t\in [\tau,\infty)}\|u\|_{L^\infty}+\langle u \rangle_{\r^3\times [\tau,\infty)}^{1/2,1/8}+\sup_{t\in [\tau,\infty)}\|\te\|_{L^\infty}+\langle \te\rangle_{\r^3\times [\tau,\infty)}^{1/2,1/8}<\infty, \ee
where we employ the usual notation for H\"{o}lder norms:
\bnn \langle w\rangle_Q^{1/2,1/8}=\sup\limits_{ \begin{subarray}{c} (x,t),(y,s)\in Q   \\ (x,t)\not=(y,s)\end{subarray}  } \frac{|w(x,t)-w(y,s)|}{|x-y|^{1/2}+|t-s|^{1/8}} ,\enn
for functions $w:Q  \subseteq \r^3\times [0,\infty)\rightarrow \r^m.$
\end{remark}

\begin{remark} Simple calculations yield  that if \be\la{te1}\sup\limits_{x\in \r^3}\te_0(x)\le \bar\te,\ee
we have \bnn \int  \n_0 (\te_0-1)^2 dx \le 2(\bar\te+1)\int  \n_0\left(\te_0- \log {\te_0} -1
\right) dx,
\enn
which implies $\tilde C_0\le C_0\le  (\bar\te+2)\tilde C_0,$
where
\bnn \ba  \tilde C_0\triangleq &\frac{1}{2}\int\n_0
|u_0|^2dx+R \int  \left( \n_0\log {\n_0}-\n_0+1 \right)dx \\&
 +\frac{R}{\ga-1}\int  \n_0\left(\te_0- \log {\te_0} -1
\right) dx \ea\enn
is the usual initial   energy.  In other words,  if we replace $C_0$ with the usual initial   energy    $\tilde{C_0},$  the $\ep$ in  Theorems \ref{th1} and \ref{th2} will also depend on the upper bound of the initial temperature.

 \end{remark}

\begin{remark}   Similar ideas can be applied to study the case
on bounded domain. This will be reported in a forthcoming paper
\cite{hlx4}.
\end{remark}

We now comment on the analysis of this paper. Note that though
 the local existence and uniqueness of strong solutions to
  (\ref{a1})  in the presence of vacuum
  was obtained by Cho-Kim (\cite{choe1}),
   the local existence of classical solutions with vacuum to (\ref{a1})   still remains unknown.
Some of the main new difficulties to obtain the classical
solutions to  (\ref{a1}) (\ref{h1}) (\ref{h2}) for initial data in the class satisfying   (\ref{co3})--(\ref{co1})   are due to the appearance of vacuum. Thus, we take the  strategy that we first  extend   the standard   local classical solutions with strictly positive initial density   (see Lemma \ref{th0}) globally in time just under the condition that the initial energy  is suitably small (see Proposition \ref{pro2}), then let the lower bound  of  the initial density go to zero. To  do so, one needs to  establish global a priori estimates, which are  independent of the lower bound of the density,  on
smooth solutions to  (\ref{a1}) (\ref{h1}) (\ref{h2}) in suitable higher norms.
 It turns out that the key issue in this paper is to derive both
  the time-independent upper bound for the density and the
   time-dependent higher norm estimates of the smooth solution $(\rho, u,\theta)$.
Compared to the isentropic case (\cite{hulx}), the first main
difficulty lies in the fact that  the basic  energy estimate  cannot
yield directly the bounds on the   $L^2$-norm (in both time and
space) of the spatial derivatives of both the  velocity and the
temperature  since the super norm of the temperature is just assumed
to satisfy the a priori   bound  $(\min\{1,t\})^{-3/2}$ (see
\eqref{z1}), which in fact could be arbitrarily large for small
time. To overcome this difficulty,  based on   careful analysis on
the basic energy estimate,  we succeed in deriving a new estimate of
the  temperature which shows that the spatial  $L^2$-norm  of the
deviation of the temperature from its far field value can be bounded
by the combination of the initial energy  with the spatial
$L^2$-norm of the spatial derivatives of the temperature (see
(\ref{a2.17})). This estimate, which will play  a crucial role  in
the analysis of this paper, together with elaborate analysis on the
bounds of the energy, then yields  the key energy-like estimate,
provided that the initial energy is suitably small (see Lemma
\ref{le3}).  We remark that one of the key issues to obtain such an
energy-like estimate lies in the positivity of the far field
density, which excludes the case of compactly supported  initial
density.

Next, the second main difficulty is to obtain the time-independent
 upper bound of the density. Based on careful initial layer analysis and  making  a full use of the structure of (\ref{a1}), we succeed  in deriving the weighted spatial mean estimates of the material derivatives of both the velocity and the temperature, which are   independent of the lower bound of density,
    provided that the initial energy    is suitably small (see Lemmas \ref{al13.4} and \ref{al13.5}).
This approach is motivated by  the basic  estimates of the material
derivatives of both the velocity and the temperature,  which are developed by Hoff
(\cite{Hof1}) in the theory of   weak solutions
with strictly positive initial density.
  Having all these estimates at hand,   we are able to obtain the desired estimates of
$L^1(0,\min\{1,T\};\,L^\infty({\r}^3))$-norm
 and the time-independent ones of
   $L^2(\min\{1,T\},T;\,L^\infty({\r}^3))$-norm
   of both the effective viscous flux (see (\ref{hj1}) for the definition)  and  the deviation of the temperature
from its far field value.
  It follows from these key estimates and a   Gronwall-type inequality (see Lemma \ref{le1}) that we are able to obtain a time-uniform upper bound of the density which is crucial for global estimates of classical solutions. This approach to estimate a uniform upper bound for the density is   new compared to our previous analysis on the
isentropic compressible Navier-Stokes equations in \cite{hulx}.

Then, the
third main step is to bound the gradients of the density, the
velocity, and the temperature. Motivated by our recent studies (\cite{hlx}) on the blow-up criteria of strong (or classical) solutions to the barotropic compressible Navier-Stokes equations, such bounds can be obtained by solving a logarithm
Gronwall inequality based on a Beale-Kato-Majda-type inequality
(see Lemma \ref{le9}) and the a priori estimates we have just
derived. Moreover, such a derivation simultaneously yields
  the bound for $L^{3/2}(0,T;L^\infty({\r}^3))$-norm of the
gradient of the velocity(see Lemma \ref{le11} and its proof).
It should be noted here that we do not require smallness of the
gradient of the initial density which prevents the appearance of
vacuum (\cite{M1}).

Finally, with these a priori estimates of
the gradients of the solutions at hand, one can obtain the desired higher order estimates  by    careful initial layer  analysis on  the time derivatives and then the spatial ones of the density, the velocity and the temperature. It should be emphasized  here that all these a priori estimates are independent of the lower bound of the density. Therefore, we can  build proper approximate solutions with strictly positive initial density then   take appropriate limits by letting  the lower bound of the initial density go to zero.  The limiting functions having  exactly the desired properties are shown to be the global classical solutions to the Cauchy problem (\ref{a1}) (\ref{h1}) (\ref{h2}).  In addition, the initial density is allowed to vanish. We can also  establish the global weak solutions   almost  the same way as we established the classical one with a new modified approximating initial data.

The rest of the paper is organized as follows: In Section 2, we collect some
elementary facts and inequalities which will be needed in later analysis. Section 3
is devoted to deriving the lower-order a priori estimates on classical solutions which
are needed to extend the local solution to all time. Based on the previous results, higher-order estimates are established in Section 4. Then finally, the main results,
Theorems \ref{th1} and \ref{th2}, are proved in Section 5.


\section{Preliminaries}\la{se2}

The following well-known  local existence theory, where the initial
density is strictly away from vacuum, can be shown by the standard
contraction mapping argument  (see for example  \cite{Na,M1}, in
particular, \cite[Theorem 5.2]{M1}).
\begin{lemma}   \la{th0} Assume  that
 $(\n_0,u_0,\te_0)$ satisfies \be \la{2.1}
(\n_0-\tn,u_0,\te_0-\tu)\in H^3, \quad \inf\limits_{x\in\r^3}\n_0(x) >0 .\ee Then there exist  a small time
$T_0>0$ and a unique classical solution $(\rho , u,\te )$ to the
Cauchy problem  (\ref{a1}) (\ref{h1}) (\ref{h2}) on
$\r^3\times(0,T_0]$ such that \be\la{mn6}
  \inf\limits_{(x,t)\in\r^3\times (0,T_0]}\n(x,t)\ge \frac{1}{2}
 \inf\limits_{x\in\r^3}\n_0(x), \ee \be\la{mn5}   \begin{cases}
 ( \rho-\tn,u,\te-\tu) \in C([0,T_0];H^3),\quad
 \n_t\in C([0,T_0];H^2),\\   (u_t,\te_t)\in C([0,T_0];H^1),
 \quad (u,\te-\tu)\in L^2(0,T_0;H^4),\end{cases}\ee  and
\be \la{mn1}\begin{cases} (\si u_{t},  \si \te_{t}) \in L^2( 0,T_0;H^3)  ,\quad (\si u_{tt},  \si\te_{tt}) \in L^2(
0,T_0;H^1), \\  (\si^2u_{tt},  \si^2\te_{tt}) \in L^2( 0,T_0;H^2), \quad
 (\si^2u_{ttt},  \si^2\te_{ttt}) \in L^2(
0,T_0;L^2) ,
\end{cases}\ee where   $\si(t)\triangleq \min\{1,t\}.$ Moreover, for any    $(x,t)\in \r^3\times [0,T_0],$ the following estimate holds
   \be\la{mn2}
\te(x,t)\ge
\inf\limits_{x\in\r^3}\te_0(x)\exp\left\{-(\ga-1)\int_0^{T_0}
 \|\div u\|_{L^\infty}dt\right\},\ee provided $ \inf\limits_{x\in\r^3}\te_0(x)\ge 0.$
 \end{lemma}

 {\it Proof. } We only have to show    (\ref{mn1}) and (\ref{mn2}),
 which are not given in
 \cite[Theorem 5.2]{M1}.

 Without loss of generality, assume that $T_0\le 1.$ We first prove $(\ref{mn1})_1.$ Differentiating  $(\ref{a1})_2$  with
respect to $t$ leads to \be\la{nt0}\ba & \n u_{tt}
+\n_tu_t+\n_tu\cdot\na u+\n u_t\cdot\na u +\n u\cdot\na u_t+\na
P_t\\&=\mu\Delta u_t+(\mu+\lambda)\na\div u_t.\ea\ee This shows that
$tu_t$ satisfies \be\la{mn7}\begin{cases} \n (tu_t)_t-\mu \Delta
(tu_t)-(\mu+\lambda)\na\div (tu_t)=F_1,\\ (tu_t)(x,0)=0,
\end{cases}\ee
 where $$F_1\triangleq\n u_t-t\n_tu_t-t\n_tu\cdot\na u-t\n u_t\cdot\na u -t\n
u\cdot\na u_t-Rt\na (\n_t\te+\n\te_t) $$ satisfies $F_1\in
L^2(0,T_0;L^2)$ due to  (\ref{mn5}). It thus follows from
(\ref{mn5}), (\ref{mn6}), and standard $L^2$-theory for parabolic
system (\ref{mn7}) that \be\la{mn9} (tu_t)_t,\na^2(tu_t)\in
L^2(0,T_0;L^2).\ee Similarly, we differentiate $(\ref{a1})_3$ with respect to $t$ to
get \be\la{eg1}\ba &-\frac{\ka(\ga-1)}{R}\Delta \te_t+\n\te_{tt}
\\&=-\n_t\te_{t}- \n_t\left(
u\cdot\na \te+(\ga-1)\te\div u\right)-\n\left( u\cdot\na
\te+(\ga-1)\te\div u\right)_t
\\&\quad+\frac{\ga-1}{R}\left(\lambda (\div u)^2+2\mu |\mathfrak{D}(u)|^2\right)_t,\ea\ee which implies that $t\te_t$ satisfies
\be\la{mn8}\begin{cases}
R\n (t\te_t)_t- {\ka(\ga-1)}\Delta (t \te_t)={R}F_2,\\
(t\te_t)(x,0)=0,
\end{cases}\ee
with \bnn\ba F_2\triangleq&\n\te_t-t\n_t\te_{t}- t\n_t\left(
u\cdot\na \te+(\ga-1)\te\div u\right)\\&-t\n\left( u\cdot\na
\te+(\ga-1)\te\div u\right)_t+
 \frac{\ga-1}{R}t\left(\lambda (\div u)^2+2\mu |\mathfrak{D}(u)|^2\right)_t.\ea\enn
One derives from (\ref{mn5}) that $F_2\in L^2(0,T_0;L^2),$ which
together with (\ref{mn5}), (\ref{mn6}), and standard $L^2$-theory for
parabolic system (\ref{mn8}) implies \be\la{mn10}
(t\te_t)_t,\na^2(t\te_t)\in L^2(0,T_0;L^2).\ee It thus follows from
(\ref{mn5}), (\ref{mn9}), and (\ref{mn10}) that $$F_1,F_2\in
L^2(0,T_0;H^1),$$ which together with (\ref{mn5}),  (\ref{mn6}),
(\ref{mn7}), and  (\ref{mn8}) gives $ (\ref{mn1})_1.$

Next, we  prove $(\ref{mn1})_2.$  Differentiating $(\ref{nt0})$ with
respect to $t$ gives \be\la{sp30}\ba &\n u_{ttt}+\n u\cdot\na
u_{tt}-\mu\Delta u_{tt}-(\mu+\lambda)\nabla{\rm div}u_{tt}\\&= 2{\rm
div}(\n u)u_{tt} +{\rm div}(\n u)_{t}u_t-2(\n u)_t\cdot\na
u_t-(\n_{tt} u+2\n_t u_t) \cdot\na u\\& \quad- \n u_{tt}\cdot\na
u-\na P_{tt}.
 \ea\ee
This together with $(\ref{mn1})_1$ and (\ref{mn5}) implies that
$t^2u_{tt}$ satisfies \be\la{mn11}\begin{cases} \n (t^2u_{tt})_t-\mu
\Delta (t^2u_{tt})-(\mu+\lambda)\na\div (t^2u_{tt})
=F_3,\\
(t^2u_{tt})(x,0)=0,
\end{cases}\ee
where \be\ba F_3\triangleq&2t\n u_{tt}-t^2\n u\cdot\na u_{tt}+2
t^2{\rm div}(\n u)u_{tt} +t^2{\rm div}(\n u)_{t}u_t\\&-2t^2(\n
u)_t\cdot\na u_t -t^2(\n_{tt} u+2\n_t u_t) \cdot\na u - t^2\n
u_{tt}\cdot\na u-t^2\na P_{tt},\ea\ee satisfies $F_3\in
L^2(0,T_0;L^2)$ due to  (\ref{mn5}) and $(\ref{mn1})_1.$ It follows
from (\ref{mn6}), (\ref{mn5}), $(\ref{mn1})_1,$ and standard
$L^2$-estimate for (\ref{mn11}) that \be\la{mn12} (t^2
u_{tt})_t,\na^2(t^2 u_{tt})\in L^2(0,T_0;L^2).\ee Similarly,
differentiating $(\ref{eg1})$ with respect to $t$ yields
\be\la{eg2}\ba &\n\te_{ttt}+\n u
\cdot\na\te_{tt}-\frac{\ka(\ga-1)}{R}\Delta \te_{tt}
\\&=2\div(\n u)\te_{tt} - \n_{tt}\left(\te_t+
u\cdot\na \te+(\ga-1)\te\div u\right)\\&\quad - 2\n_t\left(
u\cdot\na \te+(\ga-1)\te\div u\right)_t\\&\quad
 - \n\left(u_{tt}\cdot\na \te+2u_t\cdot\na\te_t+(\ga-1)(\te\div u)_{tt}\right)\\&\quad
+\frac{\ga-1}{R}\left(\lambda (\div u)^2+2\mu |\mathfrak{D}(u)|^2 \right)_{tt}.\ea\ee We thus obtain from $(\ref{mn1})_1,$
(\ref{mn5}), and (\ref{eg2})  that $t^2\te_{tt}$ satisfies
\be\la{mn13}\begin{cases} R\n (t^2\te_{tt})_t- {\ka(\ga-1)}  \Delta
(t^2\te_{tt}) =RF_4,\\
(t^2\te_{tt})(x,0)=0,
\end{cases}\ee
with $$\ba F_4\triangleq&2t\n\te_{tt}-t^2\n u\cdot\na \te_{tt}+
2t^2\div(\n u)\te_{tt} - t^2\n_{tt}\left(\te_t+ u\cdot\na
\te+(\ga-1)\te\div u\right)\\&  - 2t^2\n_t\left( u\cdot\na
\te+(\ga-1)\te\div u\right)_t
 - t^2\n u_{tt}\cdot\na \te-2t^2\n u_t\cdot\na\te_t
\\&-(\ga-1)t^2\n(\te\div u)_{tt}
+\frac{\ga-1}{R}t^2\left(\lambda (\div u)^2+2\mu |\mathfrak{D}(u)|^2\right)_{tt}.\ea$$ It thus follows from  (\ref{mn5})  and
$(\ref{mn1})_1 $ that  $F_4\in L^2(0,T_0;L^2),$ which together with
(\ref{mn6}), (\ref{mn5}), $(\ref{mn1})_1,$ and standard
$L^2$-estimate for (\ref{mn13}) gives that \be\la{mn14}
(t^2\te_{tt})_t,\na^2(t^2\te_{tt})\in L^2(0,T_0;L^2).\ee One thus
obtain $(\ref{mn1})_2$ directly from (\ref{mn5}), $(\ref{mn1})_1,$
 (\ref{mn12}), and (\ref{mn14}).

Finally, we will show the lower bound of $\te, $ (\ref{mn2}),
  by   maximum principle. In fact, it   follows from $(\ref{a1})_3$ and $(\ref{h1})$ that \be\la{mn4}  \n\te_t+\n u\cdot\na
\te-\frac{\ka(\ga-1)}{R}\Delta\te+(\ga-1)\n \te \div u\ge 0,\quad
\te\rightarrow \tu \,\,\mbox{ as }\,\,|x|\rightarrow \infty,\ee
 where we have used \be \la{a2.56} 2\mu  |\mathfrak{D}(u)|^2  +\lambda (\div u)^2
 \ge 0.\ee By (\ref{mn5}), we have
 \bnn
\int_0^{T_0}\|\div u\|_{L^\infty} dt
  <\infty,\enn
  which together with  the standard maximum principle
    thus gives  (\ref{mn2}). The proof of Lemma \ref{th0}
   is completed.

Next, the following well-known Gagliardo-Nirenberg-Sobolev-type inequality
  will be used later frequently (see \cite{nir}).

\begin{lemma}
 \la{l1} For  $p\in(1,\infty)$ and $ q\in
(3,\infty),$ there exists some generic
 constant $C>0$ which may depend  on $p$ and $q$ such that for   $f\in D^1 ({\r^3}),$
 $g\in L^p({\r^3})\cap D^{1,q}({\r^3}), $ and $\varphi,\psi\in H^2({\r^3}),$ we have\be
\la{g1}\|f\|_{L^6} \le C \|\na f\|_{L^2} ,\ee    \be
\la{g2}\|g\|_{C\left(\overline{\r^3}\right)} \le C
\|g\|_{L^p}^{p(q-3)/(3q+p(q-3))}\|\na g\|_{L^q}^{3q/(3q+p(q-3))},
\ee    \be\la{hs} \|\varphi\psi\|_{H^2}\le C\|\varphi\|_{H^2}\|\psi\|_{H^2}.\ee
\end{lemma}

Then, the following inequality   is an   easy consequence of (\ref{g1}) and  will be used frequently later.

\begin{lemma} Let the function $g(x) $ defined in $\r^3$  be  non-negative and satisfy  $g(\cdot)-1\in L^2(\r^3) .$   Then there exists a
universal positive constant $C$  such that for
$r\in [1,2]  $  and any open set $\Sigma\subset \r^3,$ the following estimate holds 
  \be \la{f}\int_\Sigma |f|^rdx\le C\int_\Sigma  g
|f|^rdx+C\|g-\tn\|_{L^2(\r^3)}^{(6-r)/3}\|\na f\|_{L^2(\r^3)}^r,\ee
 for  all
 $f\in \left\{f\in D^1(\r^3)\left|\,\,g |f|^r\in L^1(\Sigma)
\right\}\right..$
\end{lemma}

 {\it Proof.} In
fact,  Sobolev's inequality,  (\ref{g1}),  yields that\bnn\ba 2 \int_\Sigma |f|^rdx&\le
2\int_\Sigma g|f|^rdx+2\int_\Sigma|g-\tn||f|^rdx\\
&\le 2\int_\Sigma
g|f|^rdx+2\|g-\tn\|_{L^2(\r^3)}\|f\|_{L^r(\Sigma)}^{r(3-r)/(6-r)}
\|f\|_{L^6(\r^3)}^{3r/(6-r)}\\&\le 2\int_\Sigma
g|f|^rdx+ \int_\Sigma
|f|^rdx+C\|g-\tn\|_{L^2(\r^3)}^{(6-r)/3}\|\na
f\|_{L^2(\r^3)}^r,\ea\enn which implies (\ref{f}) directly.

Next, it follows from $(\ref{a1})_2$ that $G,\o$ defined in (\ref{hj1}) satisfy
 \be\la{h13}
\triangle G = \div (\rho\dot{u}),\quad\mu \triangle \o =
\nabla\times(\rho\dot{u}).  \ee Applying the  standard $L^p$-estimate  to the  elliptic systems (\ref{h13})   together with (\ref{g1})
 yields  the following elementary estimates (see \cite[Lemma 2.3]{hulx}).

\begin{lemma} \la{le4}
  Let $(\rho,u,\te)$ be a smooth solution of
   (\ref{a1}) (\ref{h1}).
    Then there exists a generic positive
   constant $C$ depending only on $\mu,$   $\lambda,$  and $R$ such that, for any $p\in [2,6],$
          \be\la{h18}
   \|\na u \|_{L^p} \le C \left( \|G\|_{L^p}  +\|\o\|_{L^p}\right)+
   C \|\n\te-1\|_{L^p} ,\ee
   \be\la{h19}\|\nabla G\|_{L^p} + \|{\nabla \o}\|_{L^p}
   \le C\|\rho\dot{u}\|_{L^p},\ee
   \be  \la{h20}\|G\|_{L^p}+\|\o\|_{L^p}
   \le C \|\rho\dot{u}\|_{L^2}^{(3p-6)/(2p) }
   \left(\|{\nabla u}\|_{L^2}
   +  \|\n\te-1\|_{L^2}\right)^{(6-p)/(2p)} ,\ee
  \be\la{h17} \|\na u\|_{L^p}\le C   \|{\nabla
u}\|_{L^2}^{(6-p)/(2p)}\left(\|\rho\dot{u}\|_{L^2}+\|\n\te-1\|_{L^6}\right)^{(3p-6)/(2p)
}.
  \ee
\end{lemma}

Next, the following Gronwall-type    inequality will be used to get the
uniform (in time) upper bound of the density $\n.$
\begin{lemma} \la{le1}   Let the function $y\in W^{1,1}(0,T)$ satisfy
\be \la{2.32}  y'(t)+\al y(t)\le  g(t)\mbox{  on  } [0,T] ,\quad y(0)=y^0, \ee
where $\al$ is a positive constant and  $ g \in L^p(0,T_1)\cap L^q(T_1,T)$   for some $p\ge 1,q\ge 1, $ and $T_1\in [0,T].$ Then
\be \la{2.34} \sup_{0\le t\le T}y(t)\le | y^0|+(1+\al^{-1})\left(\|g\|_{L^p(0,T_1)}+\|g\|_{L^q(T_1,T)}
\right)  .\ee
   \end{lemma}

{\it Proof.} Let $p'$ and $q'$ denote the conjugate numbers of $p$ and $q$ respectively.
 Multiplying  (\ref{2.32}) by $e^{\al t}$ and integrating the resulting inequality over $(0,t)$ yield  that
\be\la{2.35}\ba  e^{\al t}y(t)&\le y^0+\int_0^te^{\al s}g(s)ds\\  &\le  y^0+\int_0^{\min\{t,T_1\}}e^{\al s}|g(s)|ds+\int_{\min\{t,T_1\}}^te^{\al s}|g(s)|ds\\ &\le |y^0|+ \|g\|_{L^p(0,\min\{t,T_1\} )}\|e^{\al s}\|_{L^{p'}(0,t) }+\|g\|_{L^{q}(\min\{t,T_1\} ,t)}\|e^{\al s}\|_{L^{q'}(0,t)}
 \\ &\le |y^0|+\left(\|g\|_{L^p(0,T_1)}+\|g\|_{L^q(T_1,T)}
\right)(1+\al^{-1}) e^{\al t},\ea\ee
where  in the last inequality we have used the following simple fact: \bnn\|e^{\al s}\|_{L^r(0,t)}\le (1+\al^{-1})e^{\al t},\enn for all $r\in [1,\infty].$
We thus derive (\ref{2.34}) directly from  (\ref{2.35}) and finish the proof of Lemma \ref{le1}.

Finally, we state the following Beale-Kato-Majda-type inequality
whose proof can be found in   \cite{hlx,bkm}   and will be
used later to estimate $\|\nabla u\|_{L^\infty}$ and
$\|\nabla\rho\|_{L^2\cap L^6}$.
\begin{lemma}[\cite{hlx,bkm}]  \la{le9}  For $3<q<\infty,$ there is a
constant  $C(q)$ such that  the following estimate holds for all
$\na u\in L^2(\O)\cap D^{1,q}({\r^3}):$ \be\la{ww7}\ba \|\na
u\|_{L^\infty({\r^3})}&\le C\left(\|{\rm div}u\|_{L^\infty({\r^3})}+
\|\na\times u\|_{L^\infty({\r^3})} \right)\log(e+\|\na^2
u\|_{L^q({\r^3})})\\&\quad+C\|\na u\|_{L^2(\O)} +C . \ea\ee
\end{lemma}

\section{\la{se3} A priori estimates (I): Lower-order estimates}

In this section,  we  will establish   a priori    bounds for the
smooth, local-in-time solution to (\ref{a1}) (\ref{h1})  (\ref{h2})
obtained in Lemma \ref{th0}.
 We thus fix a smooth solution $(\n,u,\te)$ of
(\ref{a1})  (\ref{h1}) (\ref{h2}) on $\r^3\times (0,T]$ for some time
$T>0,$ with  initial data $(\n_0,u_0,\te_0)$ satisfying
(\ref{2.1}).

  For
$\si(t)\triangleq\min\{1,t\}, $  we  define
$A_i(T)(i= 1,\cdots,4)$ as follows:
  \be\la{As1}
  A_1(T) = \sup_{t\in[0,T] }\|\nabla u \|_{L^2}^2
   + \int_0^{T}\int  \rho|\dot{u}|^2dxdt,
   \ee
 \be
  A_2(T) = \frac{R}{2(\ga-1)}\sup_{t\in[0,T] }\int\n (\te-\tu)^2dx
  +\int_0^T\left( \|\na u\|_{L^2}^2+\|\na \te\|_{L^2}^2\right)dt,
  \ee
 \be\ba
  A_3(T) = &\sup_{t\in(0,T]}\left(\si \|\na
u\|_{L^2}^2+\si^2\int\n |\dot u|^2dx +
  \si^2\|\na\te\|_{L^2}^2 \right)\\ &
  + \int_0^T\int\left(\si\n
  |\dot u|^2 +\sigma^2|\nabla\dot{u}|^2
  +\sigma^2\n(\dot \te)^2 \right)dxdt,
  \ea\ee
   \be\la{As2}
  A_4(T) = \sup_{t\in(0,T] }\sigma^4\int\rho|\dot{\te}|^2dx
  + \int_0^{T}\int
  \sigma^4|\nabla\dot{\te}|^2dxdt.
  \ee

We have the following key a priori estimates on $(\n,u,\te)$.
\begin{pro}\la{pr1} For  given numbers $M>0$ (not necessarily small),
   and $\on> 2,$ assume that $(\rho_0,u_0,\te_0) $  satisfies   \be
 \la{3.1} 0<\inf_{x\in \r^3}\rho_0(x)\le\sup_{x\in \r^3}\rho_0(x)<\bar{\rho},\quad \inf_{x\in \r^3}\te_0(x)> 0, \quad \|\na u_0\|_{L^2} \le M.
   \ee     Then there exist  positive constants   $K $  and $\ep_0 $ both  depending only   on  $\mu,\lambda, \ka, R, \ga, \on,$ and $M$  such that if  $(\rho,u,\te)$  is a smooth solution of (\ref{a1}) (\ref{h1}) (\ref{h2})  on $\r^3\times (0,T] $
 satisfying \be \la{z1} 0<  \rho\le 2\bar{\rho},
      \,\,\,   A_1(\si(T))\le 3 K,
\,\,\,   A_i(T)  \le 2C_0^{1/(2i)} \,\,\,
  (i=2,3,4)     ,
   \ee
    the following estimates hold
  \be  \la{zs2}0< \rho\le 3\bar{\rho}/2,\,\,\,       A_1(\si(T))\le 2 K,
\,\,\,    A_i(T)  \le  C_0^{1/(2i)} \,\,\,
  (i=2,3,4) ,
  \ee
  provided \be\la{z01}C_0\le \ve_0.\ee      \end{pro}

{\it Proof.}    Proposition \ref{pr1} is an easy consequence of
the following Lemmas \ref{le2}, \ref{le3}, and \ref{le6}--\ref{le8},  with $\ve_0$ as in (\ref{ve}).

In this section, we let $C$ denote some generic positive constant depending
 only on $\mu,\lambda,\ka, R,$ $ \ga,   $  $\on,$ and $M,$
  and we write $C(\al)$ to emphasize that $C$
may depend  on $\al.$

The following elementary    $L^2$  bounds  are crucial for   deriving the desired estimate on $A_2(T)$   (see Lemma \ref{le3} below).

\begin{lemma}\la{al3.1} Under the conditions of Proposition \ref{pr1},
 there exists a positive constant $C=C(\on) $
depending only on
  $\mu,\lambda, \ka, R, \ga,  $ and $\on$ such
that  if $(\rho,u,\te)$ is a smooth solution of (\ref{a1})
(\ref{h1}) (\ref{h2}) on $\r^3\times (0,T] $ satisfying  $0< \n\le 2\on,$ the following estimates hold \be \la{a2.12}\sup_{0\le t\le T}\int\left( \n |u|^2+(
\n-\tn)^2\right)dx\le C(\on) C_0 , \ee and \be  \la{a2.17}  \ba
\|(\te-1)(\cdot,t)\|_{L^2} \le C(\on)C_0^{1/2} +C(\on)
C_0^{1/3}\|\na\te(\cdot,t)\|_{L^2}, \ea\ee for all $t\in (0,T].$
\end{lemma}

{\it Proof.} First, it follows from (\ref{3.1}) and  (\ref{mn2}) that, for all $ (x,t)\in \r^3\times(0,T),$
\be \la{3.2}\te(x,t)>0 .\ee Adding $(\ref{a1})_2$ multiplied by $u$ to $(\ref{a1})_3$ multiplied by $1-\te^{-1},$  we obtain
after integrating the resulting equality over $\r^3$ and   using $(\ref{a1})_1$    that\be\la{a2.7}\ba &\frac{d}{dt} \int
\left( \frac{1}{2}\n |u|^2+R(1+\n\log
\n-\n)+\frac{R}{\ga-1}\n(\te-\log \te-1)\right)dx\\&=\int \left[
-\mu |\na u|^2-(\lambda+\mu)(\div u)^2-\ka
\te^{-2}|\na\te|^2\right.\\&\left.\quad+(1-\te^{-1})(\lambda (\div
u)^2+2\mu |\mathfrak{D}(u)|^2)\right]dx
\\&=-\int \left( {\te}^{-1}(\lambda(\div u)^2+2\mu
 |\mathfrak{D}(u)|^2)+\ka
 \te^{-2}|\na \te|^2  \right)dx,\ea\ee
where in the second  equality we have used
 \be \la{3.12} 2\int | \mathfrak{D}(u)|^2 dx=\int \left(|\na u|^2+(\div u)^2\right)dx.\ee Direct calculations yield that  \be\la{a2.9}\ba \n\log\n
-\n+1&=(\n-1)^2\int_0^1\frac{1-\al}{\al (\n-1)+1}d\al \\&\ge
\frac{(\n-1)^2}{\on}\int_0^1(1-\al)d\al \\&=
\frac{1}{ 2\on  }(\n-1)^2,\ea\ee and \be\la{a2.57}\ba
 \te-\log\te-1 &=(\te-1)^2\int_0^1\frac{\al}{\al (\te-1)+1}d\al \\
&\ge \frac{1}{8} (\te-1)1_{(\te(\cdot,t)>2)
}+\frac{1}{12}(\te-1)^21_{(\te(\cdot,t)<3)} ,\ea\ee where we denote
\bnn (\te(\cdot,t)> 2)\triangleq \left.\left\{x\in
\r^3\right|\te(x,t)> 2\right\},\quad (\te(\cdot,t)< 3)\triangleq
\left.\left\{x\in \r^3\right|\te(x,t)<3\right\}.\enn
Integrating (\ref{a2.7}) with respect to $t$ over $(0,T)$ yields
that
 \be\la{a2.8}\ba &\sup_{0\le t\le T} \int \left(
\frac{1}{2}\n
|u|^2+R(1+\n\log \n-\n)+\frac{R}{\ga-1}\n(\te-\log \te-1)\right)dx\\
&+\int_0^T\int \left(\frac{1}{\te}(\lambda(\div u)^2+2\mu
|\mathfrak{D}(u)|^2 )+\ka \frac{|\na \te|^2}{\te^2}\right)dxdt \le
C_0,\ea\ee which together with
(\ref{a2.56}), (\ref{3.2}), (\ref{a2.9}), and (\ref{a2.57})
leads to \be \la{a2.11}\ba &\sup_{0\le t\le T} \int\left( \n |u|^2+(
\n-1)^2\right)dx\\&+\sup_{0\le t\le T}\int
\left(\n(\te-1)1_{(\te(\cdot,t)>
2)}+\n(\te-1)^21_{(\te(\cdot,t)<3)}\right)dx \le C(\on)C_0.\ea\ee
This fact gives (\ref{a2.12})  directly.

Next, we shall prove (\ref{a2.17}). Taking $g(x)=\n(x,t),f(x)=\te(x,t)-1,$ $r=2$
and $\Sigma=(\te(\cdot,t)<3)$ in (\ref{f}), we conclude after using (\ref{a2.11}) that
  \be \la{a2.19}\ba
\|\te(\cdot,t)-1\|_{L^2(\te(\cdot,t)<3)} \le C(\on)C^{1/2}_0+
 C(\on)C_0^{1/3}\|\na\te(\cdot,t)\|_{L^2(\r^3)}. \ea\ee
  Similarly,
taking $g(x)=\n(x,t),f(x)=\te(x,t)-1,$  $r=1$ and $\Sigma=(\te(\cdot,t)>2)$  in (\ref{f}), we obtain after using  (\ref{a2.11})  that
\bnn\ba \|\te(\cdot,t)-1\|_{L^1(\te(\cdot,t)> 2)} \le C(\on)C_0
+C(\on)C_0^{5/6} \|\na\te(\cdot,t)\|_{L^2(\r^3)},\ea\enn which
together with H\"older's inequality and (\ref{g1}) leads to \be
\la{a2.18}\ba &\|\te(\cdot,t)-1\|_{L^2(\te(\cdot,t)> 2)}\\ &\le
\|\te(\cdot,t)-1\|^{2/5}_{L^1(\te(\cdot,t)> 2)} \|
\te(\cdot,t)-1\|_{L^6(\r^3)}^{3/5}
\\ &\le C(\on)\left(C_0^{2/5}+C_0^{1/3}
\|\na \te(\cdot,t)\|_{L^2}^{2/5}\right) \|\na \te(\cdot,t)\|_{L^2}^{3/5}\\ &\le
C(\on)C_0^{1/2}+C(\on)C_0^{1/3}\|\na \te(\cdot,t)\|_{L^2} .\ea\ee Because of $\r^3=(\te(\cdot,t)<3)\cup (\te(\cdot,t)> 2),$ the
combination of (\ref{a2.19}) with (\ref{a2.18}) yields (\ref{a2.17})
directly. We finish the proof of Lemma \ref{al3.1}.

  Next, the following lemma will give an estimate on the term
$A_1(\si(T)).$
\begin{lemma}\la{le2}  Under the conditions of Proposition \ref{pr1},
  there exist   positive constants  $K\ge M+1 $  and
   $\ep_1 \le 1 $    both depending only  on
  $\mu,\lambda, \ka, R, \ga, \on,$ and $M$ such
that if  $(\rho,u,\te)$ is a smooth solution of (\ref{a1}) (\ref{h1})
(\ref{h2}) on $\r^3\times (0,T] $ satisfying   \be\la{3.q1}  0<\n\le 2\on ,\quad A_2(\si(T))\le 2C_0^{1/4},\ee  the following estimate holds
  \be\la{h23} A_1(\si(T))\le 2K ,  \ee
  provided  $A_1(\si(T))\le 3K$ and $C_0\le \ep_1.$
\end{lemma}


{\it Proof.} First, multiplying $(\ref{a1})_2$  by $2u_t,$  integrating the resulting equality over $\r^3,$  we obtain after integration by
parts that \be\ba \la{hh17}  &\frac{d}{dt}\int \left(  {\mu}
|\na u|^2+ (\mu+\lambda)(\div u)^2\right)dx+ \int\rho
|u_t|^2dx \\ &\le-2\int \na P\cdot
u_tdx+ \int \n|u\cdot \na u|^2dx\\&=  2R\frac{d}{dt}\int  (\n\te-1) \div
u  dx-2\int P_t
\div udx+\int \n|u\cdot \na u|^2dx\\&= 2R\frac{d}{dt}\int  (\n\te-1) \div
u  dx-\frac{R^2}{2\mu+\lambda}\frac{d}{dt}\int (\n\te-1)^2 dx\\&\quad-\frac{2}{2\mu+\lambda}\int P_t Gdx+ \int \n|u\cdot \na u|^2dx ,\ea\ee
where in the last equality, we have used \be  \la{ez1}{\rm
div}u=\frac{1}{2\mu+\lambda}(G +R(\n \te-1)),\ee
due to (\ref{hj1}).

Next, assume that $C_0\le 1.$ It follows from H\"older's inequality, (\ref{3.q1}), (\ref{g1}), and (\ref{a2.12})
that for $p\in [2,6],$  \be\la{p}\ba  \|\n\te-1 \|_{L^p}&= \|\n(\te-
1 )+ (
 \n -1)\|_{L^p}\\&\le \|\n(\te-1  )\|_{L^2}^{(6-p)/(2p)}
 \|\n(\te- 1 )\|_{L^6}^{ 3(p-2)/(2p)}+  \|\n-1\|_{L^p}
 \\&\le C(\on)C_0^{(6-p)/(16p)}
 \|\na\te \|_{L^2}^{ 3(p-2)/(2p)}+ C(\on)C_0^{1/p},
  \ea\ee
which together with (\ref{h17}) yields
\be \la{3.30}  \|\na
u\|_{L^6} \le C(\on) \left(  \|\n^{1/2} \dot
u\|_{L^2}+ \|\na \te\|_{L^2}+C_0^{1/6}\right).
\ee
Noticing that (\ref{a1}) implies \be \la{op3} \ba P_t=&-\div (Pu)
-(\gamma-1) P\div u+(\ga-1)\ka
\Delta\te\\&+(\ga-1)\left(\lambda
 (\div u)^2+2\mu |\mathfrak{D}(u)|^2\right), \ea\ee we obtain after integration by
parts and using  (\ref{3.q1}), (\ref{g1}), (\ref{h19}), (\ref{3.30}),
 (\ref{p}), and (\ref{a2.12}) that
\be\la{a16}\ba  &\left|\int   P_t Gdx\right| \\
& \le C\int P(|G||\na u|+ |u||\na G|)dx+ \int\left( |\na\te||\na
G|+|\na u|^2|G|\right)dx
\\ &  \le  C\int \rho (|G||\na u|+|u||\na G|)dx
+C\int \rho|\te-1|(|G||\na u|+|u||\na G|)dx
\\&\quad+ C \|\na G\|_{L^2} \|\na \te\|_{L^2}
+ C \|\na G\|_{L^2} \|\na u\|_{L^2}^{3/2}  \|\na u\|_{L^6}^{1/2}
\\ & \le C(\on) ( \|\na u\|_{L^2}+\|\n \te-1 \|_{L^2} )
 \|\na u\|_{L^2}+C\|\rho u\|_{L^2}\|\na G\|_{L^2}\\&
  \quad+C(\on)\|\rho(\te-1)\|_{L^2}^{1/2}\|\na\te\|_{L^2}^{1/2}
  \|\na G\|_{L^2}\|\na u\|_{L^2}+ C \|\na G\|_{L^2} \|\na \te\|_{L^2}
\\&\quad+ C (\on)\|\na G\|_{L^2} \|\na u\|_{L^2}^{3/2} \left(
 \|\n\dot u\|_{L^2}^{1/2}+\|\na\te\|_{L^2}^{1/2}+C_0^{1/12} \right)
  \\ &  \le  C(\de,\on) C_0^{1/4} +C(\on,\de)\|\na u\|_{L^2}^2
   + \de \|\na G\|_{L^2}^2 +C(\on,\de)
     \|\na\te\|_{L^2}\|\na u\|^2_{L^2}
      \\& \quad+C(\de,\on) \|\na \te\|_{L^2}^2+\de \|\rho^{1/2} \dot u\|^2_{L^2}
      +C(\de,\on)   \|\na u\|^6_{L^2}
       \\&  \le  C(\on)\de\|\rho^{1/2}
           \dot u\|^2_{L^2}     +C(\de,\on)
       \left( \|\na \te\|_{L^2}^2
  +\|\na u\|_{L^2}^2+1\right)  +C(\de,\on)  \|\na  u\|^6_{L^2}.
      \ea\ee

Finally, it follows from (\ref{g1})
 and (\ref{3.30}) that\be\la{op1}\ba &\int \n|u\cdot \na u|^2dx\\&\le
C(\bar\n)\|u\|_{L^6}^2 \|\na u\|_{L^2} \|\na u\|_{L^6}\\& \le
C(\bar\n) \|\na
u\|^3_{L^2}\left(\|\n \dot u\|_{L^2}+\|\na\te\|_{L^2}+C_0^{1/6} \right) \\
&\le \de\|\rho^{1/2} \dot u\|_{L^2}^2+ C(\bar\n,\de)\|\na
u\|_{L^2}^6+C(\bar\n,\de)\left(\|\na
u\|_{L^2}^2+\|\na\te\|_{L^2}^2\right),\ea\ee
where we have used
$$
\|\na u\|_{L^2}^3 \le \|\na u\|_{L^2}^2 + \|\na u\|_{L^2}^6 .
$$
 Substituting (\ref{a16}) and (\ref{op1})
into (\ref{hh17}),
  choosing $\de$ suitably small, we get after integrating (\ref{hh17}) over
 $(0,\si(T))$ and using (\ref{3.q1}) that
 \be\la{h81} \ba& \sup_{0\le t\le \si(T)}\|\na u\|_{L^2}^2+
\int_0^{\si(T)}\int\rho|\dot{u}|^2dxdt\\ &\le CM+C(\on)C_0^{1/4}
 + C(\bar\n ) C_0^{1/4}\sup_{0\le t\le \si(T)}\|\na u\|_{L^2}^4
  \\&\le K + C(\on)C_0^{1/4}\sup_{0\le t\le \si(T)}\|\na u\|_{L^2}^4 ,
 \quad\quad
\ea \ee where    $K $ is
defined by \be\la{3.q2}K\triangleq CM+C(\on) +1,\ee    depending only on $\mu,\lambda, \ka, R, \ga, \on, $ and $M.$
    We  thus finish the proof of (\ref{h23}) by choosing $\ep_1
\triangleq \min\left\{1,(9C(\on)K)^{-4}\right\}.$ The proof of Lemma
\ref{le2} is completed.

Next, the following  energy-like estimate of the local smooth
solution will play a   key role  in obtaining further estimates.
\begin{lemma}\la{le3} Under the conditions of Proposition \ref{pr1},  there exists a positive constant $\ve_2$
depending only on $\mu,\lambda, \ka, R, \ga, \on,$ and $M$
such that if $(\rho,u,\te)$ is a smooth solution of (\ref{a1}) (\ref{h1}) (\ref{h2}) on $\r^3\times (0,T] $ satisfying  (\ref{z1})   with $K$
as in Lemma \ref{le2}, the following estimate holds \be\la{a2.34} A_2(T) \le C_0^{1/4},\ee provided $C_0\le \ve_2.$
\end{lemma}

{\it Proof.} First, assume that   $C_0\le 1.$ Multiplying $(\ref{a1})_2$ by $u$ and
integrating the resulting equality over $\r^3$  give \be \la{a2.22}
\ba &\frac{d}{dt}\int\left(\frac{1}{2}\n |u|^2+R(1+\n\log \n-\n)
\right)dx\\&\quad+\mu \int|\na u|^2dx+(\mu+\lambda)\int (\div u)^2dx
\\ &\le C(\on)\left( \|\te-1\|_{L^2}  +\|\n-1\|_{L^2}\right) \|\na u\|_{L^2}
\\ &\le C(\on)\left(C_0^{1/2}+ C_0^{1/3}\|\na \te\|_{L^2}\right)
\|\na u\|_{L^2}\\ &\le C(\on) C^{2/3}_0 +
 C(\on)C_0^{1/3} \left(\|\na \te\|_{L^2}^{2}+  \|\na
u\|_{L^2}^{2}\right),\ea\ee
where in the second inequality
we have used (\ref{a2.12}) and  (\ref{a2.17}).

Multiplying $(\ref{a1})_3$ by $\te-1$ and integrating the resulting
equality over $\r^3 $  lead to\be
 \la{a2.23} \ba & \frac{R}{2(\ga-1)} \frac{d}{dt}\int \n
(\te-1)^2dx+\ka \|\na\te\|_{L^2}^2\\ &\le C(\on)\int \te|\te-1||\div
u|dx+C\int |\na u|^2|\te-1|dx  .\ea\ee
For the first term on the righthand side of   (\ref{a2.23}),  we have
  \be\la{a3.16}\ba &\int \te|\te-1||\div u|dx\\
& \le \int (\te-1)^2|\div
u|dx+\int  |\te-1||\div u|dx\\
&\le C\| \te-1 \|_{L^2}^{1/2}\|\te-1\|_{L^6}^{3/2}\|\na u\|_{L^2}+
C\|\te-1\|_{L^2}\|\na u\|_{L^2} \\
&\le C(\on,M)\left(C_0^{1/4}  +
C_0^{1/6}\|\na\te\|^{1/2}_{L^2}\right)\|\na
\te\|_{L^2}^{3/2}\\&\quad+ C(\on)\left(C_0^{1/2}  + C_0^{1/3}
\|\na\te\|_{L^2}\right)\|\na u\|_{L^2} \\
&\le C(\on,M)C_0^{1/2} + C(\on,M)C_0^{1/6}\left(\|\na
\te\|_{L^2}^{2}+ \|\na u\|_{L^2}^{2}\right),\ea\ee where we have used
(\ref{g1}), (\ref{a2.17}), (\ref{z1}), and the following simple fact:
\be\la{3.35}
\sup_{t\in[0,T]}\|\na u \|_{L^2}\le A_1(\si(T))+A_3(T)\le C(\on,M), \ee due to (\ref{z1}). For the second one on the
righthand side of   (\ref{a2.23}),  in light of (\ref{a2.17}),
(\ref{g1}),  (\ref{h17}),  (\ref{3.30}),  and (\ref{z1}), we have
\be\la{a3.17}\ba  &\int |\na u|^2|\te-1|dx\\
&\le C \|\te-1\|_{L^2}^{1/2}\|\te-1\|_{L^6}^{1/2}\|\na u\|_{L^2}
\|\na u\|_{L^6}\\
&\le C(\on,M)\left( C_0^{1/4}\|\na\te\|_{L^2}^{1/2}
+C_0^{1/6}\|\na\te\|_{L^2}\right)\left(\|\n\dot u\|_{L^2}
+\|\na\te\|_{L^2}+C_0^{1/6}\right)  \\
&\le C (\on,M,\de)C_0^{1/3}\left(\|\rho^{1/2} \dot
u\|_{L^2}^2+1\right) + C(\on,M)\left( \de +C_0^{1/6} \right)
\|\na\te\|_{L^2}^2  .\ea\ee
Substituting  (\ref{a3.16}) and (\ref{a3.17}) into  (\ref{a2.23})
leads to
 \be
 \la{a2.24} \ba & \frac{R}{2(\ga-1)} \frac{d}{dt}\int \n
(\te-1)^2dx+\ka \|\na\te\|_{L^2}^2 \\ &\le    C(\on,M)\left( \de
+C_0^{1/6} \right)\left( \|\na\te\|_{L^2}^2+ \|\na u\|_{L^2}^2
\right)\\&\quad+C (\on,M,\de)C_0^{1/3}\left(\|\rho^{1/2} \dot
u\|_{L^2}^2+1\right) .\ea\ee

The combination of (\ref{a2.22}) with (\ref{a2.24}) yields
\be \la{a2.25}\ba &\frac{d}{dt}\int\left(\frac{1}{2}\n
|u|^2+R(1+\n\log \n-\n)+
 \frac{R}{2(\ga-1)} \n (\te-1)^2\right)dx
\\ &+\mu \int|\na u|^2dx+(\mu+\lambda) \int(\div u)^2dx +\ka \int|\na \te|^2dx\\ &\le
 C(\on,M)\left( \de +C_0^{1/6} \right) \left(\|\na\te\|_{L^2}^2+\|\na
u\|_{L^2}^2\right)\\&\quad+C (\on,M,\de)C_0^{1/3}\left(\|\rho^{1/2}
\dot u\|_{L^2}^2+1\right) .  \ea\ee  We assume that $$C_0\le
\ve_2^1\triangleq\min\left\{1,\left((4C(\on,M))^{-1}\min\{\mu,\ka\}
\right)^6\right\}.$$ Choosing $\de\le
 (4C(\on,M))^{-1}\min\{\mu,\ka\}  $  and integrating
  (\ref{a2.25}) over $(0,\si(T)) ,$ we obtain after using (\ref{z1})  that
\be\la{a3.19}\ba &\sup_{0\le t\le \si(T)}\int \left( \n|u|^2+(\n-1)^2+\frac{R}{2(\ga-1)}\n
(\te-1)^2\right)dx\\& +\int_0^{\si(T)} \left(\|\na
u\|_{L^2}^2+\|\na\te\|_{L^2}^2\right)dt \le C(\on,M)C_0^{1/3}
.\ea\ee

Next, applying the standard $L^2$-estimate to  the following elliptic problem
\be\la{3.29}\begin{cases}\ka\Delta \te=\frac{R}{\ga-1}\n\dot\te +R\n\te\div
u-\lambda (\div u)^2-2\mu |\mathfrak{D}(u)|^2,\\
\te\rightarrow 1\quad \mbox{ as }|x|\rightarrow
\infty,\end{cases}\ee gives    \be  \la{op4}\ba\|\na^2\te\|_{L^2}^2
&\le C \left(\|\n\dot \te\|_{L^2}^2+ \|\na u\|_{L^4}^4+\|\te\na
u\|_{L^2}^2\right) \\ &\le C(\on) \left(\|\na\te\|_{L^2}^2
  + \|\na u\|_{L^2}^2\right)\left( \|\n^{1/2}\dot u\|^2_{L^2} +
\|\na\te\|^2_{L^2} +1\right)\\&\quad +C \left(\|\n\dot
\te\|_{L^2}^2+ \|\na u\|_{L^4}^4 \right),\ea\ee where we have used
  \be \la{m20}\ba &  \int \te^2|\na u|^2dx \\&\le C
\|\te-1\|_{L^6}^2 \|\na u\|_{L^2} \|\na u\|_{L^6} +C  \|\na
u\|_{L^2}^2 \\&\le C(\on) \left(\|\na\te\|_{L^2}^2
  + \|\na u\|_{L^2}^2\right)\left( \|\n^{1/2}\dot u\|^2_{L^2} +
\|\na\te\|^2_{L^2} +1\right),\ea\ee
due to (\ref{g1}) and  (\ref{3.30}). Note that (\ref{3.30}) and   (\ref{z1}) give
\be \la{3.42} \sup_{0\le t\le T}\si\|\na
u\|_{L^6} \le C(\on) C_0^{1/12}.
\ee This combining with (\ref{op4}) and
 (\ref{z1})  leads to
  \be \la{3.26}\ba& \sup_{0< t\le T}\si^4 \|\na^2\te\|_{L^2}^2\\
&\le C(\on) \sup_{0< t\le T}\si^2\left(\|\na\te\|_{L^2}^2
  + \|\na u\|_{L^2}^2\right)\sup_{0< t\le T}\si^2\left( \|\n^{1/2}\dot u\|^2_{L^2} +
\|\na\te\|^2_{L^2} +1\right)\\& \quad+C \sup_{0< t\le T}
\left(\si^4\|\n\dot \te\|_{L^2}^2+ \left(\si\|\na u\|_{L^2}\right)
\left(\si\|\na u\|_{L^6} \right)^3 \right), \\ &\le
C(\on)C_0^{1/8},\ea\ee which together with (\ref{g1}), (\ref{g2}),
and  (\ref{z1}) yields that \be \la{a3.23}\ba\sup_{0< t\le
T}\si^2\|\te-1\|_{L^\infty}&\le \sup_{0< t\le T} \si^2\left(\|\na
\te\|_{L^2}+\|\na^2 \te\|_{L^2}\right)\\ &\le C(\on)C_0^{1/16}\le
1/2,\ea\ee provided $C_0\le \ve_2^2\triangleq
\min\left\{1,(2C(\on))^{-16}\right\}.$ We  assume that $C_0\le\min\{
\ve_2^1, \ve_2^2\}.$ It follows from (\ref{a3.23}) that, for all
$(x,t)\in \r^3\times [\si(T),T],$ \be\la{a2.02} 1/2\le \te(x,t)\le
3/2,\ee  which as well as (\ref{a2.56}) and (\ref{3.12})--(\ref{a2.8})   gives
\be\la{a3.25}\ba &\sup_{\si(T)\le t\le T}\int \left(
\n|u|^2+(\n-1)^2+\frac{R}{2(\ga-1)}\n (\te-1)^2\right)dx\\&
+\int_{\si(T)}^T\left(\|\na u\|_{L^2}^2+\|\na\te\|_{L^2}^2\right)dt
\le C(\on)C_0.\ea\ee

Finally, the combination of (\ref{a3.19}) with (\ref{a3.25}) yields
  \bnn\ba &\sup_{0\le t\le T}\int \left( \n|u|^2+(\n-1)^2 +\frac{R}{2(\ga-1)}\n
(\te-1)^2\right)dx\\& +\int_0^T\left(\|\na
u\|_{L^2}^2+\|\na\te\|_{L^2}^2\right)dt  \\ &\le \max\left\{C(\on)C_0,
 C(\on,M)C_0^{1/3}\right\}\le
C_0^{1/4},\ea\enn provided \bnn C_0\le \ve_2\triangleq\min
\left\{\ve_2^1,\ve_2^2, (C(\on))^{-4/3},
(C(\on,M))^{-12}\right\}.\enn The proof
of Lemma \ref{le3} is completed.

Next, to estimate $A_3(T),$  we first establish the following  Lemmas \ref{al13.4} and \ref{al13.5} concerning some elementary estimates on $\dot u$ and $\dot\te$    for the case that the density may contain  vacuum states. This approach is motivated by  the basic estimates on $\dot u$ and $\dot\te$    developed by Hoff \cite{Hof1} when the density is strictly away from vacuum. The estimate of $A_3(T)$ will be postponed to Lemma \ref{le6}.

\begin{lemma}\la{al13.4} In addition to the conditions of Proposition \ref{pr1}, assume that   $C_0\le 1.$ Let $(\rho,u,\te)$ be a smooth solution of (\ref{a1}) (\ref{h1})
(\ref{h2}) on $\r^3\times (0,T] $ satisfying    (\ref{z1}) with $K$ as
in Lemma \ref{le2}.   Then  there exist positive constants $C$ and
$C_1$ both depending only on $\mu,\lambda, \ka, R, \ga, \on,$
 and $M$ such that, for any $\beta\in (0,1],$
the following estimates hold 
  \be\la{an1}\ba& (\si B_1)'(t)+\frac{3}{2} \int \sigma \rho|\dot{u}|^2dx\\&\le C
C_0^{1/4}\si' +2\beta \si^{2}\| \n^{1/2}\dot\te\|_{L^2}^2 +C
\beta^{-1}\left(\|\na\te\|_{L^2}^2+\|\na u\|_{L^2}^2\right)  + C
\sigma^2  \|\na u\|_{L^4}^4 ,\ea\ee  and \be\la{ae0} \ba & \left(
{\sigma^2} \int\rho|\dot{u}|^2dx \right)_t + \frac{3\mu}{2} \int
\sigma^2|\nabla\dot{u}|^2dx \\& \le 2\sigma \int\rho|\dot{u}|^2dx
+C_1 \si^2\| \n^{1/2}\dot\te\|_{L^2}^2
  +C \left( \|\na \te\|_{L^2}^2+  \|\na
u\|_{L^2}^2\right) +C\si^2\|\na u\|_{L^4}^4,\ea\ee
where \be \la{an2}
B_1(t)\triangleq \mu\|\nabla u\|_{L^2}^2+ (\lambda+\mu)
 \|\div u\|_{L^2}^2+2R\int \div u(\n\te-1) dx. \ee
\end{lemma}

{\it Proof.}     Multiplying
$(\ref{a1})_2 $ by $\sigma \dot{u}$ and integrating the resulting
equality over $\r^3 $ lead  to \be\la{m0} \ba  \int \sigma
\rho|\dot{u}|^2dx & = \int (-\sigma \dot{u}\cdot\nabla P + \mu\sigma
\triangle u\cdot\dot{u} + (\lambda+\mu)
 \sigma  \nabla\div u\cdot\dot{u})dx \\
& \triangleq \sum_{i=1}^{3}M_i. \ea \ee

 Using $(\ref{a1})_1,$   we get
after integration by parts that, for any $\beta\in
(0,1],$  \be\la{m1} \ba
M_1 & = - \int \sigma  \dot{u}\cdot\nabla Pdx \\
& = \int\sigma  \left(P\div u  \right)_tdx +\int \sigma
\left(-P_t\div u+  P\div(u\cdot\nabla u)
\right)dx   \\
&   = \int\sigma  \left(P \div u \right)_tdx  - R \int \sigma \div
u\left(\n \dot\te
  -\n u\cdot\na\te+\te  \n_t\right)dx \\& \quad
     + \int \sigma   P u\cdot\nabla \div u dx + \int \sigma
  P\pa_iu^j\pa_ju^i dx  \\
& = \int\sigma  \left(P \div u  \right)_tdx
 +R  \int \sigma    \div u\left( \n u\cdot\na\te
  +\te u\cdot\na\n+\te \n\div u\right)dx
     \\& \quad  + \int \sigma   P u\cdot\nabla \div u dx
    -R \int \sigma
\div u \n \dot\te dx + \int \sigma
  P\pa_iu^j\pa_ju^i dx  \\
& = R\left(\int\sigma (\n\te-1)  \div u dx \right)_t
 -   R\si'\int (\n\te-1)\div u  dx
     -R \int \sigma \div u \n \dot\te dx\\& \quad + \int \sigma
  P\pa_iu^j\pa_ju^i dx \\&\le R \left(\int\sigma(\n\te-1) \div u
   dx \right)_t+C\si'\|\na u\|_{L^2}\|\n\te-1\|_{L^2}
     \\& \quad+C(\on)\si \|\na u\|_{L^2}  \|\n^{1/2} \dot
\te\|_{L^2}  + C(\on)\si\int \te|\na u|^2  dx
  \\&\le  R\left(\int\sigma(\n\te-1)\div u dx
\right)_t+C(\on)C_0^{1/4}\si' +\beta \si^2  \|\n^{1/2} \dot
\te\|_{L^2}^2\\&\quad+ C(\on) \de \|\n^{1/2} \dot
u\|_{L^2}^2 + C(\on, \de,M)\beta^{-1} \left( \|\na
u\|_{L^2}^2+\|\na\te\|_{L^2}^{2}\right), \ea \ee where in the last inequality we have used   (\ref{p}) and
the following simple fact:
\be \la{2.48}\ba&  \int
\te |\na u|^2dx \\ & \le
 \int|\te-1||\na u|^2dx+\int |\na u|^2dx\\ &\le C
   \|\te-1\|_{L^6}\|\na u\|_{L^2}^{3/2}
 \|\na u\|_{L^6}^{1/2}+   \|\na
u\|_{L^2}^2    \\ &\le C
   \|\na\te\|_{L^2}\|\na u\|_{L^2}^{3/2}
\left(   \|\n  \dot
u\|_{L^2}+ \|\na\te\|_{L^2}+1\right)^{1/2}+   \|\na
u\|_{L^2}^2\\ &\le \de
\left(  \|\na\te\|^2_{L^2}  + \|\n^{1/2}  \dot
u\|_{L^2}^2 \right) + C(\on,\de,M)  \|\na
u\|_{L^2}^2  ,  \ea\ee
due to (\ref{g1}), (\ref{3.35}),   and (\ref{3.30}).
Integration by parts   gives \be\la{m2} \ba
M_2 & =  \int \mu\sigma  \triangle u\cdot\dot{u}dx \\
& = -\frac{\mu }{2}\left(\sigma \|\nabla u\|_{L^2}^2\right)_t +
\frac{\mu }{2}\si'  \|\nabla u\|_{L^2}^2
-\mu \sigma  \int \p_iu^j\p_i(u^k\p_ku^j)dx \\
& = -\frac{\mu }{2}\left(\sigma \|\nabla u\|_{L^2}^2\right)_t +
\frac{\mu }{2}\si'  \|\nabla u\|_{L^2}^2
-\mu \sigma  \int \p_iu^j\p_i u^k\p_ku^j dx \\&\quad+ \frac{\mu}{2}\si
\int |\na u|^2\div udx\\
& \le  -\frac{\mu }{2}\left(\sigma \|\nabla u\|_{L^2}^2\right)_t + C
\|\na u\|_{L^2}^2 + C \int\sigma  |\nabla u|^3dx\\
& \le  -\frac{\mu }{2}\left(\sigma \|\nabla u\|_{L^2}^2\right)_t + C
\|\na u\|_{L^2}^2 +   C\sigma^2 \|\na u\|_{L^4}^4  . \ea \ee
Similar to (\ref{m2}), we have \be\la{m3}\ba  M_3 &= - \frac{\lambda+\mu}{2}\left(
\sigma \|\div u\|_{L^2}^2\right)_t+ \frac{ \lambda+\mu }{2}\si'
 \|\div u\|_{L^2}^2 \\ &\quad- (\lambda+\mu)\sigma  \int\div
u\div(u\cdot\na u)dx\\& \le -\frac{\lambda+\mu}{2}\left(\sigma
\|\div u\|_{L^2}^2\right)_t+ C  \|\na u\|_{L^2}^2 + C\sigma^2 \|\na u\|_{L^4}^4 .\ea \ee
Substituting (\ref{m1}),  (\ref{m2}), and (\ref{m3}) into (\ref{m0}), we obtain  (\ref{an1}) after choosing $\de$ suitably small.

 It remains to prove (\ref{ae0}). For     $m \ge 0,$ operating $ \si^m\dot u^j[\pa/\pa t+\div
(u\cdot)]$ to $ (\ref{a1})_2^j$ and integrating the resulting
equality over $\r^3 ,$ we obtain after integration by parts that
\be\la{m4} \ba & \left(\frac{\sigma^m}{2}\int\rho|\dot{u}|^2dx
\right)_t
-\frac{m}{2}\sigma^{m-1}\si'\int\rho|\dot{u}|^2dx\\
& =   -\int  \sigma^m \dot{u}^j[\p_jP_t +\div (\p_jPu)]dx +
\mu\int\sigma^m\dot{u}^j[\triangle u_t^j +
\div (u\triangle u^j)] dx\\
&\quad  + (\lambda+\mu)\int\sigma^m\dot{u}^j[\p_t\p_j
 \div u  +\div (u\p_j \div u )] dx \\
& \triangleq\sum_{i=1}^{3}N_i. \ea \ee
We get  after integration by parts and
using  the equation $(\ref{a1})_1$ that\be\la{m5} \ba
N_1 & = - \int \sigma^m\dot{u}^j[\p_jP_t + \div (\p_jPu)]dx \\
& =R\int \sigma^m \p_j\dot{u}^j\left(\n \dot\te-\n u\cdot\na\te-\te
 u\cdot\na\n-\te  \n\div u\right) dx \\&\quad+ \int \sigma^m  \p_k\dot{u}^j\p_jPu^k dx  \\
& =R\int \sigma^m \p_j\dot{u}^j\left(\n \dot\te
 -\te  \n\div u\right)dx
 +\int \sigma^m P \div\dot{u}  \div u dx
 \\&\quad-  \int \sigma^m  P\pa_k\dot{u}^j\p_ju^k dx
 \\&\le \frac{ \mu}{8} \int \sigma^m|\nabla\dot{u}|^2dx
 +C(\on) \si^m \left( \|\n \dot\te\|_{L^2}^2
+\int\te^2|\na u|^2dx \right) \\&\le \frac{ \mu}{8} \int
\sigma^m|\nabla\dot{u}|^2dx
 +C(\on) \si^m  \|\n^{1/2} \dot\te\|_{L^2}^2 \\ &\quad
+C(\on,M)\left( \|\na u\|_{L^2}^2+  \|\na\te\|_{L^2}^2\right)
\left(\si^m\|\n^{1/2}\dot u\|^2_{L^2} +\si^m \|\na\te\|^2_{L^2}+1
\right),\ea \ee where in the last inequality we have used (\ref{m20}).
Integration by parts leads to \be\la{m6} \ba N_2 & =  \mu\int
\sigma^m\dot{u}^j[\triangle u_t^j
 + \div (u\triangle u^j)]dx \\
& = - \mu\int \sigma^m\left(\p_i\dot{u}^j\p_iu_t^j +
\triangle u^ju\cdot\nabla\dot{u}^j\right)dx \\
& = -  \mu\int\sigma^m\left(|\nabla\dot{u}|^2 -
\p_i\dot{u}^ju^k\p_k\p_iu^j - \p_i\dot{u}^j\p_iu^k\p_ku^j +
\triangle u^ju\cdot\nabla\dot{u}^j\right)dx \\
& = - \mu\int \sigma^m\left(|\nabla\dot{u}|^2 + \p_i\dot{u}^j
\p_iu^j\div u - \p_i\dot{u}^j\p_iu^k\p_ku^j - \p_iu^j\p_iu^k\p_k\dot{u}^j
\right)dx \\
&\le -\frac{7\mu}{8} \int \sigma^m|\nabla\dot{u}|^2dx  + C \int
\sigma^m|\nabla u|^4dx  . \ea \ee Similarly, we have\be\la{m7}\ba
 N_3&  \le -\frac{7(\mu+\lambda)}{8}
 \int  \sigma^m({\rm div} \dot u)^2dx
 + C \int
\sigma^m|\nabla u|^4dx  \\ &\le   C \int \sigma^m|\nabla u|^4dx,\ea
\ee where in the second inequality we have used $\mu+\lambda\ge 0$ due to (\ref{h3}).

Substituting (\ref{m5})--(\ref{m7}) into (\ref{m4}) yields
    that there exists some $C_1$ depending only on $\mu,\lambda, \ka, R, \ga, \on,$
 and $M$ such that\be\la{e0}
\ba & \left( {\sigma^m} \int\rho|\dot{u}|^2dx \right)_t +
\frac{3\mu}{2} \int \sigma^m|\nabla\dot{u}|^2dx \\& \le
m\sigma^{m-1}\si' \int\rho|\dot{u}|^2dx  +C_1 \si^m\|
\n^{1/2}\dot\te\|_{L^2}^2+C\si^m\|\na u\|_{L^4}^4
\\& \quad+C(\on,M)\left( \|\na \te\|_{L^2}^2+  \|\na
u\|_{L^2}^2\right) \left(\si^m\|\n^{1/2}\dot u\|^2_{L^2} +\si^m
\|\na\te\|^2_{L^2} +1\right).\ea\ee Taking $m=2$ in (\ref{e0})
together with  (\ref{z1}) gives  (\ref{ae0}) directly. We thus
finish the proof of Lemma \ref{al13.4}.

\begin{lemma}\la{al13.5} In addition to the conditions of Proposition \ref{pr1}, assume that   $C_0\le 1.$  Let $(\rho,u,\te)$ be a smooth solution of (\ref{a1}) (\ref{h1})
(\ref{h2}) on $\r^3\times (0,T] $ satisfying   (\ref{z1}) with $K$ as
in Lemma \ref{le2}.   Then there exists a positive constant  $C$
  depending only on   $\mu,\lambda, \ka, R,
 \ga, \on,$
 and $M$ such that  the following
estimate holds \be\la{ae7}\ba&
 \left(\sigma^2\varphi\right)'(t) + \sigma^2 \int\left(\mu |\nabla\dot{u}|^2
  +\n(\dot \te)^2\right)dx \\ &\le C \left(\|\na u\|_{L^2}^2+\|\na
\te\|_{L^2}^2\right) +2\si \int\n|\dot u|^2 dx +C \si^2\|\na u\|_{L^4}^4,\ea
  \ee
where  $\varphi(t)$ is defined by
\be\la{wq3} \varphi(t) \triangleq   \int\n
|\dot u|^2(x,t)dx +(C_1+1)  B_2(t),\ee
with $C_1$ as in Lemma \ref{al13.4} and
\be\la{e6}
B_2(t)\triangleq\frac{\ga-1}{R}\left(\ka \|\na
\te\|_{L^2}^2-2\lambda  \int (\div u)^2\te dx-4\mu \int |\mathfrak{D}(u)|^2\te
dx\right).\ee
\end{lemma}

{\it Proof.}        For $m\ge 0,$
multiplying $(\ref{a1})_3 $ by $\sigma^m \dot\te$ and integrating
the resulting equality over $\r^3 $ yield  that
 \be\la{e1} \ba &\frac{R\sigma^m}{\ga-1} \int\rho|\dot{\te}|^2dx +\frac{\ka
{\sigma^m}}{2}\left( \|\na\te\|_{L^2}^2\right)_t
\\&=-\ka\sigma^m\int\na\te\cdot\na(u\cdot\na\te)dx
+\lambda\sigma^m\int  (\div u)^2\dot\te dx\\&\quad
+2\mu\sigma^m\int |\mathfrak{D}(u)|^2\dot\te dx-R\si^m\int\n\te \div
u\dot\te dx\\&\triangleq \sum_{i=1}^4I_i . \ea\ee

 First, it follows from  (\ref{g1}) and (\ref{z1}) that
\be\la{e2} \ba  |I_1|&\le C\sigma^m \int|\na u||\na\te|^2dx\\
&\le C\sigma^{m}   \|\na u\|_{L^2}\|\na\te\|^{1/2}_{L^2}
\|\na\te\|^{3/2}_{L^6} \\
&\le \de\sigma^{m} \|\na^2\te\|^2_{L^2} +C(\on,\de,M)\sigma^{m}
\|\na\te\|^2_{L^2}  \\
&\le     C(\on,\de,M)
\left(\|\na\te\|_{L^2}^2
  + \|\na u\|_{L^2}^2\right)\left(\si^{m}\|\n^{1/2}\dot u\|^2_{L^2} +\si^{ m}
\|\na\te\|^2_{L^2}+1 \right)\\ &  \quad+ C(\on)\de\sigma^{m} \|\n^{1/2}\dot\te\|^2_{L^2}  +C\de\si^m\|\na u\|_{L^4}^4  , \ea\ee
where in the last inequality we have used (\ref{op4}).

Next, integration by parts yields that, for any $\eta\in (0,1],$
\be\la{e3}\ba I_2 =&\lambda\si^m\int (\div u)^2 \te_t
dx+\lambda\si^m\int (\div u)^2u\cdot\na\te
dx\\=&\lambda\si^m\left(\int (\div u)^2 \te
dx\right)_t-2\lambda\si^m \int \te \div u \div (\dot u-u\cdot\na u)
dx\\&+\lambda\si^m\int (\div u)^2u\cdot\na\te
dx\\=&\lambda\si^m\left(\int (\div u)^2 \te
dx\right)_t-2\lambda\si^m\int \te \div u \div \dot
udx\\&+2\lambda\si^m \int \te \div u \div ( u\cdot\na u) dx
+\lambda\si^m\int (\div u)^2u\cdot\na\te dx\\=&\lambda\si^m
\left(\int (\div u)^2 \te dx\right)_t-2\lambda\si^m\int \te \div u
\div \dot udx\\&+2\lambda\si^m\int \te \div u \pa_i u^j\pa_j  u^i dx
+ \lambda\si^m\int u \cdot\na\left(\te   (\div u)^2 \right)dx
 \\
\le &\lambda\left(\si^m\int (\div u)^2 \te dx\right)_t-\lambda
m\si^{m-1}\si'\int (\div u)^2 \te dx\\& +\eta\si^m\|\na \dot
u\|_{L^2}^2+C\eta^{-1}\si^m\int \te^2 |\na u|^2dx+\si^m\|\na u\|_{L^4}^4
\\ \le & \lambda\left(\si^m\int (\div u)^2 \te dx\right)_t- \lambda
m\si^{m-1}\si'\int (\div u)^2 \te dx
 \\& +C(\on)\eta^{-1} \left( \|\na u\|_{L^2}^2
 +\|\na\te\|^2_{L^2}\right)
 \left(\si^m \|\n^{1/2}\dot u\|_{L^2}^2+\si^m \|\na
\te\|_{L^2}^2+1\right)\\& + \eta\si^{m}\|\na\dot u\|_{L^2}^2+ \si^{
m }\|\na u\|_{L^4}^4 ,\ea\ee where in the last inequality  we have
used  (\ref{m20}).

 Then, similar to (\ref{e3}), we have that, for
any $\eta\in (0,1],$\be \la{e5}\ba I_3&\le 2\mu\left(\si^m\int
|\mathfrak{D}(u)|^2 \te dx\right)_t-2\mu m\si^{m-1}\si'\int
|\mathfrak{D}(u)|^2 \te dx
 \\&\quad+C(\on)\eta^{-1} \left( \|\na u\|_{L^2}^2
 +\|\na\te\|^2_{L^2}\right)
 \left(\si^m \|\n^{1/2}\dot u\|_{L^2}^2+\si^m \|\na
\te\|_{L^2}^2+1\right)\\&\quad+C\eta\si^{m}\|\na\dot
u\|_{L^2}^2+C\si^m\|\na u\|_{L^4}^4  . \ea\ee

Finally, it follows from  (\ref{m20}) that
 \be\la{e39}\ba
 |I_4|   & \le C(\on)\si^m \int
\te^2|\na u|^2dx+ \frac{R}{4(\ga-1)} \si^m \int \n |\dot\te|^2dx  \\ & \le
C(\on )  \left(\|\na\te\|^2_{L^2}+\|\na u \|^2_{L^2}\right)\left(
\si^m\|\n^{1/2}\dot u\|_{L^2}^2 +\si^m\|\na\te\|_{L^2}^2 +1\right)  \\ &
\quad+ \frac{R}{4(\ga-1)}\si^m  \int \n |\dot\te|^2dx.  \ea\ee

Substituting  (\ref{e2})--(\ref{e39}) into (\ref{e1}), we obtain
after using (\ref{a2.56}),  (\ref{3.2}), and choosing $\de$ suitably
small that, for any $\eta\in (0,1],$\be\la{e7}\ba &(\si^mB_2 )'(t)
+\si^m \int\n(\dot \te)^2dx\\
&\le   C(\on,M)\eta^{-1}\left(\|\na u\|_{L^2}^2+\|\na
\te\|_{L^2}^2\right)\left(\si^m\|\n^{1/2}\dot u\|_{L^2}^2+\si^m\|\na
\te \|_{L^2}^2 +1\right)\\ & +C \eta \si^m\|\na\dot
u\|_{L^2}^2+Cm\si^{m-1}\si'\|\na
\te \|_{L^2}^2+C\si^m \|\na u\|_{L^4}^4,\ea\ee with $B_2$ as in
(\ref{e6}).
For $C_1$ as in Lemma \ref{al13.4} (see also (\ref{e0})),
 adding  (\ref{e7}) multiplied by $C_1+1$ to (\ref{e0}),
  we obtain after choosing $ \eta$ suitably small that, for $\varphi$ as in (\ref{wq3}) and for $m\ge 0,$
 \be\la{e8}\ba&
 \left(\sigma^m\varphi\right)'(t) + \sigma^m \int\left(\mu |\nabla\dot{u}|^2
  +\n(\dot \te)^2\right)dx
  \\ &\le C(\on, M)\left(\|\na u\|_{L^2}^2+\|\na
\te\|_{L^2}^2\right)\left(\si^m\|\n^{1/2}\dot u\|_{L^2}^2+\si^m\|\na
\te \|_{L^2}^2+1 \right)\\ &\quad +m\si' \si^{m-1}  \int\n|\dot u|^2
dx+C(\on,M)m\si^{m-1}\si'\|\na \te \|_{L^2}^2\\&\quad +C(\on,M)\si^m\|\na
u\|_{L^4}^4.\ea  \ee Taking $m=2$ in (\ref{e8}) together with
(\ref{z1}) gives (\ref{ae7}).  The proof of Lemma \ref{al13.5} is
completed.

Next, we will use Lemmas \ref{al13.4} and \ref{al13.5} to obtain the
following estimate on $A_3(T).$

\begin{lemma}\la{le6} Under the conditions of Proposition \ref{pr1},   there exists a positive constant $\ve_3$
depending only on   $\mu,\lambda, \ka, R, \ga, \on,$ and $M$
such that if $(\rho,u,\te)$ is a smooth solution of (\ref{a1}) (\ref{h1}) (\ref{h2}) on $\r^3\times (0,T] $ satisfying      (\ref{z1}) with $K$ as
in Lemma \ref{le2}, the following estimate holds  \be\la{b2.34}  A_3(T) \le C_0^{1/6},\ee provided $C_0\le
\ve_3.$
\end{lemma}

 {\it Proof.} First, assume that $C_0\le 1.$ It follows from (\ref{h18}), (\ref{h20}),
 (\ref{g1}), (\ref{3.35}),   and (\ref{z1}) that
\be\la{ae9}\ba  \|\na u\|_{L^4}^4&\le C \|G\|_{L^4}^4+C \|\o\|_{L^4}^4
 +C \|\n\te-1\|_{L^4}^4\\ &\le
C(\on) \left(\|\na u\|_{L^2}+1\right)\|\n^{1/2} \dot
u\|_{L^2}^3 +C\|\n(\te-1)\|_{L^4}^4 +C \|\n-1\|_{L^4}^4
\\&\le
C(\on,M) \|\n^{1/2} \dot u\|_{L^2}^3 +C(\on)
\|\n(\te-1)\|_{L^2}\|\na \te\|_{L^2}^3+C \|\n-1\|_{L^4}^4
 \\ &\le
C(\on,M) \|\n^{1/2} \dot u\|_{L^2}^3 +C(\on) \|\na
\te\|_{L^2}^3  +C \|\n-1\|_{L^4}^4,\ea\ee
which together with   (\ref{z1})  yields
\be \la{m22}\ba \si \|\na u\|_{L^4}^4   &\le C(\on,M) C_0^{1/12}
 \|\n^{1/2} \dot u\|_{L^2}^2  +C(\on)   \|\na \te\|_{L^2}^2+C\si \|\n-1\|_{L^4}^4 .\ea\ee
This fact combining with (\ref{ae7})  gives  that, for $\varphi(t)$ as in \eqref{wq3}, \be\la{e25}\ba&
 \left(\sigma^2\varphi\right)'(t) + \sigma^2 \int\left(\mu |\nabla\dot{u}|^2
  +\n(\dot \te)^2\right)dx\\ &\le C(\on, M)\left(\|\na u\|_{L^2}^2+\|\na
\te\|_{L^2}^2\right)   +\left(C(\on,M) C_0^{1/12}+2\right)\si \int\n|\dot u|^2 dx \\ & \quad+C(\on,M)\si^2\|\n-1\|_{L^4}^4\\ &\le C(\on, M)\left(\|\na u\|_{L^2}^2+\|\na
\te\|_{L^2}^2\right)  +3\si \int\n|\dot u|^2 dx +C(\on,M)\si^2\|\n-1\|_{L^4}^4,\ea
  \ee provided $C_0\le
\ve_3^1\triangleq\min\left\{1,\left( C(\on,M)\right)^{-12}
\right\}.$

Next,  to estimate the second term on the righthand side of (\ref{e25}),   we substitute (\ref{m22}) into (\ref{an1}) to obtain that, for $B_1(t)$ as in \eqref{an2},
 \be\la{n1}\ba& (\si B_1)'(t)+ \int \sigma
\rho|\dot{u}|^2dx\\&\le C(\on,M) C^{1/4}_0\si' + 2\beta \si^{2}\|
\n^{1/2}\dot\te\|_{L^2}^2 +C(\on,M )\beta^{-1}\left(
\|\na\te\|_{L^2}^2+\|\na u\|_{L^2}^2\right)  \\&\quad+ C(\on,M )
 \sigma^2  \|\n-1\|_{L^4}^4 ,\ea\ee
provided $C_0\le
\ve_3^2\triangleq\min\left\{1,(2C(\on,M))^{-12}\right\}.$
From now on, we assume that
   $C_0\le\min\left\{\ve_3^1,\ve_3^2\right\}.$
It  follows from (\ref{wq3}), (\ref{e6}), and
 (\ref{2.48}) that \be\la{wq2}  \varphi(t) \ge  \frac{1}{2}\int\n
|\dot u|^2dx +\frac{\ka(\ga-1)}{2R}\|\na\te\|_{L^2}^2-C_2(\on,M)\|\na u\|_{L^2}^2, \ee
which together with (\ref{3.35}) directly gives\be \la{ae25}
  \int\n |\dot u|^2(x,t)dx+ \|\na\te(\cdot,t)\|_{L^2}^2\le
2\left(\frac{R}{\ka(\ga-1)}+1\right) \varphi(t)+C(\on,M).\ee
For   $C_2$ as in  (\ref{wq2}), adding (\ref{n1})    multiplied by $ 2(C_2+2\mu+1)/\mu
 $  to (\ref{e25}), we obtain
  after    choosing $\beta$ suitably small  that
 \be\la{ee9}\ba&
 B_3'(t) + \frac{1}{2} \int
  \left(  \si\n  |\dot u|^2 +\mu\sigma^2|\nabla\dot{u}|^2
  +\sigma^2\n(\dot \te)^2 \right)dx
  \\&\le C(\on,M)C_0^{1/4}\si'+C(\on, M)\left(\|\na \te\|^2_{L^2}
  +\|\na u\|^2_{L^2}\right)
+C(\on, M)\si^2 \|\n-1\|_{L^4}^4, \ea
  \ee
 where
\be\la{wq03}B_3(t) \triangleq
\sigma^2\varphi+\frac{2(C_2+2\mu+1) }{\mu}\si B_1. \ee
Note that    (\ref{an2}) and (\ref{p}) lead to
\be\la{nn2}\ba B_1(t) &\ge \mu \|\nabla u\|_{L^2}^2+ (\lambda+\mu)
 \|\div u\|_{L^2}^2-\left( \frac{\mu  }{2}\|\na
u\|_{L^2}^2+C(\on) \|\n\te-1\|^2_{L^2}\right)\\
&\ge \frac{\mu  }{2}\|\nabla u\|_{L^2}^2-C(\on) C_0^{1/4}, \ea\ee
which together with (\ref{wq2}) and  (\ref{wq03}) gives \be\la{e35} B_3(t)\ge\frac{\si^2 }{2}
  \int\n |\dot u|^2dx +\frac{\ka(\ga-1)}{2R}
  \si^2\|\na\te\|_{L^2}^2+ \si \|\na u\|_{L^2}^2
  -C (\on,M) C_0^{1/4}. \ee
We claim that
\be\la{e36}\int_0^T\si^2 \|\n-1\|_{L^4}^4
dt\le C(\on,M) C_0^{1/4},\ee which combining with
    (\ref{ee9}), (\ref{e35}), and (\ref{z1})  yields
\be\la{h27}\ba A_3(T)\le C(\on,M)C_0^{1/4}\le C_0^{1/6},
  \ea\ee
provided
 $C_0\le\ve_3\triangleq\min\left\{\ve_3^1,\ve_3^2,
 (C(\on,M))^{-12} \right\}.$ 

Finally, it remains to prove (\ref{e36}). In fact, it follows from $(\ref{a1})_1$  and (\ref{ez1}) that $\n-1$ satisfies
 \be
\la{a95}\ba &(\n-1)_t+\frac{R}{2\mu+\lambda}(\n-1)\\&=-u\cdot\nabla (\n-1)-(\n-1){\rm div}u-\frac{G}{2\mu+\lambda}- \frac{R\n(\te-1) }{2\mu+\lambda} .\ea\ee  Multiplying (\ref{a95}) by $4(\n-1)^3$ and integrating
the resulting equality over $\r^3 ,$  we obtain  that
 \be\ba \la{ae96} &\left(\|\n-1\|_{L^4}^4 \right)_t+\frac{4R}{2\mu+\lambda}
 \|\n-1\|_{L^4}^4 \\& = -3
 \int (\n-1)^4{\rm div}udx
 -\frac{4}{2\mu+\lambda}\int (\n-1)^3Gdx  \\&
\quad - \frac{4R}{2\mu+\lambda}\int (\n-1)^3\n(\te-1)dx
  \\& \le \frac{2R}{2\mu+\lambda}
 \|\n-1\|_{L^4}^4+C(\on)\|\na u\|_{L^2}^2
 +C\|\n-1\|_{L^4}^3 \|G\|_{L^2}^{1/4}\|\na G\|_{L^2}^{3/4}
\\& \quad +C(\on)\|\n-1\|_{L^4}^3
 \|\n(\te-1)\|_{L^2}^{1/4}\|\na \te\|_{L^2}^{3/4}
  \\&
\le \frac{3R}{2\mu+\lambda} \|\n-1\|_{L^4}^4 +C(\on) \|\nabla u\|_{L^2}^2
 +C(\on,M)\left(\|\n^{1/2}\dot u
 \|_{L^2}^3+\|\na\te\|_{L^2}^3\right)
,\ea\ee where in the last inequality,
 we have used (\ref{z1}), (\ref{3.35}), (\ref{p}), (\ref{h19}),  and (\ref{a2.12}).
It thus follows from (\ref{ae96}) that
\be\ba \la{a96}
 &\left(\|\n-1\|_{L^4}^4 \right)_t+\frac{ R}{2\mu+\lambda}
 \|\n-1\|_{L^4}^4 \\ &\le C(\on,M)\left(\|\n^{1/2}\dot u
 \|_{L^2}^3+\|\na\te\|_{L^2}^3\right)
 +C(\on) \|\nabla u\|_{L^2}^2 .\ea\ee
 Multiplying (\ref{a96}) by $\si^n$ with $n\ge 1,$  integrating
 the resulting inequality over $(0,T),$   we obtain by using  (\ref{a2.12}) and  (\ref{z1}) that
\be\la{wq1}\ba &\int_0^T\si^n  \|\n-1\|_{L^4}^4 dt\\& \le
C(\on,M) A_3^{1/2}(T)
   \int_0^T \si^{n-1}\left( \|\n^{1/2}\dot u\|_{L^2}^2+
  \|\na\te\|_{L^2}^2\right)dt\\& \quad+
C(\on)C_0^{1/4} +C\int_0^{\si(T)}\|\n-1\|_{L^4}^4  dt
 \\& \le C(\on,M)C_0^{1/4}+C(\on,M)C_0^{1/12}
  \int_0^T \si^{n-1}  \|\n^{1/2}\dot u\|_{L^2}^2 dt,\ea\ee
which together with  (\ref{z1}) directly gives \eqref{e36}. We thus complete the proof of Lemma
 \ref{le6}.

We now proceed to derive a uniform (in time) upper bound for the
density, which turns out to be the key to obtain all the higher
order estimates and thus to extend the classical solution globally.

\begin{lemma}\la{le7}Under the conditions of Proposition \ref{pr1},
    there exists a positive constant $\ve_4$
depending only on   $\mu,\lambda, \ka, R, \ga, \on,$ and $M$
such that if $(\rho,u,\te)$ is a smooth solution of (\ref{a1}) (\ref{h1}) (\ref{h2}) on $\r^3\times (0,T] $ satisfying    (\ref{z1}) with $K$ as
in Lemma \ref{le2}, the following estimate holds
      \be \la{a3.7}\sup_{0\le t\le T}\|\n(\cdot,t)\|_{L^\infty}  \le
\frac{3\bar \n }{2}  ,\ee
      provided $C_0\le \ve_4 . $

   \end{lemma}

{\it Proof.} First, assume $C_0\le 1.$ Taking $n=1$ in (\ref{wq1}) as well as  (\ref{z1}) yields
 \be\la{e37}\int_0^T\si  \|\n-1\|_{L^4}^4 dt\le
C(\on,M) .\ee
      Choosing $m=1$  in (\ref{e8}) together with (\ref{ae25})  and (\ref{m22}) yields that, for $\varphi(t)$ as in \eqref{wq3},
   \bnn  \ba &(\si\varphi)'(t)+\si \int\left(\mu |\na\dot
u|^2+\n(\dot\te)^2\right)dx\\ &\le C(\on,M)\left(\|\na
u\|_{L^2}^2+\|\na \te\|_{L^2}^2\right)(\si\varphi)+C(\on,M)\left(\|\na
u\|_{L^2}^2+\|\na \te\|_{L^2}^2\right)\\&\quad +C(\on,M)\int \n|\dot
u|^2dx+C(\on,M)\si\|\n-1\|_{L^4}^4,\ea\enn
which combining with
    (\ref{z1}),   (\ref{e37}),   and Gronwall's
  inequality yields that
  \be \la{ae36}\sup_{0\le t\le\si(T)}\si\varphi(t)+\int_0^{\si(T)}\si \int\left(\mu |\na\dot
u|^2+\n(\dot\te)^2\right)dxdt\le C(\on,M).\ee
The combination of
(\ref{ae25}) with (\ref{ae36}) thus directly gives
\be\la{ae26}\ba &\sup_{0\le t\le \si(T)}\si \left(
\int\rho|\dot{u}|^2dx+
  \|\na\te\|_{L^2}^2\right) \\&+\int_0^{\si(T)} \sigma \int
  ( |\nabla\dot{u}|^2+\n(\dot \te)^2)dxdt \le C(\on,M).\ea\ee

 Next, it follows from (\ref{op4}), (\ref{ae26}),
  (\ref{m22}), (\ref{z1}), and (\ref{e37}) that
 \bnn\ba  &\int_0^{T }\si \|\na^2\te\|_{L^2}^2dt\\
&\le   C(\on,M)\int_0^{T } \left(\si \|\n \dot\te\|_{L^2}^2+
\|\n^{1/2}\dot u \|_{L^2}^2+\| \na u\|_{L^2}^2+\| \na\te\|_{L^2}^2+\si\|\n-1\|_{L^4}^4\right) dt \\
&\le C(\on,M),\ea\enn which together with  (\ref{g2}),   (\ref{g1}),
and (\ref{z1}) gives \be\la{3.88}\ba&
\int_0^{\si(T)}\|\te-1\|_{L^\infty}dt
 \\ &\le
C\int_0^{\si(T)}\| \te-1\|_{L^6}^{1/2}\|\na\te\|_{L^6}^{1/2}dt\\
&\le C\int_0^{\si(T)}\|\na\te\|_{L^2}^{1/2}
\left(\si\|\na^2\te\|^2_{L^2}\right)^{1/4}\si^{-1/4}dt\\
&\le C\left(\int_0^{\si(T)} \|\na \te\|_{L^2}^2dt\right)^{1/4}
\left(\int_0^{\si(T)}\si\|\na^2\te\|_{L^2}^2dt\right)^{1/4}
 \left(\int_0^{\si(T)}\si^{-1/2}dt\right)^{1/2}\\ &\le
 C(\on,M)C_0^{1/16},
 \ea\ee
and  \be\la{3.89}\ba& \int_{\si(T)}^T\|\te-1\|^2_{L^\infty}dt \\
&\le C\int_{\si(T)}^T\|\na\te\|_{L^2} \|\na^2\te\|_{L^2}dt\\ &\le
C\left(\int_{\si(T)}^T\|\na\te\|^2_{L^2}dt\right)^{1/2}\left(\int_{\si(T)}^T
\|\na^2\te\|^2_{L^2}dt\right)^{1/2}\\ &\le C(\on,M) C_0^{1/8}.
 \ea\ee

Next, (\ref{g2}), (\ref{h19}), (\ref{ae26}), and (\ref{z1}) lead to
 \be\la{3.90}\ba &\int_0^{\si(T)}\|G\|_{L^\infty}dt
\\ &\le C\int_0^{\si(T)}\|\na G\|_{L^2}^{1/2}\|\na G\|_{L^6}^{1/2}dt
 \\ &\le C(\on)\int_0^{\si(T)}\|\n \dot u\|_{L^2}^{1/2}\|\na\dot u\|_{L^2}^{1/2}dt
 \\ &\le C(\on)\int_0^{\si(T)}\left(\si\|\n
  \dot u\|_{L^2}\right)^{1/4}
 \left(\si\|\n \dot u\|^{2}_{L^2}\right)^{1/8}
 \left(\si\|\na \dot u\|^2_{L^2}\right)^{1/4}\si^{-5/8}dt
  \\ &\le C(\on,M)C_0^{1/48}\int_0^{\si(T)}
 \left(\si\|\na \dot u\|^2_{L^2}\right)^{1/4}\si^{-5/8}dt\\ &\le C(\on,M)C_0^{1/48}\left(\int_0^{\si(T)}
 \si\|\na \dot u\|^2_{L^2} dt\right)^{1/4}\left(\int_0^{\si(T)}
 \si^{-5/6}dt\right)^{3/4}\\ &\le C(\on,M)C_0^{1/48},
    \ea\ee
and
 \be\la{3.91}\ba  \int_{\si(T)}^T\|G\|^2_{L^\infty}dt
  &\le C\int_{\si(T)}^T\|\na G\|_{L^2} \|\na G\|_{L^6} dt
 \\ &\le C(\on)\int_{\si(T)}^T\|\n \dot u\|_{L^2}\|\na\dot u\|_{L^2}dt
 \\ &\le C(\on)\int_{\si(T)}^T\left(\|\n^{1/2} \dot u\|^2_{L^2}+
 \|\na\dot u\|_{L^2}^2\right)dt\\ &\le C(\on)C_0^{1/6}.
    \ea\ee

Finally, denoting $ D_t\n=\n_t+u \cdot\nabla \n $   and
expressing $(\ref{a1})_1$ in terms of the Lagrangian coordinates, we obtain by (\ref{ez1}) that
\bnn\ba (2\mu+\lambda) D_t \n&=-R\n(\n-1)-R\n^2(\te-1)-\n G\\ &\le
-R (\n-1)+C(\on)\|\te-1\|_{L^\infty}+C(\on)\| G\|_{L^\infty},
\ea\enn which gives
 \be\la{3.92}\ba  D_t (\n-1)+\frac{R}{2\mu+\lambda} (\n-1)\le C(\on)\|\te-1\|_{L^\infty}+C(\on)\| G\|_{L^\infty}.
\ea\ee

Taking $$y=\n-1, \quad\al=\frac{R}{2\mu+\lambda} ,\quad g(t)=C(\on)\|\te-1\|_{L^\infty}+C(\on)\| G\|_{L^\infty},\quad T_1=\si(T),$$ in Lemma \ref{le1}, we thus deduce from (\ref{3.92}), (\ref{3.88})--(\ref{3.91}), and (\ref{2.34})  that
 \bnn\ba\n& \le \on+1 +C\left(\|g\|_{L^1(0,\si(T))}+\|g\|_{L^2(\si(T),T)}\right)\\&\le \on+1 +C(\on,M)C_0^{1/48}\le \frac{3\on}{2},\ea\enn
 provided $$C_0\le \ve_4\triangleq\min\left\{ 1,
  \left(\frac{\bar \n-2 }{2C(\on,M) }\right)^{48}\right\}.$$
We thus complete the proof of Lemma \ref{le7}.

Next, the following Lemma \ref{le8} will give an estimate on
$A_4(T),$ which together with Lemmas \ref{le2}, \ref{le3}, \ref{le6} and \ref{le7} finishes the proof of Proposition \ref{pr1}.
\begin{lemma}\la{le8} Under the conditions of Proposition \ref{pr1},
  there exists a positive constant $\ve_0$
depending only on   $\mu,\lambda, \ka, R, \ga, \on,$ and $M$ such
that  if $(\rho,u,\te)$ is a smooth solution of (\ref{a1})
(\ref{h1}) (\ref{h2}) on $\r^3\times (0,T] $ satisfying (\ref{z1})
with $K$ as in Lemma \ref{le2}, the following estimate holds  \be
\la{ae3.7}A_4(T)\le C_0^{1/8} ,\ee
      provided $C_0\le \ve_0 . $

   \end{lemma}

{\it Proof.} It follows from (\ref{op4}),   (\ref{z1}), (\ref{m22}), and (\ref{e36}) that
 \be\la{3.97}\ba  &\int_0^{T }\si^2 \|\na^2\te\|_{L^2}^2dt\\
&\le  C(\on,M)\int_0^{T }
\left(\si^2 \|\n \dot\te\|_{L^2}^2+\si\|\n\dot u \|_{L^2}^2
+\| \na u\|_{L^2}^2
+\| \na\te\|_{L^2}^2+\si^2\|\n-1\|_{L^4}^4\right) dt \\
&\le C(\on,M)C_0^{1/6}.\ea\ee
Applying the operator $\pa_t+\div(u\cdot) $ to (\ref{a1})$_3,$ we
use  (\ref{a1})$_1$ to get\be\la{3.96}\ba
&\frac{R}{\ga-1} \n \left(\pa_t\dot \te+u\cdot\na\dot \te\right)\\
&=\ka \Delta \dot \te+\ka\left( \div u\Delta \te  -
\pa_i\left(\pa_iu\cdot\na \te \right)- \pa_iu\cdot\na \pa_i\te
\right)\\&\quad +\left( \lambda (\div u)^2+2\mu |\mathfrak{D}(u)|^2\right)\div u
+R\n \te  \pa_ku^l\pa_lu^k
\\&\quad -R\n \dot\te \div u-R\n \te\div \dot u
 +2\lambda \left( \div\dot u-\pa_ku^l\pa_lu^k\right)\div u\\&\quad
+ \mu (\pa_iu^j+\pa_ju^i)\left( \pa_i\dot u^j+\pa_j\dot u^i-\pa_iu^k\pa_ku^j
-\pa_ju^k\pa_ku^i\right).\ea\ee

Multiplying (\ref{3.96}) by
$\dot \te,$  we obtain
after integration by parts that \be\la{3.99}\ba &
\frac{R}{2(\ga-1)}\left(\int \n
|\dot\te|^2dx\right)_t + \ka   \|\na\dot\te\|_{L^2}^2 \\&\le
 C  \int|\na u|\left(|\na^2\te||\dot\te|+ |\na \te| |\na
\dot\te|\right)dx +C(\on)  \int|\na
u|^2|\dot\te|\left(|\na u|+|\te-1| \right)dx
 \\&\quad +C (\on)  \int\left( |\na u|^2|\dot \te|+\n  |\dot
\te|^2|\na u|\right)dx  +C   \int |\na\dot u|\n|\dot \te| dx \\
&\quad+C(\on)\int  \n|\te-1| |\na\dot u| |\dot
\te|dx +C(\on)  \int  |\na u| |\na\dot u| |\dot
\te|dx \\&\le C  \|\na u\|^{1/2}_{L^2}\|\na
u\|^{1/2}_{L^6}\|\na^2\te\|_{L^2}\|\na\dot \te\|_{L^2} \\ &
\quad+C(\on)  \|\na u\|_{L^2}\|\na u\|_{L^6}\left(\|\na
u\|_{L^6}+\|\na \te\|_{L^2}\right)
\|\na\dot\te\|_{L^2}  \\
&\quad+C(\on)  \|\na u\|^{1/2}_{L^6}\|\na u\|^{1/2}_{L^2}
\|\na\dot\te\|_{L^2}\left(\|\na u\|_{L^2}
+\|\n\dot\te\|_{L^2}\right)  \\
&\quad+C(\on)    \|\na\dot u\|_{L^2} \|\n\dot\te\|_{L^2}
 +C(\on)\|\n^{1/2}(\te-1)\|^{1/2}_{L^2}\|\na\te\|_{L^2}^{1/2}
 \|\na\dot u\|_{L^2} \|\na\dot\te\|_{L^2}  \\&\quad+C(\on)   \|\na
u\|^{1/2}_{L^2}\|\na u\|^{1/2}_{L^6}
 \|\na\dot u\|_{L^2} \|\na\dot\te\|_{L^2} .   \ea\ee
Multiplying (\ref{3.99}) by
$\si^4$ and integrating the resulting inequality over
$(0,t),$  we obtain
after integration by parts and using (\ref{3.42}), (\ref{z1}), (\ref{p}),
  and (\ref{3.97}) that\bnn\ba & \frac{R}{2(\ga-1)} \si^4\int \n
|\dot\te|^2dx + \ka\int_0^t\si^4 \|\na\dot\te\|_{L^2}^2ds  \\&\le
C\int_0^{\si(t)}\si^2 \int \n |\dot\te|^2dxds+C(\on ) \int_0^t\si^{3}
 \|\na^2\te\|_{L^2}\|\na\dot
\te\|_{L^2}ds \\
& \quad+C(\on ) \int_0^t\si^{2}\|\na u\|_{L^2}
\|\na\dot\te\|_{L^2}ds +C(\on ) \int_0^t\si^{3}
\|\na\dot\te\|_{L^2}\|\n\dot\te\|_{L^2} ds \\
&\quad+C(\on ) \int_0^t \si^{3} \|\na\dot u\|_{L^2}
\|\n^{1/2}\dot\te\|_{L^2} ds +C(\on ) \int_0^t\si^{3}
 \|\na\dot u\|_{L^2}   \|\na\dot\te\|_{L^2} ds\\ &\le C(\on)
 \int_0^t\left( \si^2\|\na^2\te\|_{L^2}^2+\|\na u\|_{L^2}^2
 +\si^2\|\n^{1/2}\dot\te\|_{L^2}^2+\si^2\|\na\dot u\|_{L^2}^2
 \right)ds\\ &\quad+\frac{\ka}{2}\int_0^t\si^4\|\na\dot\te\|_{L^2}^2ds
 \\ &\le  C(\on,M)C_0^{1/6}+ \frac{\ka}{2}\int_0^t\si^4\|\na\dot\te\|_{L^2}^2ds, \ea\enn
which  yields that
\bnn \sup_{0\le t\le T} \si^4\int \n |\dot\te|^2dx +
\int_0^T\si^4 \|\na\dot\te\|_{L^2}^2ds \le C(\on,M)C_0^{1/6}\le
C_0^{1/8},\enn provided \be\la{ve} C_0\le \ve_0 \triangleq
\min\limits_{1\le i\le 5}\ve_i, \quad\mbox{ with }\quad \ve_5\triangleq\min\left\{1,
  \left( C(\on,M)\right)^{-24}\right\}.\ee We thus finish the proof of
  Lemma \ref{le8}   with $\ve_0 $ as in (\ref{ve}).

Finally, in the following Corollary \ref{cor1}, we summarize some  estimates on $(\n,u,\te)$ which will be useful for higher order ones in the next section.
\begin{cor} \la{cor1} In addition to the conditions of Proposition \ref{pr1}, assume that $(\rho_0,u_0,\te_0)$ satisfies     (\ref{z01}) with $\ve_0$ as in Proposition \ref{pr1}. Then there exists a  positive constant    $C $     depending only   on  $\mu,\lambda, \ka, R, \ga, \on,$ and $M$  such that if
         $(\rho,u,\te)$  is a smooth solution of
       (\ref{a1}) (\ref{h1}) (\ref{h2})  on $\r^3\times (0,T] $
 satisfying (\ref{z1}) with $K$ as
in Lemma \ref{le2},    the following estimate holds
 \be\la{vu15}\ba &\sup_{t\in (0,T]}\left(  \si^2\|\na u \|^2_{L^6}+\si^4\| \te-1\|^2_{H^2}\right)\\& +\int_0^T\si^2(\|\n -1\|_{L^4}^4+\|\na u \|_{L^4}^4+\|\na\te \|_{H^1}^2+\|u_t\|^2_{L^2}+\si^2\|\te_t\|^2_{H^1})dt\le CC_0^{1/8}. \ea\ee
 \end{cor}

  {\it Proof. } It follows from (\ref{z1}), (\ref{f}), (\ref{a2.17}), (\ref{3.42}), (\ref{3.26}), (\ref{m22}), (\ref{e36}), and  (\ref{3.97}) that \be \la{vu02}\ba &\sup_{t\in (0,T]}\left(  \si^2\|\na u \|^2_{L^6}+\si^4\| \te-1  \|^2_{H^2}\right)\\& +\int_0^T \si^2(\|\na u  \|_{L^4}^4+\|\na\te \|_{H^1}^2+\|\n -1\|_{L^4}^4 )dt\le CC_0^{1/8}, \ea\ee  which together with  (\ref{z1})   and  (\ref{f}) gives that
 \be\la{vu12}\ba   &\int_0^T  \si^2 \|  u _t\|_{L^2}^2dt  \\&\le
 C\int_0^T  \si^2\| \dot u \|_{L^2}^2  dt+ C\int_0^T  \si^2\|u
 \cdot\na  u \|_{L^2}^2dt\\ &\le C\int_0^T  \si^2\int\n |\dot u |^2dx
  dt+C\int_0^T  \si^2\|\na\dot u \|_{L^2}^2dt \\&\quad+C \int_0^T  \si^2\|u \|_{L^6}^2\|\na  u \|_{L^3}^2dt\\ &\le C C_0^{1/6} ,\ea\ee
 \be\la{vu11}\ba   &\int_0^T  \si^4 \|  \te _t\|_{L^2}^2dt
 \\&\le C\int_0^T  \si^4\| \dot \te \|_{L^2}^2  dt+ C\int_0^T  \si^4\|u \cdot\na  \te \|_{L^2}^2dt\\ &\le C\int_0^T  \si^4\int\n |\dot \te |^2dx dt+C\int_0^T  \si^4\|\na\dot \te \|_{L^2}^2dt \\&\quad+C \int_0^T  \si^4\|u \|_{L^6}^2\|\na  \te \|_{L^3}^2dt\\ &\le CC_0^{1/8} +C \int_0^T   \|\na u \|_{L^2}^2 dt \\ &\le CC_0^{1/8} ,\ea\ee
 and\be\la{vu01}\ba   &\int_0^T  \si^4 \|  \na\te _t\|_{L^2}^2dt
  \\&\le C\int_0^T  \si^4\|\na \dot \te \|_{L^2}^2  dt+ C\int_0^T  \si^4\|\na(u \cdot\na  \te )\|_{L^2}^2dt\\ &\le CC_0^{1/8} +C\int_0^T  \si^4\left(\|\na u \|_{L^3}^2+\|  u \|_{L^\infty}^2\right)\|\na^2 \te \|_{L^2}^2dt  \\ &\le C C_0^{1/8}.\ea\ee
 We thus obtain (\ref{vu15}) directly from (\ref{vu02})--(\ref{vu01})  and finish the proof of Corollary \ref{cor1}.

 \section{\la{se4}A priori estimates (II): Higher-order estimates}

In this section,  we will derive
 the higher order estimates of a smooth solution $(\rho, u, \te)$
of (\ref{a1}) (\ref{h1}) (\ref{h2}) on $ \r^3\times (0,T]$ with smooth $(\n_0 ,u_0,\te_0) $ satisfying (\ref{co3}) and (\ref{3.1}). Moreover, we shall always assume that $(\n,u,\te)$ and   $(\n_0 ,u_0,\te_0)$ satisfy respectively (\ref{z1}) and (\ref{z01}).
To proceed,
we  define $\tilde{g_1} $ and $\tilde{g_2} $     as \be \la{co12}
 \tilde{g_1}\triangleq\n_0^{-1/2}\left( -\mu \Delta u_0-(\mu+\lambda)\na\div u_0+R\na (\n_0\te_0)\right)
   \ee and  \be \la{co13} \tilde{g_2}\triangleq \n_0^{-1/2}\left( \ka\Delta \te_0+\frac{\mu}{2}|\na
u_0+(\na u_0)^{\rm tr}|^2+\lambda (\div u_0)^2\right), \ee
respectively. It thus follows from (\ref{co3}) and (\ref{3.1}) that
 \be\la{wq01}\tilde{g_1}\in L^2, \quad \tilde{g_2}\in L^2  .\ee
 From now on,
  the generic constant $C $ will depend only  on \bnn
T, \,\,\| \tilde{g_1}\|_{L^2},    \, \| \tilde{g_2}\|_{L^2},\, \,
 \|  u_0\|_{H^2},  \ \,
    \|\n_0-1\|_{H^2\cap W^{2,q}}  ,   \,  \,\| \te_0-1\|_{H^2} , \enn
besides  $\mu,\lambda, \ka, R, \ga, \on,$ and $M.$

We begin with the important estimates on the spatial gradient of
the smooth solution $(\rho,u,\te).$

\begin{lemma}\la{le11}
 The following estimates hold
  \be\label{lee2}\ba &
\sup_{0\le t\le T} \left(\|\rho^{1/2}\dot u\|_{L^2}
+\|\te-1\|_{H^1} \right)+\ia\int \n(\dot\te)^2 dxdt\\&\quad  +\ia
 \left(\|\nabla\dot
u\|_{L^2}^2+\|\nabla^2\te\|_{L^2}^2+ \|\div u\|^2_{L^\infty}+\|\o\|^2_{L^\infty} \right) dt\le C,\ea\ee
 \be\la{qq1}
  \sup_{0\le t\le T}\left(\|\n-1\|_{H^1\cap W^{1,6}}
   + \|u\|_{H^2}
  \right) + \int_0^{T}\|\nabla u\|^{3/2}_{L^{\infty}}
   dt\le C .
  \ee
\end{lemma}

{\it Proof. } We first prove (\ref{lee2}).
  Taking $m=0$ in (\ref{e8}) gives that, for $\varphi(t)$ as in \eqref{wq3},
 \be\la{ae8}\ba&
 \varphi'(t) +   \int\left(\mu |\nabla\dot{u}|^2
  +\n(\dot \te)^2\right)dx
  \\ &\le C \left(\|\na u\|_{L^2}^2+\|\na
\te\|_{L^2}^2\right)\left( \|\n^{1/2}\dot u\|_{L^2}^2+ \|\na
\te \|_{L^2}^2+1 \right)   +C\|\na u\|_{L^2}\|\na u\|_{L^6}^3\\ &\le C  \left( \|\n^{1/2}\dot u\|_{L^2}^4
+ \|\na\te\|_{L^2}^4 \right) +C\\ &\le C   \left( \|\n^{1/2}\dot u\|_{L^2}^2
+ \|\na\te\|_{L^2}^2\right) \varphi(t)+C \left( \|\n^{1/2}\dot u\|_{L^2}^2
+ \|\na\te\|_{L^2}^2 \right) +C
,\ea
  \ee  due to   (\ref{z1}),
(\ref{3.30}),  and (\ref{ae25}). It follows from $(\ref{a1})_2,$ (\ref{3.1}), and (\ref{co12}) that   \be \ba
\la{mm6} \lim_{t\rightarrow 0^+}\sqrt{\n} \dot u(x,t)=
 \n_0^{-1/2}\left( \mu \Delta u_0+(\mu+\lambda)\na\div u_0-R\na
 (\n_0\te_0)\right)=-\tilde{g_1},\ea\ee which together with
  (\ref{wq3}),   (\ref{e6}),   (\ref{2.48}), and (\ref{wq01}) yields that
  \be \la{wq02}  \lim_{t\rightarrow 0^+}|\varphi(t)|\le
  C\|\tilde g_1\|_{L^2}^2+C\le C.\ee

 Gronwall's
inequality together with (\ref{ae8}), (\ref{wq02}),
 and (\ref{z1})  leads to
 \bnn
\sup_{0\le t\le T} \varphi(t) +
 \ia\int\left(|\nabla\dot
u|^2+\n(\dot\te)^2\right)dxdt\le C ,\enn which as well as
 (\ref{ae25})   and (\ref{a2.17}) implies
 \be\label{lee3}
\sup_{0\le t\le T} \left(\|\rho^{1/2}\dot u\|_{L^2}
+\|\te-1\|_{H^1} \right) +
 \ia\int\left(|\nabla\dot
u|^2+\n(\dot\te)^2\right)dxdt\le C.\ee One thus deduces from
(\ref{op4}), (\ref{lee3}),    (\ref{ae9}), and    (\ref{z1})
 that
\be \la{4.10}\ba \int_0^T\|\na^2\te\|_{L^2}^2dt \le C+ C\int_0^T
\|\n^{1/2}\dot\te\|_{L^2}^2 dt \le C,\ea\ee which together with
  (\ref{hj1}),  (\ref{h19}), (\ref{g1}), (\ref{lee3}), and    (\ref{z1})   gives
 \bnn \ba    & \int_0^T\left(\|\div u\|^2_{L^\infty}+\|\o\|^2_{L^\infty} \right)dt
 \\&\le C\int_0^T\left(\|G\|^2_{L^\infty}+\|\n\te-1\|^2_{L^\infty}
 +\|\o\|^2_{L^\infty}\right)dt+C\\&\le
C\ia\left(\| G\|^2_{  L^6}+\|\nabla G\|^2_{  L^6}+\|\te-1\|_{L^\infty}^2
+\| \o\|^2_{L^6}+\|\nabla \o\|^2_{  L^6}\right)dt+C
\\&\le C\ia\left(\|\na
G\|^2_{ L^2}+\|\n\dot u\|^2_{  L^6}+\|\na^2\te\|^2_{L^2}+\|\na
\o\|^2_{ L^2}\right)dt+C     \\ &\le C\ia\left(\|\n \dot u\|^2_{L^2}
+\|\na^2\te\|^2_{L^2}+\|\nabla \dot
u\|_{L^2}^{2}\right) dt+C \\
&\le C.\ea\enn This fact combining with (\ref{lee3}) and (\ref{4.10}) yields (\ref{lee2})  directly.

Next, we will      prove the key estimate
 (\ref{qq1}).
 For $ 2\le p\le 6,$ $|\nabla\rho|^p$ satisfies
\bnnn \ba
& (|\nabla\rho|^p)_t + \div (|\nabla\rho|^pu)+ (p-1)|\nabla\rho|^p\div u  \\
 &+ p|\nabla\rho|^{p-2}(\nabla\rho)^t \nabla u (\nabla\rho) +
p\rho|\nabla\rho|^{p-2}\nabla\rho\cdot\nabla\div u = 0.\ea \ennn
Thus,  \be\la{L11}\ba
 \partial_t\norm[L^p]{\nabla\rho} &\le
 C(1+\norm[L^{\infty}]{\nabla u} )
\norm[L^p]{\nabla\rho}+C\|\na^2u\|_{L^p}
\\ &\le
 C\left(1+\|\na^2\te\|_{L^2}+\|\na u\|_{L^{\infty}}\right)
\norm[L^p]{\nabla\rho}\\&
 \quad +C\left(1+\|\na\dot u\|_{L^2}+\|\na^2\te\|_{L^2}
\right), \ea\ee where we have used \be\ba
 \la{ua1}\|\na^2 u\|_{L^p}&\le   C\left(\|\n\dot u\|_{L^p}+
  \|\nabla P\|_{L^p}\right)\\ & \le   C\left(\|\n\dot u\|_{L^2}+\|\na\dot u\|_{L^2}+
  \|\na\te\|_{L^p}+  \|\te \|_{L^\infty} \|\nabla\n\|_{L^p}\right)\\ & \le   C\left(1+\|\na\dot u\|_{L^2}+
  \|\na^2\te\|_{L^2}
+ (\|\na^2\te \|_{L^2} + 1)\|\nabla\n\|_{L^p}\right) ,\ea\ee which comes from the standard $L^p$-estimate of the following
elliptic system:
 \be \la{zp202}  \mu\Delta
u+(\mu+\lambda)\na {\rm div}u=\n \dot u+\na P,\quad \, u\rightarrow
0\,\,\mbox{ as } |x|\rightarrow \infty. \ee
  It follows from    Lemma \ref{le9} and (\ref{ua1})  that
    \be\la{u13}\ba   \|\na
u\|_{L^\infty }  &\le C\left(\|{\rm div}u\|_{L^\infty }+
\|\o\|_{L^\infty } \right)\log(e+\|\na^2 u\|_{L^6 }) +C\|\na
u\|_{L^2} +C   \\&\le C\left(\|{\rm div}u\|_{L^\infty }+ \|\o\|_{L^\infty }
\right)\log(e+\|\na\dot u\|_{L^2 }+ \|\na^2\te\|_{L^2 })\\&\quad
+C\left(\|{\rm div}u\|_{L^\infty }+ \|\o\|_{L^\infty } \right)
\log\left(e  +\left(e+\|\na^2\te\|_{L^2} \right)\|\na
\n\|_{L^6}\right) +C\\&\le C\left(\|{\rm div}u\|_{L^\infty }+ \|\o\|_{L^\infty }
\right)\log(e+\|\na\dot u\|_{L^2 }+ \|\na^2\te\|_{L^2 })\\&\quad
+C\left(\|{\rm div}u\|_{L^\infty }+ \|\o\|_{L^\infty } \right)
\log\left(e  + \|\na \n\|_{L^6}\right) +C. \ea\ee
Set \be\la{gt}\begin{cases}
 f(t)\triangleq  e+\|\na
\n\|_{L^6},\\ g(t)\triangleq  1+\left(\|{\rm div}u\|_{L^\infty }+
\|\o\|_{L^\infty } \right)\log(e+\|\na \dot u\|_{L^2}+
 \|\na^2\te\|_{L^2})\\ \quad\quad\quad
+\|\na \dot
u\|_{L^2}+ \|\na^2\te\|_{L^2}.\end{cases}\ee
The combination of (\ref{u13}) with (\ref{L11}), where we set   $p=6,$ gives
\bnn f'(t)\le C g(t) f(t)+ C g(t) f(t)\ln f(t)+Cg(t),\enn
 which yields \be \la{hb1} (\ln f(t))'\le Cg(t)+Cg(t)\ln f(t),\ee
due to $f(t)>1.$
 It thus follows from (\ref{gt}), (\ref{hb1}),  (\ref{lee2}),  and Gronwall's
 inequality that \bnn \sup\limits_{0\le t\le T}
   f(t)\le C,\enn
which shows \be \la{u113} \sup\limits_{0\le t\le T}\|\nabla
\rho\|_{L^6}\le C.\ee  Therefore, we deduce from
 (\ref{u13}),   (\ref{u113}), and (\ref{lee2}) that
  \be \la{v6}\ia\|\nabla u\|^{3/2}_{L^\infty}dt \le C.\ee

  Next, taking $p=2$ in (\ref{L11}), we get by  using (\ref{v6}),
  (\ref{lee2})
and  Gronwall's inequality that\be  \la{aa93}
 \sup\limits_{0\le t\le T}\|\nabla \rho\|_{L^2}\le C,\ee
which together with (\ref{lee2}) and
 (\ref{u113}) gives \be\la{aa94}\ba
  \sup\limits_{0\le t\le T}\|\nabla P\|_{L^2}  &\le
 C\sup\limits_{0\le t\le T}\left(\|\na\te\|_{L^2}+\left(\|\na\n\|_{L^2}
  +\|\te-1\|_{L^6}\|\na\n\|_{L^3}\right)\right)\\ &\le C.
    \ea\ee
    This fact combining with (\ref{ua1}) and (\ref{lee2})  leads to
      \be\ba\la{aa95}\sup\limits_{0\le t\le T} \|\na^2 u\|_{L^2}&\le C \sup\limits_{0\le t\le T}\left(\|\n\dot u\|_{L^2}+
  \|\nabla P\|_{L^2}\right)\\&\le C.\ea\ee Hence, (\ref{qq1}) follows directly from (\ref{a2.12}),
(\ref{f}),  (\ref{aa95}),    and (\ref{v6}).
  The proof of Lemma \ref{le11} is completed.

\begin{lemma}\la{le9-1}
 The following estimates hold
\be\la{va5}\ba&
   \sup\limits_{0\le t\le T}\left(
   \|\n_t\|_{H^1} +\|\te-1\|_{H^2}+\|\n-1\|_{H^2}
   +\|u\|_{H^2}\right)\\&
 +\int_0^T\left(\|  u_t\|_{H^1}^2+\| \te_t\|_{H^1}^2+\| \n u_t\|_{H^1}^2+\|\n \te_t\|_{H^1}^2
 \right)dt\le C,
 \ea  \ee
\be\la{vva5}\ba
  \int_0^T\left( \|(\n u_t)_t\|_{H^{-1}}^2+\|(\n \te_t)_t\|_{H^{-1}}^2
 \right)dt\le C.\ea\ee

\end{lemma}

{\it Proof. } First, it follows from  (\ref{hj1}), (\ref{lee2}),
(\ref{qq1}),  (\ref{h13}), and (\ref{hs})   that \be\ba
\la{va2}\|\nabla u\|_{H^2}&\le C\left( \|{\rm div}u\|_{H^2}+ \|
\o\|_{H^2}  \right)\\&\le C\left( \|G\|_{H^2}+ \|\o\|_{H^2}
+\|\n\te-1\|_{H^2}\right)\\&\le C+C\|\na(\n\dot u)\|_{L^2}+C
\|(\n-1)(\te-1)\|_{H^2}\\&\quad+ C\|\n-1\|_{H^2} +C\|\te-1\|_{H^2}
\\ &\le  C + C(\|\na\n\|_{L^3}\|\dot u\|_{L^6}+\|\na \dot u\|_{L^2})+C  \|\n-1\|_{H^2}  \|\te-1\|_{H^2} \\&\quad+C  \|\na^2\n
\|_{L^2}  +C\|\na^2\te\|_{L^2}\\ &\le  C + C (1+\|\na^2\te\|_{L^2} ) \|\na^2\n
\|_{L^2} +C\|\na\dot u \|_{L^2} +C\|\na^2\te\|_{L^2}. \ea\ee
  This fact combining with simple computations
 and $ (\ref{a1})_1$ gives
 \be\la{ua2}
 \ba
\frac{d}{dt}\|\na^2\n\|^2_{L^2} &
 \le C(1+\|\na u\|_{L^{\infty}}
 )\|\na^2\n\|_{L^2}^2+C\|\na u\|^2_{H^2}\\ &
 \le C(1+\|\na u\|_{L^{\infty}}+\|\na^2\te\|_{L^2}
  )\|\na^2\n\|_{L^2}^2\\&\quad  +C\|\na\dot u \|^2_{L^2}
 +C\|\na^2\te\|_{L^2}^2+C, \ea\ee
   which, together with
    (\ref{v6}), (\ref{lee2}),  and
Gronwall's inequality, gives directly \be\la{ja3} \sup_{0\le t\le T}
 \|\na^2\n \|_{L^2}  \le C. \ee

 Next, it follows from  $(\ref{a1})_3,$  (\ref{3.1}), and (\ref{co13})  that   \be \ba
\la{ex1}&\frac{R}{\ga-1} \lim_{t\rightarrow 0^+}\sqrt{\n} \dot
 \te(x,t)\\&=-R\n_0^{1/2}\te_0\div u_0+ \n_0^{-1/2}
 \left( \ka\Delta \te_0+\frac{\mu}{2}|\na
u_0+(\na u_0)^{\rm tr}|^2+\lambda (\div u_0)^2\right)
\\&=-R\n_0^{1/2}\te_0\div u_0+\tilde{g_2}.\ea\ee Integrating
 (\ref{3.99}) over $(0,T)$ together with (\ref{qq1}),
  (\ref{lee2}), (\ref{3.30}),   and  (\ref{ex1}) leads to
\bnn  \ba  &\sup\limits_{0\le t\le T}\int \n(\dot\te)^2dx+\int_0^T
\|\na\dot\te\|_{L^2}^2dt\\ &\le C
 \int_0^T\left(  \|\na^2\te\|_{L^2}^2+\|\na u\|_{L^2}^2
 + \|\n^{1/2}\dot\te\|_{L^2}^2+ \|\na\dot u\|_{L^2}^2
 \right)dt\\ &\quad+\frac{1}{2}\int_0^T \|\na\dot\te\|_{L^2}^2dt+C\left(\|\te_0-1\|^2_{L^6}
\|\na u_0\|^2_{L^3}+\|\na u_0\|_{L^2}^2\right)+C\|\tilde{g_2}\|_{L^2}^2
 \\ &\le C+\frac{1}{2}\int_0^T \|\na\dot\te\|_{L^2}^2dt,\ea\enn
 which shows \be\la{a5}  \sup\limits_{0\le t\le T}\int \n(\dot\te)^2dx+\int_0^T
\|\na\dot\te\|_{L^2}^2dt \le C . \ee One thus deduces  from
(\ref{op4}), (\ref{a5}),
 (\ref{qq1}),  and (\ref{lee2}) that
\be \la{va6} \sup\limits_{0\le t\le T}\|\na^2\te\|_{L^2} \le C. \ee

It   follows from  (\ref{lee2}),  (\ref{qq1}), and
 (\ref{a5})   that
\be\label{va1}\ba &\sup_{0\le t\le T}\int\n\left( |  u_t|^2 +
\te_t^2\right)dx +\int_0^T \left(\|\na
u_t\|_{L^2}^2+\|\na\te_t\|_{L^2}^2\right)dt\\ &\le
C\sup\limits_{0\le t\le T}
 \int \n\left(|\dot u|^2+(\dot\te)^2 \right)dx+C
 \sup\limits_{0\le t\le T}\int \n\left(|u\cdot\na u|^2
 +|u\cdot\na\te|^2\right)dx\\ &\quad+C\int_0^T
\left(\|\na\dot u\|_{L^2}^2+\|\na\dot\te\|_{L^2}^2+\left(\|\na u\|_{L^3}^2
+\|u\|_{L^\infty}^2\right)\left(\|\na^2 u\|_{L^2}^2+\|\na^2 \te\|_{L^2}^2\right)\right)dt
\\ &\le C,\ea\ee   which together with (\ref{lee2}) and (\ref{qq1}) gives that \be\la{vva1} \ba& \int_0^T\left(\|\na(\n u_t)\|_{L^2}^2+\|\na(\n\te_t)\|_{L^2}^2\right)dt\\ & \le C\int_0^T\left(\|  \na u_t \|_{L^2}^2+\|  \na \n\|_{L^3}^2\| u_t \|_{L^6}^2+\|  \na \te_t \|_{L^2}^2+\|  \na \n\|_{L^3}^2\| \te_t \|_{L^6}^2 \right)dt\\&\le C.\ea\ee

Next, one deduces from $(\ref{a1})_1$ and (\ref{qq1}) that \be
\la{sp1} \|\n_t\|_{L^2}\le C\|u\|_{L^\infty}\|\nabla
\n\|_{L^2}+C\|\nabla u\|_{L^2}\le C.\ee Applying $\na$ to
$(\ref{a1})_1$   yields that \bnn \nabla \n_t+u^i\pa_i\nabla \n+\nabla
u^i\pa_i \n+\nabla \n {\rm div}u+\n \nabla{\rm div}u=0,\enn which
leads to\be\la{sp2} \|\nabla \n_t\|_{L^2}\le
C\|u\|_{L^\infty}\|\nabla^2 \n\|_{L^2}+C\|\nabla u\|_{L^3}\|\nabla
\n\|_{L^6}+C\|\nabla^2 u\|_{L^2}\le C,\ee due to (\ref{qq1}). The
combination of (\ref{sp1}) with (\ref{sp2}) implies \bnn
\la{sp3}\sup_{0\le t\le T}\|\n_t\|_{H^1}\le C,\enn which together
with (\ref{lee2}), (\ref{ja3}),
  (\ref{qq1}),  (\ref{va1}),  (\ref{vva1}),  (\ref{va6}), and (\ref{f})  gives    (\ref{va5}).

   Finally, differentiating $(\ref{a1})_3$ with respect to
    $t$ yields that\be   \la{va7}\ba \frac{R}{\ga-1}(\n \te_t)_t
    =&-\frac{R}{\ga-1}(\n u\cdot\na \te)_t -R(\n\te\div u)_t+\ka\Delta \te_t\\ &+ \lambda((\div u)^2)_t+2\mu   ( |\mathfrak{D}(u)|^2)_t .\ea\ee
   It follows from (\ref{va5}) that
      \be   \la{va9}\ba &\|(\n u\cdot\na \te)_t  \|_{L^{2}}\\&=\| \n_t u\cdot\na \te+ \n u_t\cdot\na \te + \n u\cdot\na \te_t  \|_{L^{2}}\\ &\le C\|\n_t\|_{L^6} \|\na\te\|_{L^3}+  C\|u_t\|_{L^6} \|\na\te\|_{L^3}+  C\|u \|_{L^\infty} \|\na\te_t\|_{L^2} \\ &\le C+    C\|u_t\|_{H^1}+C \| \te_t\|_{H^1},\ea\ee  \be   \la{va10}\ba &\|(\n\te \div u)_t  \|_{L^2} \le C+    C\|u_t\|_{H^1}+C \| \te_t\|_{H^1},\ea\ee
  and \be  \la{va8} \ba &\|((\div u)^2)_t\|_{L^{6/5}}+ \|  ( |\mathfrak{D}(u)|^2)_t   \|_{L^{6/5}} \\ &\le C\|\na u\|_{L^3}\|\na u_t\|_{L^2} \\ &\le C+    C\|u_t\|_{H^1}.\ea\ee

 Thus, it follows directly from (\ref{va7})-(\ref{va8}) and   (\ref{va5}) that \be\la{vva04}\ba
  \int_0^T \|(\n \te_t)_t\|_{H^{-1}}^2
 dt\le C.\ea\ee Similarly, we have \bnn
  \int_0^T  \|(\n u_t)_t\|_{H^{-1}}^2
 dt\le C,  \enn which combining with (\ref{vva04}) implies
  (\ref{vva5}). The proof of Lemma \ref{le9-1} is completed.

\begin{lemma}\la{pe1}
The following estimate holds: \be\la{nq1}
   \sup\limits_{0\le t\le T} \si\left(\|\nabla u_t\|^2_{L^2}+\|\n_{tt} \|^2_{L^2}\right)
    + \int_0^T\si\int\rho |u_{tt}|^2dxdt
\le C.
  \ee

\end{lemma}
{\it Proof.}
 Multiplying (\ref{nt0}) by
$u_{tt},$   one gets after integrating  the resulting equality  by parts  that \be\la{sp9} \ba
& \frac{1}{2}\frac{d}{dt}\int \left(\mu|\nabla u_t|^2 + (
\mu +\lambda)({\rm div}u_t)^2\right)dx+\int \rho| u_{tt}|^2dx
\\&=\frac{d}{dt}\left(-\frac{1}{2}\int_{ }\rho_t |u_t|^2 dx- \int_{
}\rho_t u\cdot\nabla u\cdot u_tdx+ \int_{ }P_t {\rm
div}u_tdx\right)\\&\quad + \frac{1}{2}\int_{ }\rho_{tt} |u_t|^2 dx+
\int_{ }(\rho_{t} u\cdot\nabla u )_t\cdot u_tdx-\int_{ }\rho
u_t\cdot\nabla u\cdot u_{tt}dx\\ &\quad- \int_{ }\rho u\cdot\nabla
u_t\cdot u_{tt}dx - \int_{ }\left(P_{tt}-\ka(\ga-1)\Delta\te_t\right)
{\rm div}u_tdx\\&\quad+\ka(\ga-1)\int_{ } \na\te_t\cdot\na
{\rm div}u_tdx \\ &\triangleq
\frac{d}{dt}I_0+ \sum\limits_{i=1}^6I_i. \ea \ee
We estimate each term on the righthand side of (\ref{sp9}) as follows:

First, it follows from $(\ref{a1})_1,$   (\ref{va5}), (\ref{va1}),
and (\ref{f}) that\be \ba \la{sp10}|I_0|&
=\left|-\frac{1}{2}\int\rho_t |u_t|^2 dx- \int \rho_t u\cdot\nabla
u\cdot u_tdx+ \int P_t {\rm div}u_tdx\right|\\ &\le \left|\int {\rm
div}(\n
u)|u_t|^2dx\right|+C\norm[L^3]{\rho_t}\norm[L^2]{u\cdot\nabla u}
\norm[L^6]{u_t}+C\|(\n\te)_t\|_{L^2}\|\nabla u_t\|_{L^2}\\ &\le C
\int  \n |u||u_t||\nabla u_t| dx
+C(1+\|\n^{1/2}\te_t\|_{L^2}+\|\n_t\|_{L^2}\|\te\|_{L^\infty})\|\nabla
u_t\|_{L^2} \\ &\le C \|u\|_{L^6}\|\n^{1/2}
u_t\|_{L^2}^{1/2}\|u_t\|_{L^6}^{1/2}\|\nabla u_t\|_{L^2} +C\|\nabla
u_t\|_{L^2}\\ &\le \frac{\mu}{4}\|\nabla u_t\|_{L^2}^2+C ,\ea\ee
 \be \la{sp11}\ba
2|I_1|&=\left|\int \rho_{tt} |u_t|^2 dx\right|\\
& \le C \|\n_{tt}\|_{L^2}\|u_t\|_{L^2}^{1/2}\|u_t\|_{L^6}^{3/2}\\
 & \le  C\|\n_{tt}\|_{L^2}\left(1+\|\na u_t\|_{L^2}\right)^{1/2}
 \|\na u_t\|_{L^2}^{3/2}\\
& \le  C\|\n_{tt}\|_{L^2}^2+C
 \|\na u_t\|_{L^2}^4+C,
   \ea \ee
   and
\be \la{sp12}\ba
  |I_2|&=\left|\int \left(\rho_t u\cdot\nabla u \right)_t\cdot u_{t}dx
 \right|\\
& = \left|  \int\left(\rho_{tt} u\cdot\nabla u\cdot u_t +\rho_t
u_t\cdot\nabla u\cdot u_t+\rho_t u\cdot\nabla u_t\cdot
u_t\right)dx\right|\\ &\le   C\norm[L^2]{\rho_{tt}}
\norm[L^3]{u\cdot\nabla
u}\norm[L^6]{u_t}+C\norm[L^2]{\rho_t}\norm[L^3]{|u_t|^2}
\norm[L^6]{\nabla u} \\
&\quad+C\norm[L^3]{\rho_t}\norm[L^{\infty}]{u}
\norm[L^2]{\nabla u_t}\norm[L^6]{u_t}\\
& \le C\norm[L^2]{\rho_{tt}}^2 + C\norm[L^2]{\nabla u_t}^2. \ea \ee
 Next, Cauchy's inequality gives
\be\ba\la{sp13} |I_3|+|I_4|&= \left| \int \rho u_t\cdot\nabla u\cdot
u_{tt} dx\right| +\left| \int \rho u\cdot\nabla u_t\cdot u_{tt}
dx\right|\\& \le   C\|\n^{1/2}u_{tt}\|_{L^2}\left(
\|u_t\|_{L^6}\|\na u\|_{L^3}+\|u\|_{L^\infty}\|\na
u_t\|_{L^2}\right) \\& \le  \frac{1}{4}
\norm[L^2]{\rho^{{1/2}}u_{tt}}^2 + C \norm[L^2]{\nabla u_t}^2.\ea\ee
Next, it follows from  (\ref{va5}) that \be \la{nt1} \ba\|\na
P_t\|_{L^2}\le
 &C\|\na(\n\te_t+\te\n_t)\|_{L^2}
 \\ \le& C\|\na\n\|_{L^3}\|\te_t\|_{L^6}+C\|\na\te_t\|_{L^2}
 +C\|\na\te\|_{L^6}\|\n_t\|_{L^3}\\ &
+C\|\te\|_{L^\infty}\|\na\n_t\|_{L^2}\\
\le &C+C\|\na\te_t\|_{L^2}.
\ea\ee
This fact together with
  (\ref{op3}) and  (\ref{va5}) gives
 \bnn\ba & \|P_{tt}-\ka(\ga-1)\Delta \te_t\|_{L^2}\\&\le
C\|(u\cdot\na P)_t\|_{L^2}+C\|(P\div u)_t\|_{L^2}
+C\||\na u||\na u_t|\|_{L^2}\\&\le
C\|u_t\|_{L^6}\|\na P\|_{L^3}+C\|u\|_{L^\infty}\|\na P_t\|_{L^2}
+C\|P_t\|_{L^6}\|\na u\|_{L^3}\\ &+C\|P\|_{L^\infty}
\|\na u_t\|_{L^2}+C\|\na u\|_{L^\infty}
\|\na u_t\|_{L^2}\\ &\le C+C\left(1+\|\na u\|_{L^\infty}\right)
\|\na u_t\|_{L^2}+C\|\na\te_t\|_{L^2},\ea\enn
which directly yields
\be\ba\la{sp15} |I_5|&=
\left|\int_{ }\left(P_{tt}-\ka(\ga-1)\Delta \te_t\right)
{\rm div}u_tdx\right|\\&\le
\norm[L^2]{P_{tt}-\ka(\ga-1)\Delta \te_t}\norm[L^2]{{\rm div}u_t}
\\ &\le C+C\left(1+\|\na u\|_{L^\infty}\right)
\|\na u_t\|_{L^2}^2+C\|\na\te_t\|^2_{L^2}. \ea\ee
Finally,
 it follows from (\ref{nt1}), (\ref{va5}), and the standard $L^2$-estimate for elliptic
system  (\ref{nt0})  that
\be\la{nt4}\ba \|\na^2u_t\|_{L^2}\le&C\|\n  u_{tt}+\n_t u_t
+\n_t u\cdot\nabla u  +\n u_t\cdot\nabla u+\n u\cdot\nabla u_t+
  \nabla P_t\|_{L^2}
\\ \le &C\|\n^{1/2} u_{tt}\|_{L^2}
+C\|\n_t\|_{L^3}\|u_t\|_{L^6}+C\|\n_t\|_{L^3}\|u\|_{L^\infty}
\|\na u\|_{L^6}\\&+C\|u_t\|_{L^6}\|\na u\|_{L^3}+C
\|u\|_{L^\infty}\|\na u_t\|_{L^2}+C\|\na P_t\|_{L^2}
\\ \le &C+C\|\n^{1/2} u_{tt}\|_{L^2}+C\|\na\te_t\|_{L^2}
+C\|\na u_t\|_{L^2}.
 \ea\ee The combination of this fact with Cauchy inequality thus leads  to\be \la{asp16}\ba  |I_6| &= \left| \ka(\ga-1)\int_{ } \na\te_t\cdot\na
{\rm div}u_tdx \right|   \\& \le  C \|\na^2u_t\|_{L^2}
\|\na\te_t\|_{L^2}
\\& \le
C\left(1+ \|\n^{1/2} u_{tt}\|_{L^2}+ \|\na\te_t\|_{L^2}
+ \|\na u_t\|_{L^2}\right)
 \|\na\te_t\|_{L^2} \\& \le
C+\frac{1}{4} \|\n^{1/2} u_{tt}\|^2_{L^2}+ C\|\na\te_t\|^2_{L^2}
+ C\|\na u_t\|^2_{L^2}
.\ea\ee
Substituting all the estimates (\ref{sp11})--(\ref{asp16}) into
(\ref{sp9}) gives
 \be\la{4.052} \ba&
  \frac{d}{dt}\int \left(\mu|\nabla u_t|^2 + (
\mu+\lambda)({\rm div}u_t)^2-2I_0\right)dx+\int \rho| u_{tt}|^2dx\\ &\le C\|\n_{tt}\|_{L^2}^2+C(1+\|\na u\|_{L^\infty}+\|\na u_t\|_{L^2}^2)\|\na u_t\|_{L^2}^2+C\|\na \te_t\|_{L^2}^2+C.\ea\ee

 Then, differentiating  $(\ref{a1})_1$  with
respect to $t$ shows \bnn\n_{tt} +   \n_t{\rm div}u +
  \n{\rm div}u_t + u_t\cdot\nabla \n + u\cdot\nabla \n_t = 0,\enn
 which combining with (\ref{va5}) implies
\be \la{s4} \ba    \|\n_{tt}\|_{L^2} & \le C
\left(\|\n_t\|_{L^6}\|\nabla u\|_{L^3}+ \|\nabla
u_t\|_{L^2}+\|u_t\|_{L^6}\|\nabla \n\|_{L^3}+\|\nabla
\n_t\|_{L^2}\right) \\ &\le C+C\|\na u_t\|_{L^2}.\ea\ee
This fact together with (\ref{va1}) yields
\be \la{s5}\int_0^T\|\n_{tt}\|_{L^2}^2dt\le C.\ee
 One thus  deduces from (\ref{4.052}), (\ref{sp10}),  (\ref{va5}),
 (\ref{qq1}),     (\ref{s5}),
 and Gronwall's inequality that
 \bnn
   \sup\limits_{0\le t\le T} \si \|\nabla u_t\|^2_{L^2}
    + \int_0^T\si\int\rho |u_{tt}|^2dxdt
\le C ,
  \enn
 which together with (\ref{s4}) gives
 (\ref{nq1}).  We  complete  the proof of
Lemma \ref{pe1}.

\begin{lemma}\la{pr3} For $q\in (3,6)$ as in Theorem \ref{th1}, it holds that \be\la{y2}\ba &
\sup_{0\le t\le T}\left( \|\n-1\|_{W^{2,q}}
+\si\|u\|^2_{H^3}\right)\\& +\int_0^T
 \left( \|u\|_{H^3}^2+ \|\na^2u\|_{W^{1,q}}^{p_0}+\si\|\na u_t\|_{H^1}^2\right) dt\le
C,\ea \ee
where  \be \la{pppppp} p_0\triangleq \frac{1}{2}\min\left\{  \frac{5q-6}{3(q-2)},\frac{9q-6 }{5q-6} \right\} \in (1,7/6).\ee
\end{lemma}

 {\it Proof.}
First, it follows from  (\ref{va2}),  (\ref{va5}), and  (\ref{nq1}) that
  \be\la{sp21} \sup\limits_{0\le
t\le T}\si\|u\|^2_{H^3}+\int_0^T  \|u\|^2_{H^3}dt \le
 C .\ee
The standard $H^1$-estimate for  elliptic problem (\ref{3.29})
together with   (\ref{va5}) leads to \be\la{ex4}\ba
\|\na^2\te\|_{H^1}\le &C\|\na (\n\te_t)\|_{L^2}+C\left(\|\na (\n
u\cdot\na \te)\|_{L^2} +\|\na(\n\te\div u)\|_{L^2}\right)\\&
+C\||\na u||\na^2u|\|_{L^2}+C\\ \le &C\left(\|\na
\n\|_{L^3}+1\right)\|\na \te_t\|_{L^2}+C
(1+\|\n-1\|_{H^2})(1+\|\te-1\|_{H^2}) \|u\|_{H^2} \\& +C\|\na
u\|_{L^6}\|\na^2u\|_{L^2}^{1/2}\|\na^2u\|_{L^6}^{1/2}+C\\ \le
&C+C\|\na \te_t\|_{L^2}+C\|\na^2u\|_{L^6}^{1/2},\ea\ee which
combining with (\ref{va5}),   (\ref{nt4}),  (\ref{nq1}), and
(\ref{sp21}) yields that
  \be\la{sp24} \int_0^T\left(\| \te-1\|_{H^3}^2+\|u\|_{H^3}^2+
\si\|\nabla u_t\|_{H^1}^2\right)dt\le C . \ee

Next, it follows from standard $W^{1,p}$-estimate for elliptic
systems \eqref{h13} that
 \be\la{a4.74}\ba   \|\na^2u\|_{W^{1,q}}\le& C\|u\|_{H^3}+C\|\na^2 \div u\|_{L^q}+C\|\na^2 \o\|_{L^q}+C \\ \le& C\|u\|_{H^3}+C\|\na(\n\dot u)\|_{L^q}+C\|\na^2 (\n\te)\|_{L^q} +C \\ \le & C
 \|   u \|_{H^3}+C\|\na(\n\dot u)\|_{L^q}
  +C \|\te\na^2\n\|_{L^q}
 +C\| \na\n\na\te\|_{L^q}\\&+C  \|\n\na^2\te\|_{L^q}+C\\ \le &C
 \|   u \|_{H^3}+ C
  \|\na(\n\dot u)\|_{L^q}  +C \| \na^2 \n\|_{L^q} +C
   \| \na^2 \te\|_{H^1} + C.\ea\ee
Applying operator $\dl$ to $(\ref{a1})_1$ gives
\be\la{4.52}  (\dl \n)_t+\div (u\dl \n)+\div (\n\dl u)+2\div(\pa_i\n\cdot\pa_i u)=0. \ee
Multiplying (\ref{4.52}) by $q|\dl \n|^{q-2}\dl \n$ and  integrating the resulting equality over $\r^3,$ we obtain after using  (\ref{va5}) and (\ref{a4.74}) that
\be\la{sp28}\ba  (\|\dl \n\|^q_{L^q})_t\le& C(1+\|\na u\|_{L^\infty} )\|\dl \n\|_{L^q}^q  +C\left(\|\na\n  \|_{L^q}+
1  \right) \|\na^2 u\|_{W^{1,q}}\|\dl
\n\|_{L^q}^{q-1}   \\ \le& C(1+\| u\|_{H^3}+\|\na(\n\dot u)
 \|_{L^q}+\|\na^2\te\|_{H^1})\left(\|\dl \n\|_{L^q}^q+1\right). \ea\ee
Note that (\ref{va5}) and  (\ref{nq1})  give
\be\la{4.49}\ba     \|\na(\n\dot u)\|_{L^q}
&\le C \|\na \n\|_{L^6}\|\na\dot u\|_{L^2}^{q/(3(q-2))}\|\na\dot u
\|_{L^q}^{2(q-3)/(3(q-2))}+C\|\na\dot u \|_{L^q}
\\&\le C \|\na\dot u\|_{L^2}+C\|\na u_t \|_{L^q}
+C\|\na(u\cdot \na u ) \|_{L^q}\\
&\le C \|\na  u_t\|_{L^2} +C+C\|\na u_t \|_{L^2}^{(6-q)/(2q)}
\|\na u_t \|_{L^6}^{3(q-2)/(2q)}\\ & \quad
+C\|\na u \|_{L^6}^{(6-q)/q}\| u \|_{H^3}^{3(q-2)/q}
+C\| u \|_{L^\infty}\|\na^2 u \|_{L^q}\\
&\le C \si^{-1/2}+C\|u\|_{H^3}^{3(q-2)/q}+C\si^{-1/2}
\left(\si\|\na u_t \|^2_{H^1}\right)^{3(q-2)/(4q)}
    \ea \ee which combining with (\ref{sp24})
    shows that, for $p_0$ as in (\ref{pppppp}),\be \la{4.53}\int_0^T \|\na(\n \dot u)\|^{p_0}_{L^q}dt\le C. \ee
  Applying  Gronwall's
inequality to (\ref{sp28}), we obtain after using   (\ref{sp24})
 and (\ref{4.53}) that  \bnn  \sup\limits_{0\le t\le
T}\|\dl \n\|_{L^q}\le C,\enn which combining with (\ref{va5}),
(\ref{sp21}),  (\ref{a4.74}),    (\ref{4.53}), and (\ref{sp24}) gives (\ref{y2}).  We finish  the proof of Lemma
\ref{pr3}.

\begin{lemma}\la{sq90} For $q\in (3,6)$ as in Theorem \ref{th1}, the following estimate holds\be \ba\la{eg17}&\sup_{ 0\le t\le
T}\si \left(\|\te_t\|_{H^1}+\|\na^3\te\|_{L^2}+\| u_t\|_{H^2}
+\| u\|_{W^{3,q}}\right)  +\int_0^T   \si^2\|\na u_{tt}\|_{L^2}^2 dt\le C.\ea  \ee

\end{lemma}

{\it Proof.} First, multiplying (\ref{sp30}) by $u_{tt}$ and
integrating the resulting equality over ${\r^3} ,$ one gets after
integration by parts that \be \la{sp31}\ba
&\frac{1}{2}\frac{d}{dt}\int \n |u_{tt}|^2dx+\int \left(\mu|\na u_{tt}|^2+(\mu+\lambda)({\rm div}u_{tt})^2\right)dx
\\&=-4\int  u^i_{tt}\n u\cdot\na
 u^i_{tt} dx-\int (\n u)_t\cdot \left[\na (u_t\cdot u_{tt})+2\na
u_t\cdot u_{tt}\right]dx\\&\quad -\int (\n_{tt}u+2\n_tu_t)\cdot\na u\cdot u_{tt}dx-\int   \n
u_{tt}\cdot\na u\cdot  u_{tt} dx+\int  P_{tt}{\rm
div}u_{tt}dx\\&\triangleq\sum_{i=1}^5J_i.\ea\ee

 H\"{o}lder's
inequality and (\ref{va5}) give\be \la{sp32} \ba |J_1|&\le
C\|\n^{1/2}u_{tt}\|_{L^2}\|\na u_{tt}\|_{L^2}\| u \|_{L^\infty}\\
&\le \frac{\mu}{8} \|\na u_{tt}\|_{L^2}^2+C \|\n^{1/2}u_{tt}\|^2_{L^2}
.\ea\ee It follows from (\ref{va1}), (\ref{va5}), (\ref{nq1}),  and (\ref{y2}) that \be \la{sp33}\ba |J_2|&\le C\left(\|\n
u_t\|_{L^3}+\|\n_t u\|_{L^3}\right)\left(\| \na u_{tt}\|_{L^2}\| u_t\|_{L^6}+\| u_{tt}\|_{L^6}\| \na
u_t\|_{L^2}\right)\\&\le
C\left(\|\n^{1/2} u_t\|^{1/2}_{L^2}\|u_t\|^{1/2}_{L^6}+\|\n_t
\|_{L^6}\| u\|_{L^6}\right) \| \na u_{tt}\|_{L^2} \| \na
u_t\|_{L^2}\\ &\le \frac{\mu}{8}
\|\na u_{tt}\|_{L^2}^2+C\| \na
u_t\|_{L^2}^{3}+C\\ &\le \frac{\mu}{8}
\|\na u_{tt}\|_{L^2}^2+C\si^{-3/2}  ,\ea\ee

\be  \la{sp34}\ba |J_3|&\le C\left(\|\n_{tt}\|_{L^2}
\|u\|_{L^6}+\|\n_{
t}\|_{L^2}\|u_{t}\|_{L^6} \right)\|\na u\|_{L^6}\|u_{tt}\|_{L^6} \\
&\le \frac{\mu}{8}\|\na u_{tt}\|_{L^2}^2+C\|\n_{tt}\|_{L^2}^2+C\| \na
u_t\|_{L^2}^2  ,\ea\ee and
\be  \la{sp36}\ba |J_4|+|J_5|\le& C\|\n u_{tt}\|_{L^2} \|\na
u\|_{L^3}\|u_{tt}\|_{L^6} +C \|(\n_t\te+\n\te_t)_t\|_{L^2}\|\na
u_{tt}\|_{L^2}\\
\le& \frac{\mu}{8} \|\na u_{tt}\|_{L^2}^2+C\|\n^{1/2}u_{tt}\|^2_{L^2}
+C\|\n_{tt}\te\|_{L^2}^2+C\|\n_{t}\te_t\|_{L^2}^2\\&
 +C\|\n^{1/2}\te_{tt}\|_{L^2}^2 \\
\le& \frac{\mu}{8} \|\na u_{tt}\|_{L^2}^2+C\|\n^{1/2}u_{tt}\|^2_{L^2}
+C\|\n_{tt}\|_{L^2}^2+C\|\na\te_t\|_{L^2}^2\\&+C\|\n^{1/2}\te_{tt}\|_{L^2}^2
 . \ea\ee
Substituting (\ref{sp32})--(\ref{sp36}) into (\ref{sp31}) yields
\be \la{ex12}\ba & \frac{d}{dt}\int \n
|u_{tt}|^2dx+ \mu\int|\na u_{tt}|^2dx \\ &\le C\si^{-3/2} +C\|\n^{1/2}u_{tt}\|^2_{L^2}+C\|\n_{tt}\|_{L^2}^2
 +C\|\na\te_t\|_{L^2}^2
+C_3\|\n^{1/2}\te_{tt}\|_{L^2}^2.\ea\ee

Then, to estimate the last term on the righthand side of (\ref{ex12}),
 we multiply  (\ref{eg1}) by $\te_{tt}$ and integrate the resulting
 equality over $\r^3$ to get
  \be\la{ex5}\ba & \left(\frac{\ka(\ga-1)}{2R}\|\na \te_t\|_{L^2}^2+H_0\right)_t
 + \int\n\te_{tt}^2dx \\&=
 \frac{1}{2}\int\n_{tt}\left( \te_t^2
+2\left(u\cdot\na \te+(\ga-1)\te\div u\right)\te_t\right)dx\\&
\quad+ \int\n_t\left(u\cdot\na\te+(\ga-1)\te\div u
\right)_t\te_{t}dx\\&
\quad-\int\n\left(u\cdot\na\te+(\ga-1)\te\div u\right)_t
\te_{tt}dx\\ &\quad-\frac{\ga-1}{R}\int
\left(\lambda (\div u)^2+2\mu |\mathfrak{D}(u)|^2\right)_{tt}\te_tdx
\\&\triangleq\sum_{i=1}^4H_i,\ea\ee
where
\bnn\ba H_0\triangleq & \frac{1}{2}\int \n_t\te_{t}^2dx+
\int\n_t\left(u\cdot\na\te+(\ga-1)\te\div u\right)
 \te_tdx\\&- \frac{\ga-1}{R}\int\left(\lambda (\div u)^2+2\mu |\mathfrak{D}(u)|^2 \right)_t\te_t
 dx  \ea\enn
satisfies
\be\la{ex6}\ba |H_0|\le & C\int \n|u||\te_{t}||\na\te_{t}|dx+
 C\|\n_t\|_{L^3}\|\te_t\|_{L^6}\left( \|\na\te\|_{L^2}
 + \|\na u\|_{L^2}\right)
 \\&+ C\|\na u\|_{L^3}\|\na u_t\|_{L^2} \|\te_t\|_{L^6}
 \\ \le &C \|\n\te_t\|_{L^2}\|u\|_{L^\infty}
 \|\na\te_t\|_{L^2}+C\|\na\te_t\|_{L^2}+C\|\na\te_t\|_{L^2} \|\na u_t\|_{L^2}\\ \le &\frac{\ka(\ga-1)}{4R}
 \|\na\te_t\|_{L^2}^2+C\si^{-1},\ea\ee
due to $(\ref{a1})_1,$  (\ref{va5}), (\ref{va1}),  and  (\ref{nq1}).
Note that (\ref{va1}) and (\ref{f}) yield
 \be\la{aa196} \|\te_t\|_{L^2}\le C+C\|\na\te_t\|_{L^2},\ee which as well as
 (\ref{va5}) gives
  \be\la{ex7}\ba |H_1|&\le C\|\n_{tt}\|_{L^2}\left(\|\te_t\|_{L^2}^{1/2}\|\te_t\|_{L^6}^{3/2}
 +\|\te_t\|_{L^6}\left(\|u\cdot\na \te\|_{L^3}+\| \te\div u\|_{L^3}
 \right)\right)\\& \le  C\|\na \te_t\|^4_{L^2}+C\|\n_{tt}\|_{L^2}^2+C.\ea\ee
It follows from   (\ref{va5})  that \be\la{eg12}\ba
&\|\left(u\cdot\na\te+(\ga-1)\te\div u \right)_t\|_{L^2}\\&\le C
\left(\|u_t\|_{L^6}
\|\na\te\|_{L^3}+\|u\|_{L^\infty}\|\na\te_t\|_{L^2}+\|\te_t\|_{L^6}
\|\na u\|_{L^3}+\|\te\|_{L^\infty}\|\na u_t\|_{L^2}\right)
\\ &\le  C\|\na \te_t\|_{L^2}+C\|\na u_t\|_{L^2},\ea\ee which together with (\ref{va5})
shows
\be\la{ex9}\ba |H_2|+|H_3|&\le C(\|\na\te_t\|_{L^2}+\|\na u_t\|_{L^2})\left(\|\n_t\|_{L^3}
\|\te_t\|_{L^6}+\|\n \te_{tt}\|_{L^2}\right)\\ &\le
 \frac{1}{2}\int\n\te_{tt}^2dx+C\|\na\te_t\|_{L^2}^2+C\|\na u_t\|^2_{L^2}. \ea\ee
One deduces from (\ref{va5}) and (\ref{nq1}) that
\be\la{ex10}\ba |H_4|&\le C\int \left(|\na u_t|^2+|\na u||\na u_{tt}|\right)
|\te_t|dx\\ &\le C\left(\|\na u_t\|_{L^2}^{3/2}\|\na u_t\|_{L^6}^{1/2}
 + \|\na u\|_{L^3} \|\na u_{tt}\|_{L^2}\right)
\|\te_t\|_{L^6}\\ &\le \de\|\na u_{tt}\|^2_{L^2}+C\|\na^2 u_t\|^2_{L^2}
+C(\de)\|\na\te_t\|_{L^2}^2+C\si^{-2}\|\na u_t\|_{L^2}^2.\ea\ee
Substituting (\ref{ex7}), (\ref{ex9}), and (\ref{ex10}) into (\ref{ex5}) gives
\be\la{ex11}\ba & \left(\frac{\ka(\ga-1)}{2R}\|\na \te_t\|_{L^2}^2+H_0\right)_t
 +\frac{1}{2}\int\n\te_{tt}^2dx
 \\ &\le \de\|\na u_{tt}\|^2_{L^2}
+C(\de)\|\na\te_t\|_{L^2}^4+C\|\na^2 u_t\|^2_{L^2}+C\si^{-2}\|\na u_t\|_{L^2}^2\\ &\quad+C\|\n_{tt}\|_{L^2}^2+C.\ea\ee

 Finally, for $C_3$ as in (\ref{ex12}), adding (\ref{ex11}) multiplied by
  $2 (C_3+1) $ to  (\ref{ex12}),
 we obtain after choosing $\de$ suitably small that

 \be\la{ex13}\ba & \left( 2 (C_3+1)
 \left(\frac{\ka(\ga-1)}{2R}\|\na \te_t\|_{L^2}^2+H_0\right)+\int_{ }\n
|u_{tt}|^2dx\right)_t\\&\quad
 + \int\n\te_{tt}^2dx+\frac{\mu}{2}\int |\na u_{tt}|^2dx
 \\ &\le C\si^{-3/2}+C \|\na\te_t\|_{L^2}^4+C\|\na^2 u_t\|^2_{L^2}+C\si^{-2}\|\na u_t\|_{L^2}^2+C\|\n_{tt}\|_{L^2}^2
\\ &\quad +C\|\n^{1/2}u_{tt}\|^2_{L^2}
  .\ea\ee
Multiplying (\ref{ex13}) by $\si^2$ and integrating the resulting
inequality over $(0,T),$ we  obtain by using (\ref{ex6}), (\ref{y2}),  (\ref{nq1}),  (\ref{va1}), and Gronwall's inequality that
\be \la{eg10}\sup_{ 0\le t\le
T}\si^2\int \left(|\na\te_t|^2+\n |u_{tt}|^2\right)dx
+\int_{0}^T\si^2\int \left(\n\te_{tt}^2+|\nabla
u_{tt}|^2\right)dxdt\le C,\ee
which together with  (\ref{nt4}), (\ref{nq1}), (\ref{ex4}),  (\ref{sp21}), (\ref{a4.74}), and (\ref{4.49}) gives
\be\la{sp20} \sup_{ 0\le t\le
T}\si \left(\|\na u_t\|_{H^1}+ \|\na^3\te\|_{L^2}+\|\na^2u\|_{W^{1,q}} \right)\le C.\ee

We thus derive (\ref{eg17}) from  (\ref{eg10}),  (\ref{aa196}),
  (\ref{sp20}),
  and (\ref{y2}). The proof of Lemma \ref{sq90} is completed.

\begin{lemma}\la{sq91} The following estimate holds \be \la{egg17}\sup_{ 0\le t\le
T}\si^2  \left(\|\na^2\te\|_{H^2}+\| \te_t\|_{H^2}
 \right)+\int_0^T\si^4\|\na \te_{tt}\|_{L^2}^2 dt\le C.\ee
\end{lemma}
{\it Proof. }
 Multiplying (\ref{eg2}) by $\te_{tt}$ and integrating the resulting
 equality over $\r^3$ yield  that
\be\la{eg3}\ba &\frac{1}{2}\frac{d}{dt}\int\n(\te_{tt})^2dx
+\frac{\ka(\ga-1)}{R}\int|\na \te_{tt}|^2dx
\\&=-4\int \te_{tt}\n u\cdot\na\te_{tt}dx  -\int
 \n_{tt}\left(\te_t+
u\cdot\na \te+(\ga-1)\te\div u\right)\te_{tt}dx\\&\quad
- 2\int\n_t\left(u\cdot\na \te+(\ga-1)\te\div u\right)_t\te_{tt}dx
\\&\quad  - \int\n\left(u_{tt}\cdot\na \te+2u_t\cdot\na\te_t
+(\ga-1)(\te\div u)_{tt}\right)\te_{tt}dx\\&\quad
+\frac{\ga-1}{R}\int
\left(\lambda (\div u)^2+2\mu |\mathfrak{D}(u)|^2\right)_{tt}\te_{tt}dx\\&
\triangleq \sum_{i=1}^5K_i.\ea\ee

 H\"{o}lder's
inequality and (\ref{va5}) give \be \la{eg4} \ba \si^4|K_1|&\le
C\si^4\|\n^{1/2}\te_{tt}\|_{L^2}\|\na \te_{tt}\|_{L^2}\| u \|_{L^\infty}\\
&\le \de \si^4\|\na \te_{tt}\|_{L^2}^2+C(\de)
\si^4\|\n^{1/2}\te_{tt}\|^2_{L^2} .\ea\ee  It follows from
(\ref{nq1}), (\ref{eg17}),  and (\ref{va5})  that
 \be \la{eg16}\ba \si^4|K_2|&\le C \si^4\|\n_{tt}\|_{L^2}\|\te_{tt}\|_{L^6}
 \left( \|\te_t\|_{H^1} +\|\na\te\|_{L^6}\|u\|_{L^6}+
  \|\na u\|_{L^3}\|\te\|_{L^\infty}\right) \\ &\le C\si^2
\|\na \te_{tt}\|_{L^2}\\ &\le C\de\si^4
\|\na \te_{tt}\|_{L^2}^2+C(\de),\ea\ee
 \be \la{eg7}\ba \si^4|K_4|&\le C\si^4\|\te_{tt}\|_{L^6}
 \left( \|\na\te\|_{L^3}\|\n u_{tt}\|_{L^2}+
  \|\na\te_t\|_{L^2}\|\n u_t\|_{L^2}^{1/2}\|u_t\|_{L^6}^{1/2}
  \right)\\ &\quad+ C\si^4\|\te_{tt}\|_{L^6}
 \left( \|\na u\|_{L^3}\|\n \te_{tt}\|_{L^2}+
  \|\na u_t\|_{L^2}\|\n \te_t\|_{L^2}^{1/2}\|\te_t\|_{L^6}^{1/2}
  \right)\\&\quad+C\si^4\|\te\|_{L^\infty}\|\n\te_{tt}\|_{L^2}
  \|\na u_{tt}\|_{L^2} \\ &\le \de\si^4
\|\na \te_{tt}\|_{L^2}^2+C(\de)\si^4\left(
 \|\n \te_{tt}\|_{L^2}^2
+\|\na u_{tt}\|_{L^2}^2  \right)+C(\de),\ea\ee
and
\be \la{eg8}\ba \si^4|K_5|&\le C\si^4\|\te_{tt}\|_{L^6}
 \left( \|\na u_t\|_{L^2}^{3/2}\|\na u_{t}\|_{L^6}^{1/2}+
  \|\na u\|_{L^3}\|\na u_{tt}\|_{L^2}
  \right)  \\ &\le \de\si^4
\|\na \te_{tt}\|_{L^2}^2+C(\de) \si^4
 \|\na u_{tt}\|_{L^2}^2 +C(\de).\ea\ee
For $K_3,$ one deduces from (\ref{eg12}),   (\ref{eg17}),   and (\ref{va5}) that
 \be \la{eg6}\ba \si^4|K_3|&\le C \si^4\|\n_t\|_{L^3}
 \|\te_{tt}\|_{L^6}
 \left( \|\na u_t\|_{L^2}+\|\na\te_{t}\|_{L^2} \right) \\ &\le C\de\si^4
\|\na \te_{tt}\|_{L^2}^2+C(\de).\ea\ee
Multiplying (\ref{eg3})  by $\si^4,$ substituting (\ref{eg4})--(\ref{eg6}) into the resulting equality
and choosing $\de$ suitably small
lead to
\bnn \ba & \frac{d}{dt}\int\si^4\n(\te_{tt})^2dx
+\frac{\ka(\ga-1)}{R}\si^4\int|\na \te_{tt}|^2dx \\ &\le
C\si^2\left(
 \|\n^{1/2} \te_{tt}\|_{L^2}^2
+\|\na u_{tt}\|_{L^2}^2  \right)+C,\ea\enn which together with
 (\ref{eg10})   gives
\be\la{eg13} \sup_{ 0\le t\le T}\si^4\int  \n |\te_{tt}|^2dx
+\int_{0}^T\si^4\int_{ } |\nabla \te_{tt}|^2 dxdt\le C.\ee Applying
  the standard $L^2$-estimate  to (\ref{eg1}), by (\ref{eg12}), (\ref{va5}), (\ref{eg13}), 
and (\ref{eg17}), we get
\be\la{eg14}\ba &\sup_{0\le t\le T}\si^2\|\na^2\te_t\|_{L^2}\\&\le
 C\sup_{0\le t\le T}\si^2\left(\|\n\te_{tt}\|_{L^2}
 +  \|\n_t\|_{L^3}\|\te_t\|_{L^6}+\|\n_t\|_{L^6}
 \left(\|\na\te\|_{L^3}+\|\na u\|_{L^3}\right)\right)
\\& \quad+C\sup_{0\le t\le T}\si^2\left(\|\na\te_t\|_{L^2}+\|\na u_t\|_{L^2}+
 \|\na u\|_{L^3} \|\na u_t\|_{L^6}\right)+C\\ &\le C.\ea\ee
It follows from  the standard $H^2$-estimate  of
$(\ref{3.29})$   that \bnn\ba \|\na^2\te\|_{H^2}&\le
C\left(\|\n\te_t\|_{H^2}+\|\n u\cdot\na\te\|_{H^2}+
\|\n\te\div u\|_{H^2}+\||\na u|^2\|_{H^2} \right)\\
&\le C\left(\left(\|\n-1\|_{H^2}+1\right)\|\te_t\|_{H^2}
+\left(\|\n-1\|_{H^2}+1\right)\| u\|_{H^2}\|\na\te\|_{H^2}
 \right)\\
&\quad+C\left(\|\n\te-1\|_{H^2}+1\right)\| \div u\|_{H^2}+C\|\na
u\|^2_{H^2}\\ &\le C+C\| \na^3 u\|_{L^2}+C\| \na^3\te \|_{L^2}+C\| \te_t\|_{H^2},\ea\enn
due to (\ref{hs})  and (\ref{va5}). This fact as well as
(\ref{y2}),  (\ref{eg17}),  (\ref{eg14}), and  (\ref{eg13}) implies (\ref{egg17}).  The proof of
Lemma \ref{sq91} is completed.

\section{\la{se5}Proofs of  Theorems  \ref{th1} and \ref{th2}}

With all the a priori estimates in Sections \ref{se3} and \ref{se4}
at hand, we are ready to prove the main results of this paper in
this section.

\begin{pro} \la{pro2}

 For  given numbers $M>0$ (not necessarily small),
  $\on> 2,$    assume that  $(\rho_0,u_0,\te_0
)$ satisfies (\ref{2.1}),  (\ref{3.1}),
and   (\ref{z01}).  Then    there exists a unique classical solution  $(\rho,u,\te) $      of (\ref{a1}) (\ref{h1}) (\ref{h2})
 in $\r^3\times (0,\infty)$ satisfying (\ref{mn5})--(\ref{mn2}) with $T_0$ replaced by any $T\in (0,\infty).$   Moreover,  (\ref{a2.12}), (\ref{z1}), and  (\ref{vu15})  hold for any $T\in (0,\infty).$


 \end{pro}

{\it Proof. }
First,  standard local existence result, Lemma \ref{th0},   applies to show that the Cauchy problem  (\ref{a1}) (\ref{h1}) (\ref{h2}) with   initial data $(\n_0 ,u_0,\te_0 )$ has   a unique local solution $(\n,u,\te), $
defined up to a positive $T_0 $ which may depend on
$\inf\limits_{x\in \r^3}\n_0(x), $  and satisfying (\ref{mn5})--(\ref{mn2}),  and $ \inf\limits_{x\in \r^3}\n_0(x)/4\le \n\le
2\overline{\n}. $
One deduces from (\ref{As1})--(\ref{3.1}) that
\bnn A_1(0)\le M,\quad  A_2(0)\le  C_{0 }\le  C_0^{1/4},\quad A_3(0)=A_4(0)=0, \quad  \n_0<
 \bar{\rho}.\enn  Then  there exists a
$T_1\in(0,T_0]$ such that (\ref{z1}) holds for $T=T_1.$
 We set \be \la{ss1}T^* =\sup\left\{T\,\left|\, \sup_{t\in [0,T]}\|(\n-1,u,\te-1)\|_{H^3}<\infty\right\},\right.\ee  and \be \la{s1}T_*=\sup\{T\le T^* \,|\,{\rm (\ref{z1}) \
holds}\}.\ee Then $ T^*\ge T_* \geq T_1>0.$
 We claim that
 \be \la{s2}  T_*=\infty.\ee  Otherwise,    $T_*<\infty.$
  Proposition \ref{pr1} implies that (\ref{zs2})
  holds for all $0<T<T_*,$ which together with (\ref{z01}) yields
    Lemmas \ref{le11}--\ref{sq91} still hold for all  $0< T< T_* .$ Note here that  all  constants $C$  in  Lemmas \ref{le11}--\ref{sq91}  depend  on $T_*  $ and $\inf\limits_{x\in \r^3}\n_0(x),$ and are in fact  independent  of  $T.$

Next,   we claim that  there
exists a positive constant $\tilde{C}$ which may  depend  on $T_* $
and $\inf\limits_{x\in \r^3}\n_0(x)$   such that, for all  $0< T<
 T_*,$  \be\la{y12}\ba \sup_{0\le t\le T}
\| \n-1\|_{H^3}   \le \tilde{C},\ea \ee which together with Lemmas \ref{pr3}--\ref{sq91}  and (\ref{3.1}) gives
 \bnn
 \|(\n(x,T_*)-1,u(x,T_*),\te(x,T_*)-1)\|_{H^3}
 \le \tilde{C},\quad\inf_{x\in \r^3}\n(x,T_*)>0.\enn
This fact as well as Lemma \ref{th0} implies that there exists some $T^{**}>T_*,$  such that
(\ref{z1}) holds for $T=T^{**},$   which contradicts (\ref{s1}).
Hence, we obtain (\ref{s2}) which together with Lemma \ref{th0} finishes the proof of   Proposition \ref{pro2}.

 Finally, it remains to prove (\ref{y12}).  It follows  from (\ref{3.1}), $(\ref{a1})_2,$
 and (\ref{mn6}) that we can define \bnn u_t(\cdot,0)\triangleq
 -u_0\cdot\na
 u_0+\n_0^{-1}\left( \mu \Delta u_0 +(\mu+\lambda)\na\div
u_0-R\na (\n_0\te_0)\right),\enn which together with  (\ref{2.1})
gives \be \la{ssp9}\|\na u_t(\cdot,0)\|_{L^2}\le \tilde{C}.\ee It
thus follows from  (\ref{4.052}),  (\ref{ssp9}), (\ref{sp10}),
(\ref{s5}), (\ref{va5}),
 (\ref{qq1}),  and Gronwall's inequality that
 \be \la{ssp1}\sup_{0\le t\le T}\|\na u_t\|_{L^2}+\int_0^T\int \n
|u_{tt}|^2dxdt\le \tilde{C},\ee which as well as  (\ref{va2}) and
(\ref{va5}) yields  \be\la{sp221} \sup\limits_{0\le t\le
T}\|u\|_{H^3} \le
 \tilde{C}.\ee  This combining with (\ref{ex4}),
(\ref{nt4}), (\ref{ssp1}),   and  (\ref{va5}) gives
   \be\la{ssp24} \ia\left(\|\na^3\te\|_{L^2}^2+
\|\nabla u_t\|_{H^1}^2\right)dt\le \tilde{C} . \ee
Applying the   $H^2$-estimate to elliptic systems
(\ref{h13})  leads to \be \la{sp38}\ba \|\na^2 u\|_{H^2}&\le
\tilde{C}\|\na \div u\|_{H^2}+\tilde{C}\|\na \o\|_{H^2} \\&\le
\tilde{C}\| \n \dot u \|_{H^2}+\tilde{C}\|\na  P
 \|_{H^2}\\ & \le \tilde{C} +\tilde{C} \|\na^2  u_t\|_{L^2}+\tilde{C}\|\na^3 \n
 \|_{L^2}+\tilde{C}\|\na^3 \te \|_{L^2},\ea\ee
where one has used  (\ref{va5})  and the
following
 simple facts: \be
\la{sp25}\ba \| \n u_t \|_{H^2}&\le \tilde{C}  \| (\n-1) u_t
\|_{H^2}+ \tilde{C}\| u_t \|_{H^2} \no & \le \tilde{C}  \|
\n-1\|_{H^2}\| u_t \|_{H^2}+ \tilde{C}\| \na^2 u_t \|_{L^2}+\tilde{C} \\ & \le
\tilde{C} +\tilde{C} \|\na^2 u_t\|_{L^2},\ea\ee
\bnn \ba \| \n u\cdot\na u \|_{H^2}&\le \tilde{C}\left(\| (\n-1) u \|_{H^2}+\|   u  \|_{H^2} \right)\|  \na u \|_{H^2} \\
&\le \tilde{C}\|\n-1\|_{H^2}\| u \|_{H^2}+\tilde{C}\\ &\le
\tilde{C},\ea\enn and \bnn\ba \|\na^3(\n \te)\|_{L^2}\le
&\tilde{C}\|\na^3 \n\|_{L^2} \|\te\|_{L^\infty}+\tilde{C}\|\na^2
\n\|_{L^6} \|\na\te\|_{L^3}\\&+\tilde{C}\|\na \n\|_{L^3}
\|\na^2\te\|_{L^6}+\tilde{C}\|\na^3\te\|_{L^2}\\\le &
\tilde{C}\|\na^3\n\|_{L^2}+\tilde{C}\|\na^3\te\|_{L^2},\ea\enn due
to (\ref{hs}),  (\ref{va5}),  (\ref{ssp1}), (\ref{nq1}), and
(\ref{sp221}). Standard calculations lead to
 \be\la{sp134}\ba &\left(\|\na^3 \n\|_{L^2}^2\right)_t\\&\le
\tilde{C}\left(\||\na^3u||\na \n|\|_{L^2}+ \||\na^2u||\na^2
\n|\|_{L^2}+ \||\na u||\na^3 \n|\|_{L^2} \right)\|\na^3\n\|_{L^2}
\\&\quad + \tilde{C}\| \na^4u \|_{L^2} \|\na^3\n\|_{L^2}\\&\le
\tilde{C}\left(\| \na^3u\|_{L^2}\|\na \n \|_{H^2}+ \|
\na^2u\|_{L^3}\|\na^2 \n \|_{L^6}+ \| \na
u\|_{L^\infty}\|\na^3 \n \|_{L^2}\right)\|\na^3\n\|_{L^2}\\
&\quad+\tilde{C}\left( 1+\|\na^2 u_t \|_{L^2}+ \| \na^3\n \|_{L^2}+
\| \na^3\te\|_{L^2}\right) \| \na^3\n \|_{L^2}
\\ &\le \tilde{C}+\tilde{C} \| \na^2 u_t
\|^2_{L^2}+\tilde{C} \| \na^3\n \|^2_{L^2}+\tilde{C} \|
\na^3\te\|^2_{L^2}, \ea\ee due to (\ref{va5}), (\ref{sp221}), and
(\ref{sp38}).
 It thus follows from (\ref{sp134}), (\ref{ssp24}), and
Gronwall's inequality that \be\la{sp26} \sup\limits_{0\le t\le
T}\|\nabla^3  \n\|_{L^2} \le \tilde{C},\ee which together with
(\ref{va5})  gives   (\ref{y12}).
The proof of Proposition \ref{pro2} is completed.

With  Proposition \ref{pro2} at hand, we are now in a position to prove our main results, Theorems \ref{th1} and \ref{th2}.

{\it Proof of  Theorem   \ref{th1}.}
 Let $(\n_0,u_0,\te_0)$ satisfying (\ref{co3})--(\ref{co1}) be initial data as described in
Theorem \ref{th1}.  Assume that  $C_0$  satisfies (\ref{co14}), where
 \be \la{de2}  \ve\triangleq \ve_0/2  ,\ee with  $\ve_0$  as in Proposition \ref{pr1}. For constants  \be \la{uv6} \de,\eta \in (0, \min\{1,\on-\sup\limits_{x\in \r^3}\n_0(x) \}),\ee   we define
\be\la{uv4}\rd_0\triangleq \frac{ j_\de*\n_0  +\eta}{1+\eta} ,
\quad \ud_{0}\triangleq j_\de*u_0, \quad \td_0 \triangleq
 \frac{ j_\de*\te_0  +\eta}{1+\eta},\ee where  $j_\de$
is the standard mollifying kernel of width $\de.$
Then, $(\rd_0,\ud_0,\td_0)$ satisfies
\be \la{de3}\begin{cases}(\rd_0-1,\ud_0, \td_0-1)\in
H^\infty ,\\
\dis \frac{\eta}{1+\eta} \le  \rd_0 \le  \frac{\on+\eta}{1+\eta}  <\bar\n  ,\quad \td_0\ge \frac{\eta}{\on+\eta} , \quad\|\na \ud_0\|_{L^2} \le M,  \end{cases}
\ee and \be \la{de03}
\begin{cases}\lim\limits_{\de+\eta \rightarrow 0}\left(\| \rd_0 - \n_0 \|_{H^2\cap W^{2,q}}+\| \ud_0-u_0\|_{H^2}+\| \td_0- \te_0  \|_{H^2}\right)=0, \\
\|\na(\rd_0,\ud_0,\td_0)\|_{H^1}\le \|\na(\n_0,u_0,\te_0)\|_{H^1},
\quad \|\na \rd_0 \|_{W^{1,q}}\le \|\na \n_0 \|_{W^{1,q}}, \end{cases}
\ee
due to (\ref{co3}) and (\ref{co4}).
 Moreover,  the initial norm $C_0^{\de,\eta}$
for $(\rd_0,\ud_{0},\td_0),$  i.e., the right hand side of (\ref{e})
with $(\n_0,u_0,\te_0)$   replaced by
$(\rd_0,\ud_{0},\td_0),$
satisfies \bnn \lim\limits_{\eta\rightarrow 0} \lim\limits_{\de\rightarrow 0}C_0^{\de,\eta}=C_0.\enn
Therefore, there exists  an $\eta_0\in (0, \min\{1,\on-\sup\limits_{x\in \r^3}\n_0(x) \}) $ such that, for any $\eta\in (0,\eta_0),$ we can find some $\de_0(\eta)>0$  such that       \be \la{de1} C_0^{\de,\eta}\le
C_0+\ve_0/2\le  \ve_0 , \ee provided that\be  \la{de7}0<\eta\le
\eta_0 ,  \quad 0<\de\le \de_0(\eta).\ee

  We assume that $\de,\eta$ satisfy (\ref{de7}). Proposition \ref{pro2} together with (\ref{de1}) and (\ref{de3}) thus yields that  there exists a smooth solution  $(\rd,\ud,\td)  $
   of (\ref{a1}) (\ref{h1}) (\ref{h2}) with  initial data $(\rd_0, \ud_0,\td_0 )$  on $\r^3\times[0,T] $ for all $T>0. $ Moreover, (\ref{a2.12}) and (\ref{z1}) both hold  with $(\n,u,\te)$ being replaced by $(\rd,\ud,\td). $

 Next, for the initial data  $(\rd_0,\ud_{0},\td_0),$  the $\tilde{g_1}$ in (\ref{co12})   in fact is
 \be \la{co5}\ba \tilde g_1 & \triangleq(\rd_0)^{-1/2}\left(-\mu \Delta\ud_0-(\mu+\lambda)\na\div
\ud_0+R\na (\rd_0\td_0)\right)\\
& = (\rd_0)^{-1/2}(j_\de*\n_0)^{1/2}g_1+(\rd_0)^{-1/2}\left(j_\de*(\sqrt{\n_0}g_1)
-\sqrt{j_\de*\n_0}g_1\right)\\&\quad+R(\rd_0)^{-1/2}\na\left(
j_\de*(\n_0\te_0)-(1+\eta)^{-2}(j_\de* \n_0)(j_\de*\te_0 )\right)\\&\quad + R\eta(1+\eta)^{-2} (\rd_0)^{-1/2}\na
(\rd_0+\td_0),\ea\ee
where in the second equality we have used (\ref{co2}).
Similarly,   the $\tilde{g_2}$ in (\ref{co13})    is
  \be
\la{co6} \ba \tilde g_2 & \triangleq (\rd_0)^{-1/2}\left(\ka\Delta \td_0+\frac{\mu}{2}|\na
\ud_0+(\na \ud_0)^{\rm tr}|^2+\lambda (\div \ud_0)^2\right)\\&=
(\rd_0)^{-1/2}(j_\de*\n_0)^{1/2}g_2
+(\rd_0)^{-1/2}\left(j_\de*(\sqrt{\n_0}g_2)
-\sqrt{j_\de*\n_0}g_2\right)\\&\quad-
\frac{\mu}{2} (\rd_0)^{-1/2}\left(j_\de*\left(|\na u_0+(\na u_0)^{\rm
tr}|^2\right)-|\na (j_\de*u_0)+(\na(j_\de*u_0))^{\rm tr}|^2  \right) \\&\quad-
\lambda(\rd_0)^{-1/2}\left(j_\de*\left((\div u_0)^2\right)- (\div
(j_\de*u_0))^2\right),\ea\ee due to   (\ref{co1}).
Since $g_1,g_2\in L^2,$ one deduces from   (\ref{co5}),  (\ref{co6}), (\ref{de3}), (\ref{de03}), and  (\ref{co3})  that there exists some positive constant $C$ independent of $\de$
 and $\eta$ such that  \be\la{de4} \begin{cases} \| \tilde g_1 \|_{L^2}  \le (1+\eta)^{1/2}\|g_1\|_{L^2}  +C\eta^{-1/2}m_1(\de)+C\sqrt{\eta} , \\   \|\tilde g_2 \|_{L^2} \le (1+\eta)^{1/2} \|g_2\|_{L^2}
 +C\eta^{-1/2}m_2(\de), \end{cases}\ee
with   $0\le m_i(\de) \rightarrow 0 \,\,(i=1,2)$ as
$\de \rightarrow 0.$ Hence,  for any   $0<\eta<\eta_0,$ there exists some $0<\de_1(\eta)\le \de_0(\eta)$ such that \be \la{de9}m_1(\de)+m_2(\de)<\eta,\ee for any $ 0<\de<\de_1(\eta).$  We thus obtain from (\ref{de4}) and (\ref{de9}) that
there exists some positive constant $C$ independent of $\de$
 and $\eta$ such that  \be\la{de14}  \|\tilde g_1 \|_{L^2}+\| \tilde g_2 \|_{L^2}\le 2\|g_1 \|_{L^2}+2\| g_2 \|_{L^2}+C,\ee provided that\be \la{de10} 0<\eta<\eta_0,\quad  0<\de <\de_1(\eta).\ee

 Now, we   assume that $\eta,$  $\de$ satisfy (\ref{de10}).
 It thus follows from (\ref{de1}), Proposition \ref{pr1}, Corollary \ref{cor1}, (\ref{de03}),    (\ref{de14}), and Lemmas \ref{le11}--\ref{sq91} that for any $T>0,$  there exists some positive constant $C$
  independent of $\de$ and $\eta$ such that  (\ref{a2.12}), (\ref{z1}),   (\ref{vu15}),  (\ref{va5}),  (\ref{vva5}), (\ref{y2}),  (\ref{eg17}), and  (\ref{egg17})  hold for  $(\rd,\ud,\td) .$   Then passing  to the limit first
  $\de\rightarrow 0,$ then $\eta\rightarrow 0,$ together with standard arguments yields that
  there exists a solution $(\n,u,\te)$ of  (\ref{a1}) (\ref{h1})
  (\ref{h2}) on $\r^3\times (0,T]$ for all $T>0,$ such that  $(\n,u,\te)$ satisfies   (\ref{a2.12}),  (\ref{z1}),   (\ref{vu15}), (\ref{va5}),  (\ref{vva5}), (\ref{y2}),  (\ref{eg17}) and  (\ref{egg17}). Hence,    $(\n,u,\te)$ satisfies  (\ref{h8}), $(\ref{h9})_2,$ $(\ref{h9})_3 ,$  and
   \be  \la{sq1}\rho-1\in L^\infty(0,T;H^2\cap W^{2,q}),\quad ( u,\te-1)  \in L^\infty(0,T;H^2).  \ee  Moreover,
 (\ref{4.52})   holds in $\mathcal{D}'(\r^3\times (0,T)).$

Next, to finish  the existence part of Theorem \ref{th1}, it remains to prove  \be \la{sq2} \rho-1 \in C([0,T];H^2\cap W^{2,q}),\quad   u,\,\,\te-1   \in C([0,T];H^2).  \ee
It follows from   $ (\ref{va5})  $  and (\ref{sq1})   that
  \be  \la{sq3}\rho-1 \in C([0,T];H^1\cap W^{1,\infty})\cap C([0,T];H^2\cap W^{2,q} \mbox{ -weak}),\ee and for all $r\in [2,6),$ \be  \la{sq5}    u,\,\,\te-1 \in C([0,T];H^1\cap W^{1,r} ).\ee

 Since  (\ref{4.52}) holds in $\mathcal{D}'(\r^3\times (0,T))$ for all $T\in (0,\infty),$
one derives from  \cite[Lemma 2.3]{L2} that, for $j_\nu(x)$ being the standard  mollifying kernel of width $\nu,$ $\n^\nu\triangleq \n*j_\nu$ satisfies \be\la{4.63}\ba  (\dl \n^\nu)_t+\div (u\dl \n^\nu)=-\div (\n\dl u)* j_\nu-2\div(\pa_i\n\cdot\pa_i u)* j_\nu+R_\nu, \ea\ee
where $R_\nu$ satisfies\be \la{5.32} \int_0^T\|R_\nu\|_{L^2\cap L^q}^{3/2}dt\le C\int_0^T\|  u \|^{3/2}_{W^{1,\infty}}\|\dl \n\|_{L^2\cap L^q}^{3/2}dt \le C ,\ee
due to (\ref{qq1}), (\ref{va5}),  and  (\ref{y2}).  Multiplying (\ref{4.63}) by $q|\dl \n^\nu|^{q-2}\dl \n^\nu,$ we obtain after  integration by parts that\bnn\ba &(\|\dl \n^\nu\|_{L^q}^q)'(t)\\&=(1-q)\int  |\dl \n^\nu|^q\div u dx-q\int (\div (\n\dl u)* j_\nu) |\dl \n^\nu|^{q-2}\dl \n^\nu dx\\ &\quad-2q\int (\div(\pa_i\n\cdot\pa_i u)* j_\nu) |\dl \n^\nu|^{q-2}\dl \n^\nu dx+q\int R_\nu|\dl \n^\nu|^{q-2}\dl \n^\nu dx,\ea\enn which together with (\ref{va5}), (\ref{y2}),  and  (\ref{5.32}) yields that, for $p_0$ as in (\ref{pppppp}), \bnn\ba&\sup_{t\in [0,T]} \|\dl \n^\nu\|_{L^q}  +\int_0^T|( \|\dl \n^\nu\|_{L^q}^q)'(t) |^{p_0}dt\\& \le  C+C\int_0^T\left(\|\na u\|_{W^{2,q}}^{p_0}+\|R_\nu\|^{p_0}_{L^2\cap L^q}\right)dt\\ &\le C.\ea\enn
This fact combining with the Ascoli-Arzela theorem thus leads to \bnn \|\dl \n^\nu(\cdot,t) \|_{L^q}\rightarrow \|\dl \n(\cdot,t) \|_{L^q} \mbox{ in }C([0,T]),\mbox{ as }\nu\rightarrow 0^+.\enn  In particular, we have \be \la{5.42}\|\na^2\n(\cdot,t)\|_{L^q}\in C([0,T]).\ee Similarly, one can obtain that\be \la{5.43}\|\na^2\n(\cdot,t)\|_{L^2}\in C([0,T]).\ee
Therefore, the continuity of $\na^2\n$ in $L^p$ $ (p=2,q),$ i.e.,
 \be \la{sq4} \na^2\rho \in C([0,T];L^2\cap L^q)
,  \ee follows directly from (\ref{sq3}), (\ref{5.42}),  and (\ref{5.43}).

It follows from  (\ref{va5})  and   (\ref{vva5})     that
  \be \la{sq6} \n u_t,\n\te_t \in
C([0,T];L^2),\ee which together with  (\ref{zp202}), (\ref{sq3}), (\ref{sq5}),  and
(\ref{sq4}) gives   \be\la{sqq1} u\in C([0,T];H^2).\ee
This fact combining with (\ref{3.29}), (\ref{sq6}), (\ref{sq4}), (\ref{sq5}), and
(\ref{va5}) leads to \bnn\la{sqq2} \te -1\in C([0,T];H^2),\enn  which as well as    (\ref{sq3}),  (\ref{sq4}),   and (\ref{sqq1})
leads to  (\ref{sq2}).

Finally, since the proof of the uniqueness of $(\n,u,\te)$ is similar to that  of \cite[Theorem 1]{cj}, to finish the proof of Theorem \ref{th1}, it remains to
prove (\ref{h11}). We will only show  \be\la{n8}\lim_{t\rightarrow \infty}\|\na
u\|_{L^2}=0,\ee
since  the other terms in (\ref{h11}) follow directly from  (\ref{lar1}).
It follows from (\ref{vu15}) and \eqref{z1} that
\bnn\ba &\int_1^\infty |(\|\na u\|_{L^2}^2)'(t)|dt \\&=2\int_1^\infty \left|\int \pa_j u^i\pa_j u^i_t dx\right|dt\\&=2\int_1^\infty \left|\int \pa_j u^i\pa_j(\dot u^i-u^k\pa_k u^i ) dx\right|dt\\&=2\int_1^\infty \left|\int (\pa_j u^i\pa_j \dot u^i-\pa_j u^i\pa_j u^k\pa_k u^i-\pa_j u^i u^k\pa_{kj} u^i ) dx\right|dt\\&=\int_1^\infty \left|\int (2\pa_j u^i\pa_j \dot u^i-2\pa_j u^i\pa_j u^k\pa_k u^i+|\na u|^2\div u ) dx\right|dt\\&\le C\int_1^\infty \left(\|\na u\|_{L^2}\|\na \dot u\|_{L^2}+\|\na u\|_{L^3}^3\right)dt\\&\le C\int_1^\infty \left(\|\na  \dot u\|_{L^2}^2+\|\na u\|^2_{L^2}+\|\na u\|_{L^4}^4\right)dt\\ &\le C,\ea\enn
 which  together with \eqref{z1}  implies (\ref{n8}).
   We finish the proof of Theorem \ref{th1}.

{\it Proof of  Theorem   \ref{th2}. } We will prove  Theorem   \ref{th2} in three steps.

 {\it Step 1. Construction  of approximate  solutions.} Let $(\n_0,u_0,\te_0)$ satisfying (\ref{co4})  be initial data as described in Theorem \ref{th2}.  Assume that  $C_0$ satisfies (\ref{cco14})   with  $\ve $  as in  (\ref{de2}). Let $\de$ and $\eta$ be as in (\ref{uv6}) and  $j_\de$
be the standard mollifier. We define
\be\la{uv5}\htn_0\triangleq \frac{ j_\de*\n_0  +\eta}{1+\eta} ,
\quad \htu_{0}\triangleq j_\de*u_0, \quad \hte_0 \triangleq
\frac{ j_\de*(\n_0\te_0)  +\eta}{   j_\de*\n_0  +\eta } .\ee
 Then, $(\htn_0 ,\htu_0, \hte_0 )$ satisfies \be \la{uv7}
 \begin{cases}(\htn_0-1,\htu_0, \hte_0-1)\in
H^\infty ,\\
\dis0<\frac{\eta}{1+\eta}\le   \htn_0 \le\frac{\on+\eta}{1+\eta}  <\bar\n  ,\quad \hte_0\ge \frac{\eta}{\on+\eta}>0, \quad\|\na \htu_0\|_{L^2} \le M, \end{cases}
\ee
due to  (\ref{co4}). Moreover, it follows from (\ref{co4})  and (\ref{cco14}) that \be \la{vua1}\lim\limits_{\eta\rightarrow 0}\lim\limits_{\de\rightarrow 0}\left(\|\htn_0-\n_0\|_{L^2}+\|\htu_0-u_0\|_{H^1}+\|\htn_0\hte_0
-\n_0\te_0\|_{L^2}\right)=0. \ee
 We claim that  the initial norm $\hat C_0^{\de,\eta}$
for $(\htn_0,\htu_{0},\hte_0),$  i.e., the right hand side of
(\ref{e}) with $(\n_0,u_0,\te_0)$  replaced by
$(\htn_0,\htu_{0},\hte_0),$   satisfies
\be \la{uv9}  \lim\limits_{\eta\rightarrow 0}
\lim\limits_{\de\rightarrow 0}\hat C_0^{\de,\eta}\le C_0,\ee
which yields  that there exists an $\hat\eta >0$ such that, for any $\eta\in
(0,\hat\eta ),$ there exists some $\hat\de (\eta)>0$ such that       \be \la{uv8}\hat
C_0^{\de,\eta}\le C_0+\ve_0/2\le \ve_0 , \ee
provided \be\la{uv01}
0<\eta\le \hat\eta  , \quad 0<\de\le \hat\de (\eta).\ee
We assume that $\de,\eta$ always satisfy (\ref{uv01}). Proposition \ref{pro2} as well as (\ref{uv7}) and (\ref{uv8}) thus yields that there exists a smooth solution  $(\htn,\htu,\hte)  $  of (\ref{a1}) (\ref{h1}) (\ref{h2}) with  initial data $(\htn_0, \htu_0,\hte_0 )$  on $\r^3\times[0,T] $ for all $T>0. $ Moreover, for any $T>0,$ $(\htn,\htu,\hte)  $  satisfies  (\ref{a2.12}), (\ref{z1}), and  (\ref{vu15}) with $(\n ,u ,\te )$   replaced by $(\htn ,\htu ,\hte ).$

It remains to prove (\ref{uv9}). In fact, we only have to show
\be\la{uv10}\lim_{\eta\rightarrow 0}\lim_{\de\rightarrow 0}\int\htn_0\left(\hte_0- \log
 \hte_0 -1 \right)dx\le \int\n_0\left(\te_0- \log
 \te_0 -1 \right)dx ,\ee since   the other terms in  (\ref{uv9}) can be proved in a  similar and even simpler way.
Noticing that\bnn\ba &\htn_0\left(\hte_0- \log \hte_0 -1
\right)\\&=\htn_0(\hte_0-1)^2 \int_0^1\frac{\al}{\al
(\hte_0-1)+1}d\al
\\&= \frac{(j_\de*(\n_0\te_0-\n_0))^2}{1+\eta}\int_0^1\frac{\al}{\al (j_\de*(\n_0\te_0)-j_\de*
\n_0 )+j_\de* \n_0 +\eta}d\al \\&\in \left[0, \, \eta^{-1}
 (j_\de*(\n_0\te_0-\n_0))^2 \right],\ea\enn
we deduce from (\ref{vua1}) and Lebesgue's dominated convergence theorem that
 \be \la{uv11}\ba &\lim_{\de\rightarrow 0}\int\htn_0\left(\hte_0- \log
 \hte_0 -1 \right)dx\\&=\int\frac{\n_0+\eta}{1+\eta}\left(\frac{\n_0\te_0+\eta}{\n_0+\eta}
 - \log\frac{\n_0\te_0+\eta}{\n_0+\eta}
-1 \right)dx\\&=(1+\eta)^{-1} \int_{(\n_0\te_0<1/2)\cup (\n_0\te_0>2)} \left(\n_0\te_0 -\n_0 -(\n_0+\eta) \log\frac{\n_0\te_0+\eta}{\n_0+\eta}
 \right)dx \\&\quad +(1+\eta)^{-1}\int_{(1/2\le \n_0\te_0\le 2)}
(\n_0+\eta)\left(\frac{\n_0\te_0+\eta}{\n_0+\eta}
 - \log\frac{\n_0\te_0+\eta}{\n_0+\eta}
-1 \right)dx \\&\triangleq (1+\eta)^{-1}I_1+(1+\eta)^{-1}I_2,  \ea\ee
 where we have used the following simple fact that, for
$f\in L^p (1\le p<\infty),$\bnn
\lim_{\de\rightarrow 0}\|j_\de*f-f\|_{L^p}=0,\quad
\lim_{\de\rightarrow 0} j_\de*f(x)=f(x),\,\,\mbox{ a.e. }
 x\in\r^3.\enn
 It follows from (\ref{cco14}) that \bnn \ba & |(\n_0\te_0<1/2)\cup (\n_0\te_0>2)|  \\ &\le 4\int(\n_0\te_0-1)^2dx\\ &\le  8\int(\n_0\te_0-\n_0)^2dx+
8\int(\n_0-1)^2dx\\ &\le C,\ea\enn which combining with
  Lebesgue's dominated convergence theorem yields
 \be \ba \la{uv12} I_1& =\int_{(\n_0\te_0<1/2)\cup (\n_0\te_0>2)}
 \left(\n_0\te_0- \n_0 \log( \n_0\te_0+\eta )-\eta\log( \n_0\te_0+\eta ) \right)dx\\&\quad+\int_{(\n_0\te_0<1/2)\cup (\n_0\te_0>2)}
   \left((\n_0+\eta)\log(\n_0+\eta) -\n_0
  \right)dx \\& \le\int_{(\n_0\te_0<1/2)\cup (\n_0\te_0>2)}
 \left(\n_0\te_0- \n_0 \log( \n_0\te_0 )-\eta\log
  \eta \right)dx\\&\quad+\int_{(\n_0\te_0<1/2)\cup (\n_0\te_0>2)}
   \left( \n_0 \log(\n_0+\eta) +\eta \log(\n_0+\eta) -\n_0
  \right)dx  \\ & \rightarrow \int_{(\n_0\te_0<1/2)\cup (\n_0\te_0>2)}
 \n_0\left(\te_0-  \log\te_0 -1 \right)dx,\quad \mbox{ as }
 \eta\rightarrow 0.\ea\ee
Noticing that  \bnn \ba&(\n_0+\eta)\left(\frac{\n_0\te_0+\eta}{\n_0+\eta}
 - \log\frac{\n_0\te_0+\eta}{\n_0+\eta}
-1 \right)\\ &= \left( \n_0\te_0-\n_0\right)^2\int_0^1\frac{\al}{\al(\n_0\te_0-\n_0)+\n_0+\eta}d\al \\ &\in \left[0, 2\left( \n_0\te_0-\n_0\right)^2\right],\ea \enn provided
$\n_0\te_0\ge 1/2, $ we deduce from Lebesgue's dominated convergence theorem
that \bnn \lim_{\eta\rightarrow 0}I_2=\int_{(1/2\le\n_0\te_0\le 2) } \n_0\left(\te_0-  \log\te_0 -1 \right)dx,\enn
which together with (\ref{uv11}) and (\ref{uv12}) gives (\ref{uv10}).

{\it Step 2. Compactness results.}  For the approximate
solutions $(\htn,\htu,\hte)$ obtained in the previous step,
 we will pass to the limit  first   $\de\rightarrow 0,$ then
  $\eta\rightarrow 0 $  and  apply (\ref{z1}) and  (\ref{vu15})
  to obtain the global existence of weak solutions.
   Since the two steps are similar, we will only   sketch the
   arguments for $\de \rightarrow 0.$    Thus, we fix
   $\eta\in (0,\hat\eta)$ and simply denote  $(\htn ,\htu ,\hte )$
   by $(\hn ,\hu ,\he ).$ For $R\in (0,\infty),$ let  $ B_R(x_0)\triangleq\{x\in \r^3||x-x_0|<R\} $   denote   a ball centered at $x_0\in \r^3$ with radius $R.$ We claim that there exists some appropriate subsequence $ \de_j \rightarrow 0$ of $\de\rightarrow 0$ such that, for any $0<\tau<T<\infty $ and $0<R<\infty, $ we have
\be \la{vu4} \begin{cases} \te^{\de_j}-1\rightharpoonup \te-1 \,\,\mbox{ weakly in }\,\, L^2(0,T; H^1(\r^3)), \\ u^{\de_j}\rightharpoonup u  \,\,\mbox{ weakly star in }\,\, L^\infty(0,T; H^1(\r^3)), \end{cases}\ee
\be \la{vu1} \begin{cases}\n^{\de_j}-1\rightarrow \n-1  \quad \mbox{ in }\,\, C([0,T];L^2(\r^3) \mbox{-weak}) , \\ \n^{\de_j}-1\rightarrow \n-1  \quad \mbox{ in }\,\, C([0,T];H^{-1}(B_R(0))), \end{cases}\ee
\be \la{vu5} \begin{cases}
   \n^{\de_j}u^{\de_j}\rightarrow \n u,\,\,\n^{\de_j}(\te^{\de_j}-1)
   \rightarrow \n (\te-1)   \quad \mbox{ in }\,\, C([0,T];L^2(\r^3) \mbox{-weak}),
   \\ \n^{\de_j}u^{\de_j}\rightarrow \n u
   \,\,\mbox{ in }\,\, C([0,T];H^{-1}(B_R(0))),\end{cases}   \ee
\be \la{vu26} \n^{\de_j} |u^{\de_j}|^2\rightarrow \n |u|^2 \,\, \mbox{ in }\,\, C([0,T];L^3 \mbox{-weak}),\ee and
\be   \la{vu18} \begin{cases} u^{\de_j}\rightarrow u, \,\,  G^{\de_j}\rightarrow G, \,\, \o^{\de_j} \rightarrow \o, \,\, \na \te^{\de_j} \rightarrow \na\te  \,\,\mbox{ in }\,\, C([\tau,T];H^1(\r^3) \mbox{-weak}) ,\\ u^{\de_j}\rightarrow u,\,\,
G^{\de_j}\rightarrow G, \,\, \o^{\de_j} \rightarrow \o, \,\, \na \te^{\de_j} \rightarrow \na\te  \,\,\mbox{ in }\,\,   C([\tau,T];L^2(B_R(0)) ).\end{cases}\ee
We thus write $(\ref{a0})$ in the weak forms for the approximate
 solutions $(\n^\de,u^\de,\te^\de),$ then let $\de=\de_j$ and
 take appropriate limits. Standard arguments as well as
   (\ref{vu4})--(\ref{vu18})
      thus yield that the limit $(\n,u,\te)$ is
   a weak solution  of   (\ref{a0})    (\ref{h1})    (\ref{hh2})
    in the sense of Definition \ref{def} and satisfies (\ref{hq1})--(\ref{hq4}) except $\n-1\in C([0,\infty),L^2)$
    which in fact can be obtained by similar arguments leading to (\ref{sq4}).  In addition, the estimates (\ref{hq5})--(\ref{hq8}) follows direct from  (\ref{a2.12}), (\ref{vu15}),  (\ref{z1}),  and (\ref{vu4})--(\ref{vu18}).

It remains to prove (\ref{vu1})--(\ref{vu18}) since (\ref{vu4}) is a
direct consequence of  (\ref{z1}). It follows from (\ref{a2.12}),
(\ref{z1}), and $(\ref{a1})_1$ that \bnn \sup_{t\in[0,
\infty)}\|\n^\de_t\|_{H^{-1}(\r^3)}\le C,\enn which as well as
(\ref{z1}), \cite[Lemma C.1]{L2}, and  the Aubin-Lions lemma   yields
that there exists a subsequence of  $\de_j \rightarrow 0,$  still
denoted by  $\de_j,$  such that (\ref{vu1})  holds.
  Moreover, one deduces from (\ref{vu15}) that (extract a subsequence) \bnn \n^{\de_j}-1\rightharpoonup \n-1,\,\, \na u^{\de_j}\rightharpoonup \na u\,\, \mbox{ weakly in }\,\, L^4(\r^3\times (1,\infty)),\enn with $\n-1$ and $\na u$ satisfying \be\la{vu016} \int_1^\infty \left(\|\n-1\|_{L^4}^4+\|\na u\|_{L^4}^4\right)dt\le C.\ee

Then, simple calculations together with  (\ref{z1}) yield  that, for any $0<T<\infty,$ there exists some $C(T)$ independent of $\de$ and $\eta$ such that \be \la{vu6}\|(\n^\de u^\de)_t\|_{L^2(0,T;H^{-1}(\r^3))}+\|(\n^\de \te^\de)_t\|_{L^2(0,T;H^{-1}(\r^3))}\le C(T),\ee which together with (\ref{z1}),
  (\ref{vu1}),   and (\ref{vu4}) gives (\ref{vu5}).

Next, to prove (\ref{vu26}), one deduces from (\ref{z1}) and $(\ref{a1})_1$ that, for any $\zeta\in H^1(\r^3),$
\bnn\ba &\left|\int (\n^\de |u^\de|^2)_t\zeta dx\right|\\ &=\left|-\int  \div(\n^\de u^\de) |u^\de|^2\zeta dx+2\int  \n^\de  u^\de\cdot u^\de_t  \zeta dx\right|\\ &=\left| \int   \n^\de u^\de\cdot\na( |u^\de|^2\zeta) dx+2\int  \n^\de  u^\de\cdot(\dot u^\de-u^\de\cdot\na u^\de)  \zeta dx\right|\\ &\le C \int   \n^\de |u^\de|^3  |\na\zeta| dx +C \int   \n^\de |u^\de|^2 |\na u^\de|| \zeta| dx+ C  \int  \n^\de  |u^\de| |\dot u^\de ||  \zeta| dx \\ &\le C \| u^\de\|_{L^6}^3\|\na\zeta\|_{L^2}+C \| u^\de\|_{L^6}^2\|\na u^\de\|_{L^2}\|\zeta\|_{L^6}+C\|u^\de\|_{L^6}\|(\n^\de)^{1/2} \dot u^\de\|_{L^2}\|\zeta\|_{L^3}\\& \le C\left(\|\na u^\de\|_{L^2}+\|(\n^\de)^{1/2}\dot u^\de\|_{L^2}\right)\|\zeta\|_{H^1},\ea\enn
which together with (\ref{z1}) gives
\be  \la{vu24}\int_0^\infty \|(\n^\de |u^\de|^2)_t\|^2_{H^{-1}}dt\le C.\ee
It follows  from (\ref{z1}) that \bnn  \sup_{t\in [0,\infty)}\|\n^\de |u^\de|^2\|_{L^1\cap L^3}\le C,\enn
which combining with (\ref{vu24}),  (\ref{vu4}),   and (\ref{vu5}) yields  (\ref{vu26}).

Finally, we prove (\ref{vu18}) which implies the strong limits of $u^\de $  and $\te^\de.$   We deduce  from  (\ref{z1}), (\ref{h19}),   (\ref{vu6}),  and (\ref{vu15})   that
\be  \la{vu16}\sup\limits_{t\in [0,\infty)} (\|  u^\de\|_{H^1}+\si^2\|  G^\de\|_{H^1}+\si^2\| \o^\de\|_{H^1}+\si^2\|\na \te^\de\|_{H^1}) \le C,\ee and  \be \la{vu17} \int_{\tau}^T\left(\|u^\de_t\|_{L^2(\r^3)}^2+\|G^\de_t\|_{H^{-1}(\r^3)}^2+
\|\o^\de_t\|_{H^{-1}(\r^3)}^2+\|\te^\de_t\|_{H^1(\r^3)}^2\right)dt\le C(\tau,T), \ee  for all $0<\tau<T<\infty.$ The Aubin-Lions lemma together with (\ref{vu16}) and  (\ref{vu17}) thus gives  (\ref{vu18}).

{\it Step 3. Proofs of (\ref{vu019}) and (\ref{lar1}).}
We first prove that $(\n,u,\te)$ satisfies (\ref{vu019}).  We rewrite the energy equation $(\ref{a1})_3$ in the form
\be   \la{vu20}\ba
&\frac{R}{\ga-1}\left((\n    \te)_t+\div(\n u\te)\right)-\ka\Delta \te \\&= G\div u -R\div u+2\mu\div(u\cdot\na u-u\div u)+\frac{\mu}{2}|\o|^2 .\ea\ee
Thus, for any $\varphi\in \mathcal{D}(\r^3\times (0,\infty)),$ we have
\be\la{vu22}\ba &\frac{R}{\ga-1}\int_0^\infty\int\n^\de\te^\de\left(\varphi_t+u^\de\cdot\na\varphi
\right)dxdt-\ka \int_0^\infty\int \na\te^\de\cdot\na\varphi dxdt\\ &=-\int_0^\infty\int G^\de\div u^\de\varphi dxdt + R \int_0^\infty\int\div u^{\de}\varphi dxdt\\&\quad +2\mu \int_0^\infty\int \left(u^\de\cdot\na u^\de -u^\de\div u^\de\right)\cdot \na \varphi dxdt -\frac{\mu}{2}\int_0^\infty\int |\o^\de|^2 \varphi dxdt.\ea\ee
Letting $\de=\de_j $ in  (\ref{vu22}) and taking appropriate limits, we thus deduce  from  (\ref{vu1}), (\ref{vu4}),  (\ref{vu5}), and  (\ref{vu18}) that
\be\la{vu30}\ba &\frac{R}{\ga-1}\int_0^\infty\int\n \te \left(\varphi_t+u \cdot\na\varphi
\right)dxdt-\ka \int_0^\infty\int \na\te \cdot\na\varphi dxdt\\ &=-\int_0^\infty\int G \div u \varphi dxdt +R \int_0^\infty\int \div u\varphi dxdt\\&\quad+2\mu \int_0^\infty\int \left(u \cdot\na u  -u \div u \right)\cdot \na \varphi dxdt -\frac{\mu}{2}\int_0^\infty\int |\o |^2 \varphi dxdt
\\ &=-\int_0^\infty\int (\lambda\div u-P) \div u \varphi dxdt-2\mu \int_0^\infty\int |\mathfrak{D}(u)|^2  \varphi dxdt\\&\quad+2\mu \int_0^\infty\int \left(\pa_ku^i\pa_i(u^k\varphi)  -\div(u\varphi)  \div u \right)    dxdt \\ &= \int_0^\infty\int P  \div u \varphi dxdt-\int_0^\infty\int \left(\lambda(\div u)^2+2\mu   |\mathfrak{D}(u)|^2  \right) \varphi dxdt,\ea\ee
where in the last equality, we have used    the following simple fact that, for   standard mollifier $j_\nu(x),$
\bnn\la{vu31}\ba &\left|\int_0^\infty\int \left(\pa_ku^i\pa_i(u^k\varphi)  -\div(u\varphi)  \div u \right)    dxdt\right|\\&=\left|\int_0^\infty\int \pa_k(u^i-u^i*j_\nu) \pa_i\left(  u^k \varphi\right)  dxdt\right. \\&\quad+\left.\int_0^\infty\int \left(\pa_k(u^i*j_\nu)\pa_i( u^k \varphi)  -\div(u\varphi)  \div u \right)    dxdt\right|\\&=\left|\int_0^\infty\int\left( \pa_k(u^i-u^i*j_\nu) \pa_i (  u^k \varphi )+\div(u\varphi)\div( u*j_\nu-u) \right)    dxdt\right|\\&\le C\int_0^\infty\int |\na (u\varphi) ||\na(u -u *j_\nu)|    dxdt\rightarrow 0, \mbox{ as }\nu\rightarrow 0, \ea\enn
 due to (\ref{hq7}).  We thus derive (\ref{vu019}) directly from (\ref{vu30}),   (\ref{vu5}), and (\ref{vu1}).

Finally, to finish the proof of Theorem \ref{th2},  it remains to prove (\ref{lar1}). Since $(\n,u)$ satisfies  (\ref{def1}),  for the standard mollifier $j_\nu(x) (\nu>0),$ $\n^\nu\triangleq \n*j_\nu$ satisfies
\be \la{vu39} \begin{cases}\n^\nu_t+\div(u\n^\nu)=r_\nu ,\\ \n^\nu(x,t=0)=\n_0*j_\nu, \end{cases}\ee
where $r_\nu $  satisfies, for any $T>0,$\be \la{vu40} \lim_{\nu\rightarrow 0^+} \int_0^T\|r_\nu\|_{L^2}^2dt=0,\ee due to \eqref{a2.12}, \eqref{z1}, and \cite[Lemma 2.3]{L2}.    Multiplying (\ref{vu39}) by $4(\n^\nu-1)^3,$ we obtain after integration by parts that, for $t\ge 1,$\be \la{vu43}\ba&(\|\n^\nu-1\|_{L^4}^4)' \\&=-4\int(\n^\nu-1)^3\div udx -3\int(\n^\nu-1)^4\div udx +4\int r_\nu(\n^\nu-1)^3dx \\ &\le C\|\n^\nu-1\|_{L^4}^4 +C\|\na u\|_{L^4}^4  + C\|r_\nu\|_{L^2}  ,\ea\ee
 which implies that, for all $1\le N\le s\le N+1\le t\le N+2,$ \be\la{vu55}\ba \|\n^\nu(\cdot,t)-1\|_{L^4}^4\le&   \|\n^\nu(\cdot,s)-1\|_{L^4}^4+C\int_N^{N+2} \left(\|\n^\nu-1\|_{L^4}^4 + \|\na u\|_{L^4}^4\right)dt\\&+C\int_N^{N+2} \|r_\nu\|_{L^2} dt.\ea \ee
 Letting $\nu\rightarrow 0^+$ in (\ref{vu55}) together with (\ref{vu40}) and  (\ref{hq1})  yields that\be\la{vu56}\ba \|\n (\cdot,t)-1\|_{L^4}^4\le&   \|\n (\cdot,s)-1\|_{L^4}^4+C\int_N^{N+2} \left(\|\n -1\|_{L^4}^4 + \|\na u\|_{L^4}^4\right)dt .\ea \ee
 Integrating (\ref{vu56}) with respect to $s$ over $[N,N+1]$ leads to \be\la{vu57}\ba \sup_{t\in [N+1,N+2]}\|\n (\cdot,t)-1\|_{L^4}^4&  \le C\int_N^{N+2} \left(\|\n -1\|_{L^4}^4 + \|\na u\|_{L^4}^4\right)dt \\& \rightarrow 0,\mbox{ as }N\rightarrow \infty,\ea \ee  due to
 (\ref{vu016}). This together with (\ref{hq4}) and  (\ref{hq7}) implies that, for
all $p\in (2,\infty),$
\be\la{hq1115} \lim_{t\rightarrow\infty}\int |\rho-1|^p dx = 0. \ee Finally, we will prove \be \la{vu36} \lim_{t\rightarrow \infty}\left(\|u\|_{L^4}+\|\na\te\|_{L^2}\right)=0,\ee
which combining with (\ref{hq1115}), (\ref{hq4}),  (\ref{hq5})--(\ref{hq8}),  and the Gagliardo-Nirenberg inequality thus gives (\ref{lar1}).
In fact, one deduces from (\ref{hq5})--(\ref{hq8}) that \be \la{vu32} \ba &\int_1^\infty\left( \|u\|_{L^4}^4+ \|\na\te\|_{L^2}^2 \right)dt \\ &\le C \int_1^\infty \|u\|_{L^2}\|\na u\|_{L^2}^3dt+\int_1^\infty \|\na \te\|_{L^2}^2dt\\ &\le C,\ea\ee
\be\la{vu33} \ba\int_1^\infty\left|\frac{d}{dt}\left(\|u(\cdot,t)\|_{L^4}^4\right)
\right|dt&= 4\int_1^\infty\left| \int |u|^2u\cdot u_tdx
\right|dt\\ &\le C\int_1^\infty \|u\|_{L^\infty}
\|u\|_{L^4}^2\|u_t\|_{L^2}dt\\ &\le C,\ea\ee
and
\be \la{vu34}\ba \int_1^\infty\left|\frac{d}{dt}\left(\|\na\te(\cdot,t)\|_{L^2}^2\right)
\right|dt&= 2\int_1^\infty\left| \int \na\te\cdot \na\te_tdx
\right|dt\\ &\le C\int_1^\infty
\|\na\te\|_{L^2} \|\na \te_t\|_{L^2}dt\\ &\le C.\ea\ee
Thus, we derive  (\ref{vu36}) easily   from  (\ref{vu32})--(\ref{vu34}).
The proof of Theorem \ref{th2} is finished.



\begin {thebibliography} {99}

\bibitem{akm} S. N. Antontsev,  A. V. Kazhikhov,  V. N. Monakhov, {\it  Boundary value problems in mechanics of nonhomogeneous fluids,} North-Holland Publishing Co., Amsterdam, 1990.


\bibitem{bkm} J. T. Beale,  T. Kato, A. Majda,
Remarks on the breakdown of smooth solutions for the 3-D Euler
equations, {\it Commun. Math. Phys.} {\bf 94} (1984), 61--66.

\bibitem{bd}D. Bresch,  B. Desjardins, On the existence of global weak solutions to the Navier-Stokes equations for viscous compressible and heat conducting fluids.{\it J. Math. Pures Appl. (9) }{\bf 87} (2007), 57--90.

\bibitem{cj} Y. Cho, B.J. Jin,   Blow-up of viscous heat-conducting
compressible flows, {\it J. Math. Anal. Appl. } {\bf 320} (2006),
 819--826.


\bibitem{K3} Y. Cho,  H.  Kim,
On classical solutions of the compressible Navier-Stokes equations
with nonnegative initial densities.{ \it Manuscript Math. }{ \bf120} (2006), 91--129.

 \bibitem{choe1}Y. Cho, H. Kim, Existence results for viscous polytropic fluids with vacuum. {\it J. Differential Equations  } {\bf 228} (2006), 377--411.



\bibitem{feireisl1}E. Feireisl, Dynamics of Viscous Compressible Fluids, Oxford Science Publication, Oxford, 2004.


\bibitem{feireisl} E. Feireisl, On the motion of a viscous, compressible, and heat conducting fluid, {\it Indiana Univ. Math. J.} {\bf 53} (2004), 1707--1740.


\bibitem{F1} E. Feireisl,  A. Novotny, H. Petzeltov\'{a},  On the existence of globally defined weak solutions to the
Navier-Stokes equations. {\it J. Math. Fluid Mech.} {\bf 3}  (2001), 358--392.




\bibitem{Hof1} D. Hoff,  Discontinuous solutions of the Navier-Stokes equations
for multidimensional flows of heat-conducting fluids. {\it Arch.
Rational Mech. Anal.}  {\bf 139} (1997), 303--354.



\bibitem{hlx4} X. D. Huang,  J. Li,
Global
  classical and weak solutions to  the  three-dimensional
   full compressible  Navier-Stokes system with vacuum on bounded domains. In preparation.


\bibitem{hlx} X. D. Huang, J. Li, Z. P. Xin,
Serrin type criterion for the three-dimensional compressible flows. {\it  Siam J. Math. Anal.} in press.




\bibitem{hulx} X. D. Huang, J. Li, Z. P. Xin,  Global well-posedness of classical solutions with large
oscillations and vacuum to the three-dimensional isentropic
compressible Navier-Stokes equations, {\it Comm. Pure Appl. Math.} in press.


\bibitem{kazh01} A. V. Kazhikhov,   Cauchy problem for viscous gas equations, {\it Siberian Math. J.} {\bf 23} (1982),   44--49.

\bibitem{Kaz} A. V. Kazhikhov, V. V.  Shelukhin,
Unique global solution with respect to time of initial-boundary
value problems for one-dimensional equations of a viscous gas,
{\it J. Appl. Math. Mech.  } {\bf 41} (1977), 273--282.






\bibitem{L2} P. L. Lions,  \emph{Mathematical topics in fluid
mechanics. Vol. {\bf 1}. Incompressible models,}  Oxford
University Press, New York, 1996.

\bibitem{L1} P. L. Lions,  \emph{Mathematical topics in fluid
mechanics. Vol. {\bf 2}. Compressible models,}  Oxford
University Press, New York,   1998.


\bibitem{M1} A. Matsumura, T.   Nishida,   The initial value problem for the equations of motion of viscous and heat-conductive
gases, {\it J. Math. Kyoto Univ. }{\bf 20}   (1980), 67--104.



\bibitem{Na} J. Nash,  Le probl\`{e}me de Cauchy pour les \'{e}quations
diff\'{e}rentielles d'un fluide g\'{e}n\'{e}ral,{\it Bull. Soc. Math.
France.} {\bf 90} (1962), 487--497.

\bibitem{nir} L. Nirenberg, On elliptic partial differential equations,  {\it Ann. Scuola Norm. Sup. Pisa (3)} {\bf 13 } (1959), 115--162.


\bibitem{R} O. Rozanova,   Blow up of smooth solutions to the compressible
Navier--Stokes equations with the data highly decreasing at
infinity, {\it J. Differ. Eqs.}  {\bf 245} (2008),  1762--1774.


\bibitem{se1} J. Serrin, On the uniqueness of compressible fluid motion, {\it
Arch. Rational. Mech. Anal.} {\bf 3 }(1959), 271--288.



\bibitem{X1} Z. P. Xin,
Blowup of smooth solutions to the compressible {N}avier-{S}tokes
equation with compact density. {\it Comm. Pure Appl. Math. }   {\bf 51} (1998),
229--240.



\end {thebibliography}
\end{document}